\documentclass[twocolumn]{aastex631}

\begin{document}
\title{Two-Dimensional Models of Microphysical Clouds on Hot Jupiters I: \\ Cloud Properties}

\author[0000-0002-4250-0957]{Diana Powell}
\affiliation{Department of Astronomy \& Astrophysics, University of Chicago, Chicago, IL 60637, USA}

\author[0000-0002-8706-6963]{Xi Zhang}
\affiliation{Department of Earth and Planetary Sciences, University of California, Santa Cruz, CA 95064, USA}

\begin{abstract}
We present a new two-dimensional, bin-scheme microphysical model of cloud formation in the atmospheres of hot Jupiters that includes the effects of longitudinal gas and cloud transport. We predict cloud particle size distributions as a function of planetary longitude and atmospheric height for a grid of hot Jupiters with equilibrium temperatures ranging from 1000-2100 K. The predicted 2D cloud distributions vary significantly from models that do not consider horizontal cloud transport and we discuss the microphysical and transport timescales that give rise to the differences in 2D versus 1D models. We find that the horizontal advection of cloud particles increases the cloud formation efficiency for nearly all cloud species and homogenizes cloud distributions across the planets in our model grid. In 2D models, certain cloud species are able to be transported and survive on the daysides of hot Jupiters in cases where 1D models would not predict the existence of clouds. We demonstrate that the depletion of condensible gas species varies as a function of longitude and atmospheric height across the planet, which impacts the resultant gas-phase chemistry. Finally, we discuss various model sensitivities including the sensitivity of cloud properties to microphysical parameters, which we find to be substantially less than the sensitivity to the atmospheric thermal structure and horizontal and vertical transport of condensible material. 
\end{abstract}


\section{Introduction}
Clouds are common in the atmospheres of exoplanets, where they are often the dominant source of atmospheric opacity such that they regulate the planetary climate and significantly impact observations and inferred planetary properties \citep[e.g.,][]{pruppyklett,rossow1978}. When clouds are observable they both obfuscate signatures of atmospheric properties such as the abundance of gaseous species \citep[e.g.,][]{sing2013, kreidberg2014clouds} while simultaneously having the potential to serve as important tracers of atmospheric physics such as mixing and transport. Recent observations have demonstrated that even planets that were previously thought to be cloud-free \citep{Wakeford2017} have atmospheres that are shaped by the detailed properties of clouds \citep{Feinstein2022}. Furthermore, even when clouds are present but not readily observable, they still regulate planetary climate through regulating radiative and chemical processes, and they alter the observed abundances of gas-phase species via gas-cloud interactions. It is therefore necessary to understand clouds when we are interpreting planetary atmospheres. 

Clouds are highly sensitive to background atmospheric conditions such as the temperature structure and the dynamical circulation \citep[e.g.,][]{2018ApJ...860...18P}. The study of how cloud properties are impacted by atmospheric conditions is cloud microphysics. Cloud microphysical studies have been conducted for a broad range of exoplanet, substellar, and planetary atmospheres \citep[e.g.,][]{2003Icar..162...94B,2004A&A...423..657H,gao2014bimodal,helling2008dust,2022ApJ...927..184M,2018ApJ...855...86G,2018ApJ...860...18P, 2015A&A...580A..12L,ohno2018,2008A&A...485..547H}. However, the most well-studied exoplanetary atmospheres to-date are hot Jupiters, which are gas-giant planets in short period orbits that are typically tidally-locked. 

Tidally-locked hot Jupiters have permanent daysides and extreme insolation gradients across the planet. The insolation gradients drive super-rotating equatorial jets that operate to reduce the day-night temperature contrast but do so less efficiently at hotter temperatures where heat re-radiation becomes more efficient \citep[e.g.,][]{komacek2016}. These planets are thus inherently 3D objects with significant differences in cloud properties across the planetary globe. Furthermore, atmospheric inhomogeneity, in which spatially non-uniform clouds play a significant role, has been shown to substantially modulate the interior cooling and planetary evolution of hot Jupiters \citep{zhang2023}.

Hot Jupiter atmospheres are complex 3D structures with conditions that vary significantly across the planet and with large-scale atmospheric flows that transport material throughout the atmosphere. However, many previous cloud studies have thus far been spatially one-dimensional (1D). These previous studies have either focused on globally averaged atmospheric conditions when modeling a large grid of hot Jupiter atmospheres \citep[e.g.,][]{sing2013,2020NatAs...4..951G}, or on simulating the vertical cloud properties for individual regions such as the east/west limbs or each longitude/latitude grid on the planetary surface \citep[e.g.,][]{2018ApJ...860...18P,2019ApJ...887..170P,Helling2022,2021ApJ...918L...7G,2019arXiv190608127H}. The latter approach enables an approximation of the atmospheric spatial diversity on a given hot Jupiter, however, it neglects the interaction between clouds and the large scale atmospheric flows such as the super-rotating equatorial jet. 

Several works have directly coupled simple cloud schemes with large-scale general circulation models (GCMs) to capture the full 3D nature of cloud formation. Due to computational limitations, GCMs that include the impact of clouds for large grids of models require simplifying cloud assumptions that neglect microphysical processes \citep[e.g.,][]{parmentier2016transitions,roman2019,komacek2022}. Similarly, a full coupling of microphysical models to GCMs across a large grid of models and over a period long enough to allow the models to reach a steady state(such that cloud and atmospheric properties are not dominated by initial conditions) is computationally challenging. Thus, it has only been done so far for individual hot Jupiters \citep{2016A&A...594A..48L,2018A&A...615A..97L,lee2023}. 

In this work, we adopt a flexible, computationally-inexpensive approach to simulate the interaction between cloud formation and both vertical and horizontal atmospheric dynamics along the equatorial regions of hot Jupiters. We use an input atmospheric structure from a grid of GCM models and evolve our bin-scheme cloud microphysical model as clouds are transported across the atmosphere on the super-rotating equatorial jet. In Section \ref{themodel} we describe the new 2D-ExoCARMA cloud microphysical model. In Section \ref{cloudprops} we present the cloud properties for our nominal model grid of hot Jupiters with equilibrium temperatures ranging from 1000 - 2100 K. In Section \ref{horiz_advec} we detail the importance of considering horizontal advection in studies of cloud formation on hot Jupiters by comparing our 2D models to 1D models for the same grid of hot Jupiter atmospheres. We present the longitudinally and vertically varying depletion of condensible gases in Section \ref{deplete}. We discuss our results in Section \ref{discuss} and present our conclusions in Section \ref{conclude}.

\section{The 2D-ExoCARMA Model}\label{themodel}

In this work, we adapt the 1D CARMA model used in \citet{2019ApJ...887..170P} and \citet{2020NatAs...4..951G} to simulate the clouds on Hot Jupiters in both vertical and longitudinal dimensions: 2D-ExoCARMA. 2D-ExoCARMA computes vertical and size distributions of aerosol particles along the planetary equator as a function of planetary longitude. 

\subsection{Base Microphysical Cloud Model}
The formation and evolution of clouds is governed by microphysical processes that depend sensitively on planetary properties, such as a planet's thermal structure, chemical composition, gravity, and the horizontal and vertical mixing \citep[e.g.,][]{2015AA...575A..11L,2015A&A...580A..12L,2017Icar..287..116G,2018ApJ...863..165G,2016MNRAS.460..855H,2018ApJ...860...18P,2019ApJ...887..170P}. We model the formation and evolution of condensible clouds in the atmospheres of hot Jupiters using an adapted version of the nonequilibrium Community Aerosol and Radiation Model for Atmospheres \citep[CARMA;][]{turco1979,1988JAtS...45.2123T} version 3.0 \citep{JGRD:JGRD14488,JGRD:JGRD15781}. CARMA was originally developed to study cloud formation on Earth and has since been applied to a diverse array of solar system and extrasolar objects where it has been used to successfully explain a diverse array of observational phenomena \citep[e.g.,][]{1993Icar..102..261M,1999JGR...104.9043C,gao2014bimodal,2003Icar..162...94B,2004GeoRL..3117S07B,2006Icar..182..230B,2017Icar..287..116G,2020NatAs...4..951G}. CARMA has previously been adapted to the study of exoplanets and substellar objects \citep{2018ApJ...860...18P,2019ApJ...887..170P,2018ApJ...863..165G,2018ApJ...855...86G,2021ApJ...918L...7G,2022ApJ...927..184M} as well as for the study of protoplanetary disks \citep{2022NatAs...6.1147P}. CARMA now includes the formation of several cloud species that are relevant in different environments, including: H$_2$O, CO, sulfuric acid, polysulfur species, Na$_2$S, KCl, ZnS, MnS, TiO$_2$, Fe, Cr, Al$_2$O$_3$, MgSiO$_3$, and Mg$_2$SiO$_4$. A more complete history of CARMA is described in several previous works \citep[e.g.,][]{2018ApJ...855...86G}. 

Except for the inclusion of horizontal advection (see Section \ref{advec}), the cloud formation in this work uses the same microphysical setup as that described in \citet{2019ApJ...887..170P,2020NatAs...4..951G,2021ApJ...918L...7G}. We consider the homogeneous nucleation of clouds comprised of KCl, TiO$_2$, Fe, and Cr. Several cloud species are also able to heterogeneously nucleate on TiO$_2$ cloud condensation nuclei (CCN). The species that can heterogeneously nucleate on TiO$_2$ are Na$_2$S, MnS, Mg$_2$SiO$_4$, Fe, Cr, and Al$_2$O$_3$. Finally, we allow ZnS clouds to heterogeneously nucleate on KCl CCN. As in previous work, we only consider mixed cloud compositions of species that heterogeneously nucleate on TiO$_2$ and KCl. Otherwise, the various condensing species do not interact with each other via condensation or coagulation. For coagulation, this assumption is likely reasonable as the bulk of the cloud particle evolution is dominated in our modeling by nucleation and condensation \citep[e.g.,][]{2018ApJ...860...18P}. A detailed exploration of co-nucleation and co-condensation is outside the scope of this paper and will be addressed in future work. We note that the species considered is likely not an exhaustive list of species that could condense in hot Jupiter atmospheres and other species, such as SiO$_2$ may be particularly important \citep{grant2023}. Future work that explores the microphysical properties and the impact of horizontal advection on a variety of additional cloud species, though beyond the scope of this work, would be valuable. Due to a lack of experimental data, two key microphysical parameters remain largely unconstrained: the desorption energy of the condensate molecule on the surface of the CCN, and the contact angle between the condensate species and the CCN. Following \citet{2020NatAs...4..951G,2021ApJ...918L...7G}, for our nominal case we assume a desorption energy of 0.5 eV and a contact angle given by $\cos \theta_c = \sigma_C/\sigma_x$ where $\theta_c$ is the contact angle, $\sigma_C$ is the surface energy of the CCN, and $\sigma_x$ is the condensate surface energy. For the reasoning behind these choices and the values for the different species' surface energies (as well as the rest of their material properties) see \citet{2020NatAs...4..951G}. We test the sensitivity to these parameters in Section \ref{micro_sense}. 

The model is initialized with a solar composition of all species of condensate vapor at the bottom boundary of the model with the exception of TiO$_2$ and KCl. For TiO$_2$ and KCl there are other atmospheric species present in significant abundance that also utilize atmospheric Ti and K, namely TiO and KOH. We thus calculate the mixing ratios for TiO$_2$ and KCl using the thermochemical equilibrium model GGChem \citep{2018A&A...614A...1W}. The initial abundance of cloud particles is zero. The minimum particle radius for the cloud species that homogeneously nucleate is $10^{-4}$ $\mu$m while all other species have a minimum particle radius of $1.26\times 10^{-4}$ $\mu$m (these bin offsets avoid numerical instability when calculating the cloud particle size distributions). There are 80 particle size bins where each bin is a factor of two larger in particle mass. The gaseous species diffuse upwards in the atmosphere until they reach a height in the atmosphere where they are sufficiently supersaturated such that clouds can form via nucleation. The cloud formation and evolution processes that are considered are: nucleation, condensational growth, evaporation, coagulation, settling, mixing, and horizontal advection (see Section \ref{advec}). The top boundary condition for both the gas and clouds is zero flux, the bottom boundary condition for the condensible gases is set to their initialized abundance while the bottom boundary condition for the cloud particles is set to a zero mixing ratio. 

\subsection{Adaptations for Two Spatial Dimensions -- Horizontal Advection}\label{advec}

We adapt CARMA to multiple dimensions in a similar way as the framework pioneered for modeling non-equilibrium kinetic chemistry in \citet{2012A&A...548A..73A,2014A&A...564A..73A} known as pseudo-2D chemical modeling. This approach is readily applicable to the class of hot Jupiter exoplanets as the dominant atmospheric flow along the equator is due to stable equatorial zonal jets, which dominate the observable region of the planetary atmosphere \citep[e.g.,][]{showman-etal-2010}. In this approach, the entire atmospheric column is advected to simulate the rotation of the column across the planet at a constant rate controlled by the equatorial zonal wind speed. While the speed of the equatorial jet is thought to vary with altitude such that a column would not be transported at a uniform rate, this scenario can be used as a first-order test of the formation and evolution of clouds in the case of significant horizontal advection and may provide an appropriate description of clouds in the observable regions of the atmosphere where advection is particularly important. Furthermore, the inclusion of horizontal advection in this form allows for the efficient inclusion of a physical transport process that has been ignored in previous 1D microphysical cloud models. Understanding the importance of horizontal advection in determining cloud properties is one of the primary goals of this study. 

In this 2D microphysical framework, as the CARMA Column is advected around the planet on the equatorial zonal jet, the background atmospheric thermal structure is also varied to account for variations in temperature with longitude. We consider both vertical and horizontal transport, although the horizontal transport into or out of the vertical column is neglected. To better characterize the longitudinal variation of atmospheric temperature and wind on a constant pressure plane from the GCMs (as a hydrostatic atmosphere behaves more like an incompressible flow in the pressure coordinate), we altered the vertical coordinate system from the altitude coordinate in the original 1D CARMA model to the log-pressure coordinate\citep[see e.g.,][]{1988JAtS...45.2123T}. The fundamental microphysical equations were adjusted self-consistently to account for this coordinate transformation following the conversions described in \citet{1988JAtS...45.2123T}.

\begin{figure*} 
   \centering
   \includegraphics[width=0.89\textwidth]{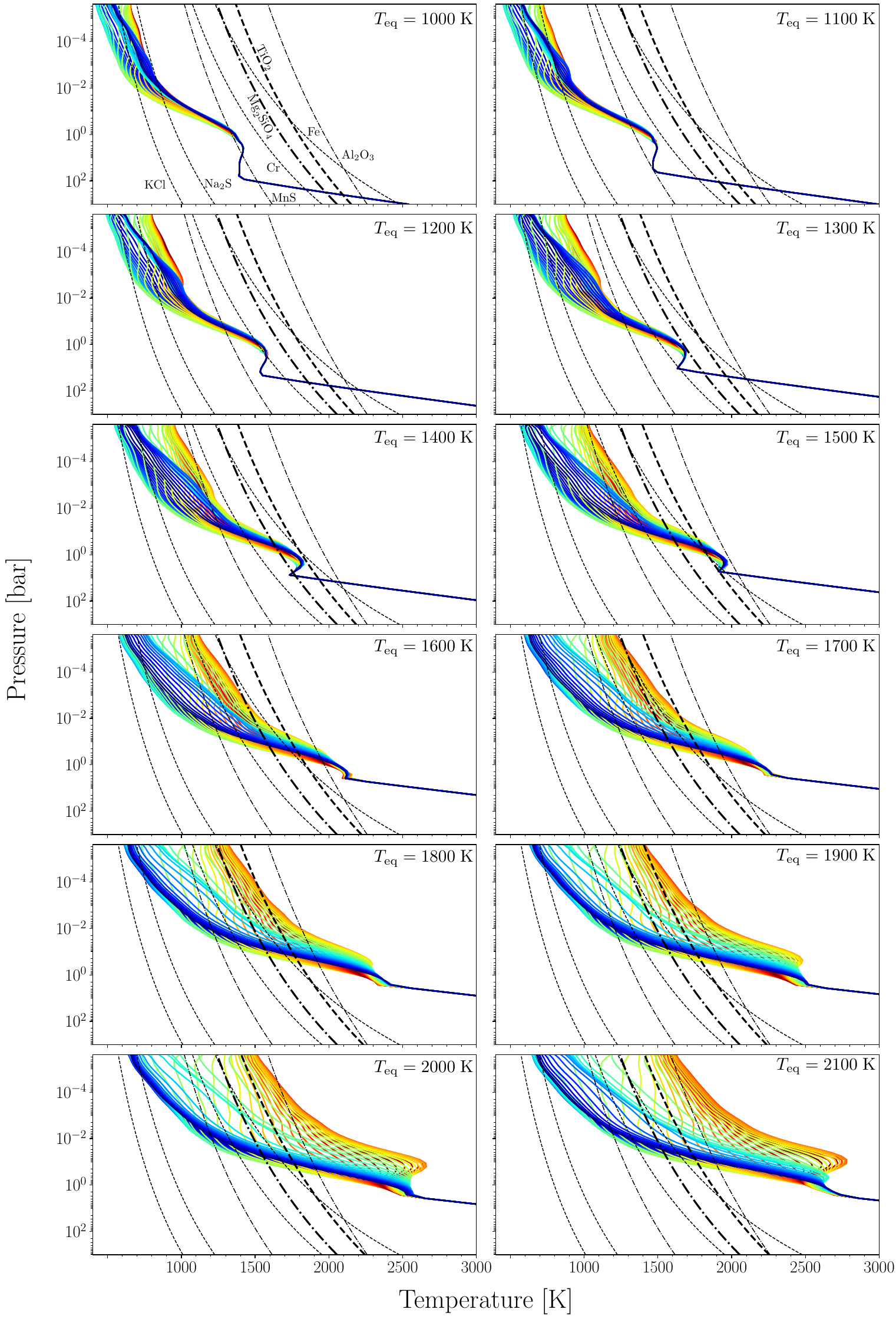}
   \caption{Temperature profiles across the equator of 12 Hot Jupiters. Dark blue colors represent the pressure/temperature profiles on the nightside, light-blue is on the western limb, red is on the dayside, yellow is on the eastern limb. The condensation curves of various cloud species are plotted in dashed or dashed-dotted lines. The condensation curves for TiO$_2$, which serves as the primary CCN, as well as for Mg$_2$SiO$_4$, which typically dominates the cloud opacity, are in bold.}
   \label{fig:PTprofs}
\end{figure*}

While this 2D framework does not capture the full 3D dynamical behavior expected for exoplanet atmospheres, we base our model on the output from a 3D general circulation model so as to capture as much of the pertinent 3D atmospheric properties as possible. Based on the 3D studies of the atmospheres of hot Jupiters, we focus our efforts on the dominant atmospheric flow pattern that is thought to primarily shape atmospheric observations: the equatorial jet. 

\subsection{Simulated Hot Jupiter Atmosphere Structure}\label{atm_struc}

The 2D-ExoCARMA model calculates cloud and vapor distributions as a function of planetary longitude and atmospheric pressure. The inputs for the atmospheric structure necessary for the 2D-ExoCARMA model are the temperature and wind distributions from a 3D general circulation model. For this study, we consider the grid of Jupiter sized planets from \citet{2016ApJ...828...22P} calculated using the SPARC/MITgcm. All of the model planets have a gravity of $g = 1000$ cm s$^{-2}$ and orbit a solar type star. These planetary profiles are calculated using an identical cloud-free GCM model and only differ due to varying the orbital distances from the host star such that T$_\mathrm{eq}$ = 1000 - 2100 K at 100 K increments for a total of 12 model atmospheres. This regime of equilibrium temperatures may have atmospheres that are free of significant haze opacity in the near infrared \citep[e.g.,][]{2020NatAs...4..951G} such that clouds dominate the aerosol species present in their atmosphere. This sample is thus particularly well-suited to study using a version of 2D-ExoCARMA that only considers the formation and evolution of condensational clouds. 

As input to 2D-ExoCARMA, for each longitude point, we average the temperature-pressure (T-P) profiles from SPARC/MITgcm within $\pm$20$^{\circ}$ of latitude from the planetary equator. The 2D-ExoCARMA column is advected over 64 discrete longitudinal grid points that vary in temperature. The T-P profiles for the 12 hot Jupiters in this study are shown in Figure \ref{fig:PTprofs}. The T-P profiles from the SPARC/MITgcm are modified such that the planetary radiative-convective boundary (RCB) corresponds to those computed in \citet{2019ApJ...884L...6T}, as higher planetary internal temperatures are required to reproduce the observed inflated radius distribution of hot Jupiters. Below the RCB, we assume that the T-P profile is described by an adiabatic gradient for molecular hydrogen \citep[from][see their Equation (13)]{2015A&A...574A..35P}. We note that this choice of deep atmosphere structure likely introduces some inconsistencies in terms of the atmospheric dynamics as the deep atmosphere thermal structure was not directly simulated in the GCMs and can impact the flow in the upper atmosphere \citep{komacek2022}. This choice of deep atmosphere structure also limits the formation of atmospheric cold traps as shown in \citep{2018ApJ...860...18P}. The T-P profiles extend to 1000 bar and we model 59 vertical atmospheric layers.

We use the same globally averaged K$_{zz}$ profiles for these planets as described in \citet{2021ApJ...918L...7G} \citep[see also][]{2016ApJ...828...22P,2019ApJ...887..170P}. These profiles have K$_{zz}$ parameter magnitudes ranging from $\sim2\times10^{7} - 10^{10}$ cm$^2$ s$^{-1}$ at the bottom of our model domain to $\sim 10^{9} - 10^{10}$ cm$^2$ s$^{-1}$ at the bottom of our model domain. In our pseudo-2D approach, we assume a vertically-constant zonal jet speed. We take the average (again within $\pm$20$^{\circ}$ of latitude from the planetary equator) horizontal wind speed from SPARC/MITgcm calculated at the pressure of the silicate cloud base for a equatorially-averaged T/P profile. The values used for each simulated planet are given in Table \ref{horizontal_winds}. We note that in the vertical profiles of the average wind speeds in our input model the wind velocities decrease monotonically as pressure decreases. For some of our input models, the wind speeds at pressures lower than $10^{-4}$ bar are nearly constant, have low magnitudes, and are sometimes negative (i.e., a different rotation direction). Thus, it is an approximation to use a constant horizontal advection wind speed across the entire vertical column of the atmosphere. Given the small vertical wind speeds and occasional shift in wind direction in the upper atmosphere, future work that incorporates vertical variation in wind speed (i.e., a real 2D approach instead of pseudo-2D) may show the most significant difference in cloud properties in these upper atmospheric regions. As we demonstrate later in this work, the bulk of our cloud mass is located at higher atmospheric pressures such that the approximation of a constant horizontal wind speed with height is likely appropriate, at least to first order. 

\begin{deluxetable}{ll} [!htb]
\tablecolumns{2}
\tablecaption{Global Averaged Horizontal Wind Speeds at the TiO$_2$ Cloud Base \label{horizontal_winds}}
\tablehead{ 
 \colhead{Equilibrium Temperature} &
 \colhead{U [Km/s]} 
}
\startdata
1000 K & 1.0 \\
1100 K & 1.2 \\
1200 K & 1.5 \\
1300 K & 1.8 \\
1400 K & 1.9 \\
1500 K & 2.4 \\
1600 K & 2.7 \\
1700 K & 2.9 \\
1800 K & 3.2 \\
1900 K & 3.3 \\
2000 K & 3.4 \\
2100 K & 3.8 \\
\enddata 
\end{deluxetable}

\section{Cloud Properties}\label{cloudprops}

Using the 2D-ExoCARMA model described in Section \ref{themodel} we calculate the distribution and abundance of cloud particles as a function of planetary longitude and atmospheric pressure. Each simulation is run for $10^9$ seconds and the results shown here were time-averaged over the last 7 column rotation periods after the models have reached a steady-state. We note, however, that there is variability in the cloud properties as a function of time due to differences in the gas transport and microphysical cloud formation timescales. This variability occurs over a myriad of timescales and the amplitude of the variability is sensitive to many of the planetary properties (i.e., mixing, atmospheric composition). A detailed description of cloud variability will be addressed in future work.

\begin{figure*} 
   \centering
   \includegraphics[width=0.65\textwidth]{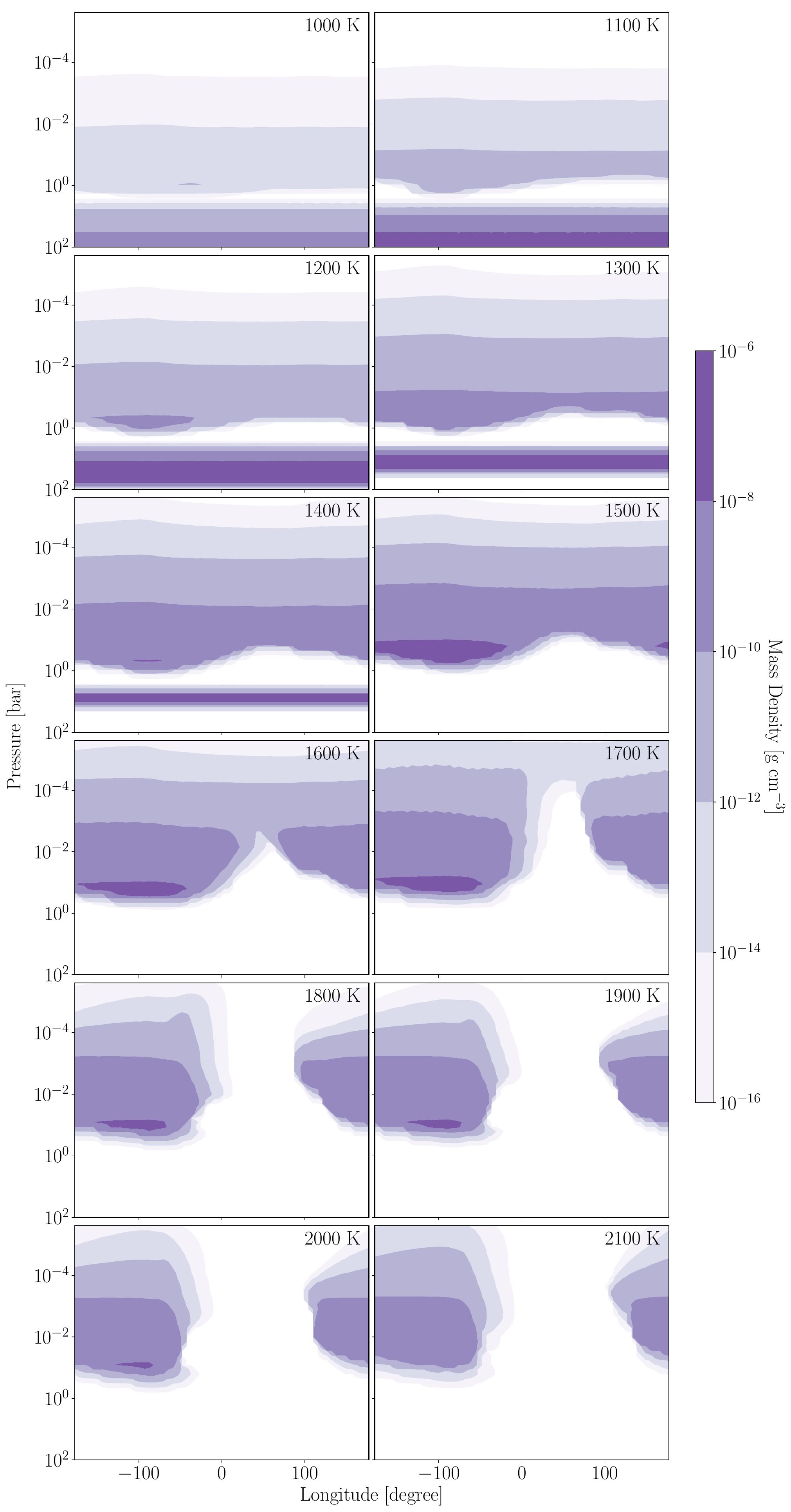}
   \caption{The distribution of condensed silicate (Mg$_2$SiO$_4$) cloud particles varies significantly as a function of planetary equilibrium temperature. The purple contours represent the time-averaged (over $\sim$7 column rotation periods) total condensed silicate cloud mass as a function of atmospheric pressure and planetary longitude where a longitude of 0$^o$ is the substellar point of the planet and 180$^o$ is the antistellar point. While 2D-ExoCARMA calculates the full cloud particle size distribution, here we have summed the cloud mass over all particle sizes.}
   \label{fig:nom_mg}
\end{figure*}

We first present the distribution of silicate bearing clouds, where each cloud particle is comprised of Mg$_2$SiO$_4$ coated TiO$_2$ particles. We initially focus on this population of clouds because they tend to dominate the cloud opacity \citep{2019ApJ...887..170P, 2020NatAs...4..951G, 2021ApJ...918L...7G} for the majority of planetary equilibrium temperatures probed in this study. The distribution of cloud mass for each planet across the range of equilibrium temperatures studied is shown in Figure \ref{fig:nom_mg}. 

At planetary temperatures of 1400 K or cooler, the majority of the cloud mass is located in the lower atmosphere at pressures higher than 1 bar. This is because cloud formation is particularly efficient at high temperatures and pressures when the condensible gas is supersaturated, as-is the case for these planets. For these same planets, there exists a region of the atmosphere where silicates are not sufficiently supersaturated (between $\sim$10 - 1 bar) for efficient nucleation and growth to occur and instead clouds undergo evaporation (for a detailed discussion of cloud formation timescales see Section \ref{horiz_advec}). In these atmospheres above 1 bar, enough gas is mixed to the highly supersaturated regions of the upper atmosphere such that a thin cloud deck also forms. As a result, a cloud-free gap forms at around $\sim$10 bar to $\sim$1 bar between the lower and upper cloud decks. These planets are generally covered in a roughly homogeneous layer of silicate clouds (see Section \ref{nom_distributions} for more detail), though we note that the cloud mass in the upper atmosphere above 1 bar is moderately more dense and located deeper in the atmosphere on the western limb than the eastern limb. We refer to the planets in our sample with equilibrium temperatures of 1400K or cooler, with deep atmospheres that are cool enough to form clouds, as planets in a ``hidden" high-cloud-mass regime. While the majority of the deep cloud mass in these planets is not observable, the clouds at depth have the potential to shape the radiative environment of the deep atmosphere and thus affect the planetary climate \citep[e.g.,][]{marley-etal-2002}. 

\begin{figure*} 
   \centering
   \includegraphics[width=0.65\textwidth]{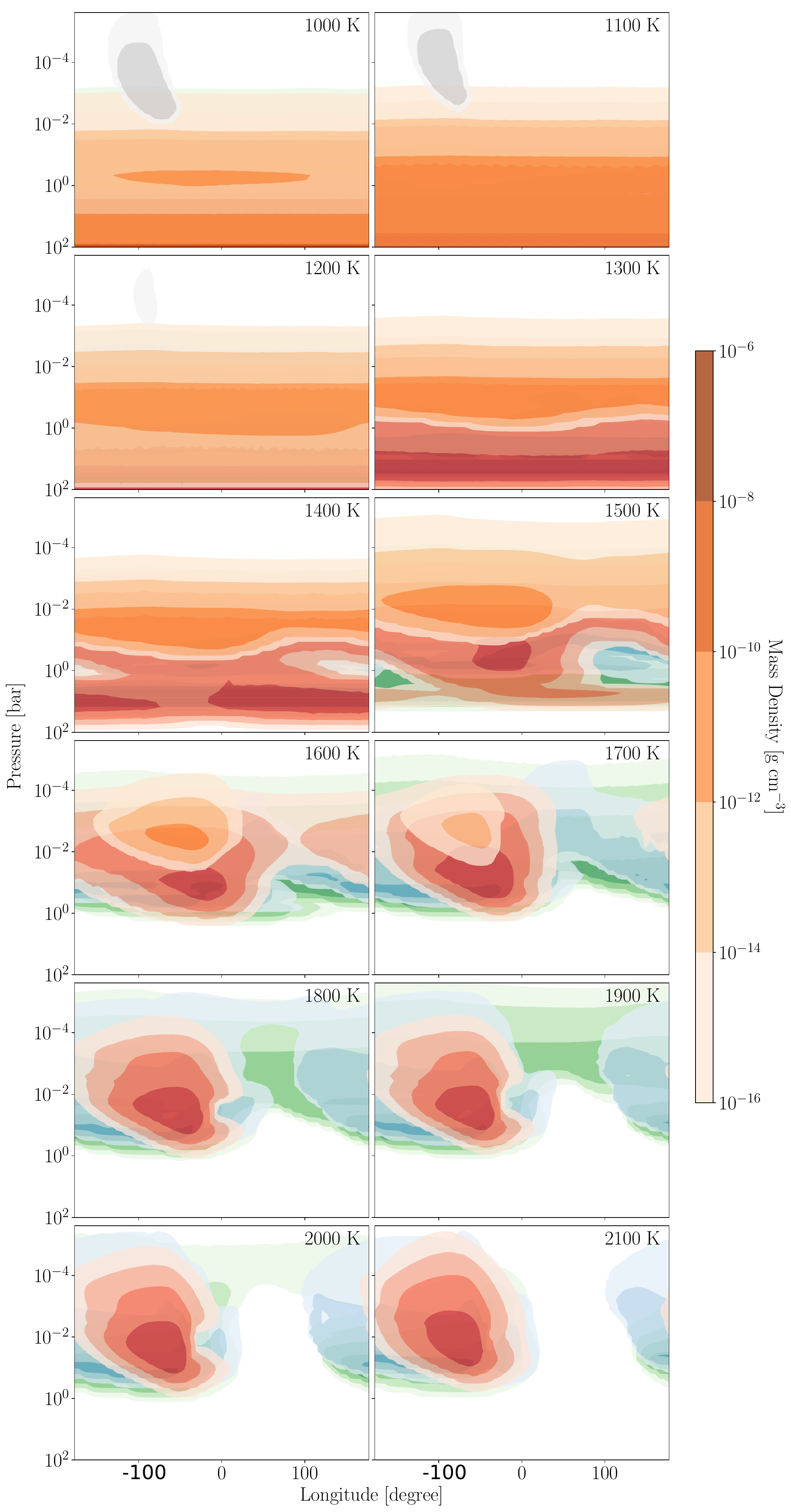}
   \caption{The distribution of a variety of cloud species varies significantly as a function of local planetary properties and equilibrium temperature. This figure is the same as Figure \ref{fig:nom_mg} except that each color indicates a different cloud species: KCl is gray, Cr is orange, Fe is red, TiO$_2$ is blue, and Al$_2$O$_3$ is green. Some species overlap on this plot. To better see the distribution of cloud particle mass for a given cloud species see Figures~\ref{fig:TiMass} - \ref{fig:KcMass}.}
   \label{fig:nom_all}
\end{figure*}

For planets with equilibrium temperatures of 1500 K and higher, the cloud mass is located entirely in the upper atmosphere at pressures lower than 1 bar as the interior becomes too hot for significant cloud formation (see Figure \ref{fig:PTprofs}. For all of the higher temperature planets in our sample, silicate cloud formation is preferentially efficient around the west limb ($\sim$90$^o$) of the planet. The overall coverage of the silicate clouds, particularly on the planetary dayside, decreases with increasing equilibrium temperature. For planets with equilibrium temperatures of $\sim$~1500-1600K, silicates clouds extend across the entirety of the planetary atmosphere, including the hotter dayside. At temperatures of $\sim$1700K and higher, the dayside is progressively more clear of silicate clouds and the east limb also decreases in total cloud cover. Despite significant increases in planetary equilibrium temperature, the silicate cloud bases remains at a roughly constant pressure level on the west limb and night side. The east limb cloud base, however, increases in height from $\sim 10^{-1}$ bar to $\sim 10^{-3}$ bar. We refer to the planets in our sample with equilibrium temperatures of 1500 K and higher as planets in an inhomogeneous cloud-dominated upper atmosphere regime. The majority of the cloud mass in these atmospheres is readily observable and the cloud opacity is likely to significantly impact inferences of observational properties \citep[e.g.,][]{2020NatAs...4..951G,2019ApJ...887..170P}. Furthermore, in this regime, the cloud coverage changes significantly with planetary location such that the planet is not necessarily cloud-dominated, or equally cloud-dominated, at all locations.

While the silicate clouds often dominate the cloud opacity, the properties of the remaining cloud species that we consider are important to consider because they can also shape the atmospheric opacity as well as determine the gas-phase composition. The distribution of other cloud species are shown in Figure \ref{fig:nom_all}. Although we consider the possibility that 9 different cloud species may form, we find that only 6 cloud species can nucleate and condense given the atmospheric conditions considered. The species that readily form are: KCl, TiO$_2$, Fe, Cr, Mg$_2$SiO$_4$, and Al$_2$O$_3$. The species that do not form clouds are: ZnS, Na$_2$S, and MnS. The planets in our sample are too warm to abundantly form ZnS, which is a species that may form more readily on cooler planets \citep[e.g.,][]{2018ApJ...855...86G}. While both Na$_2$S and MnS are supersaturated on many locations in our model atmospheres (see Figure \ref{fig:PTprofs}), these species do not form abundantly due to their high surface energies, which serve as a significant barrier to nucleation. For more of a discussion of the lack of MnS and Na$_2$S cloud formation see \citet{2020NatAs...4..951G}. We find that each cloud species is uniquely distributed in the atmospheres of the planets in our sample and each species demonstrates unique behaviors and features. 

For the coolest planets in our sample, with equilibrium temperatures ranging from 1000 - 1200 K, all 6 cloud species form in our simulated domain. KCl clouds only form near the cooler western limb of these planets above $\sim10^{-2}$ bar in the atmosphere where KCl is marginally supersaturated (see also Figures \ref{fig:KcMass} and \ref{fig:KcNum}). For these planets, KCl is the only abundant condensible cloud species in the uppermost atmospheric regions. The fact that KCl clouds are able to form in particular regions of these planetary atmospheres is an interesting consequence of modeling planets as multi-dimensional objects. Previous work that modeled planets using a globally averaged thermal profile found that the bulk of KCl cloud formation occurs on planets with equilibrium temperatures less than 950 K where it is observationally obscured by the abundant production of photochemical hazes \citep{2020NatAs...4..951G}. In contrast, in the 2D-ExoCARMA modeling presented here, we find that KCl clouds can form in relatively cool regions on planets with equilibrium temperatures higher than 950 K. These planets also have similar behavior in terms of cloud coverage for TiO$_2$, Cr, Fe, and Al$_2$O$_3$ (see also Figures \ref{fig:CrMass}, \ref{fig:FeMass}, \ref{fig:AlMass}, \ref{fig:CrNum}, \ref{fig:FeNum}, and \ref{fig:AlNum}) cloud species. For these species, most of the cloud mass is present in the lower atmosphere, with the exception of T$_\mathrm{eq}$ = 1200 K where the bulk of the Cr cloud mass has moved to the upper atmosphere above 1 bar. These cloud species also extend throughout the atmosphere from high pressures of $\sim 10^2$ bar to less than $\sim 10^{-2}$ bar or even less than $10^{-4}$ bar in the case of TiO$_2$. 

For planets with equilibrium temperatures of 1300 K and higher the cloud mass of all species increasingly moves into the upper atmosphere and the deep atmosphere becomes progressively more clear. The exception to this case are KCl clouds, which are not present anywhere on the planet at equilibrium temperatures of 1300K or greater. After planets reach equilibrium temperatures of $\sim 1600$ K the atmospheric cloud cover becomes inhomogeneous as particular cloud species are no longer stable at all planetary longitudes. At planetary equilibrium temperatures higher than 1700K, Cr clouds are no longer able to form at any location in the planetary atmosphere despite being supersaturated at specific locations throughout all modeled atmospheres (see Figure \ref{fig:PTprofs} and discussion in Section \ref{hotdiff}). The majority of the various cloud species are no longer stable on the dayside/east limb at equilibrium temperatures of 1800 K or higher. Once the planetary equilibrium temperature reaches 2100 K there are no longer any cloud species present on the planetary dayside or the hotter eastern limb. For all planets, the cooler western limb is a preferred location for efficient cloud formation for all cloud species. 

\begin{figure} 
   \centering
   \includegraphics[width=0.49\textwidth]{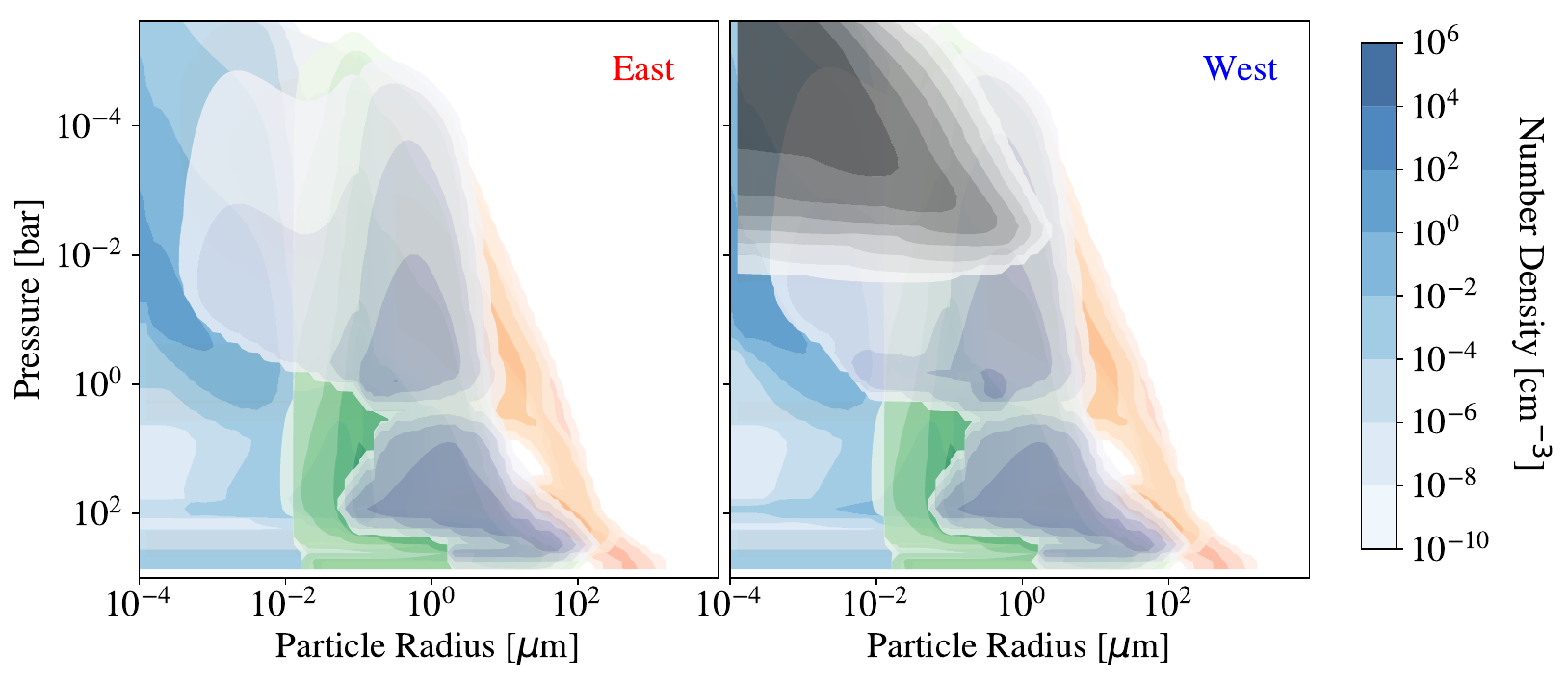}
   \caption{Even for planets with a near homogeneous covering of clouds, the cloud particle size distributions vary as a function of planetary location. Here we show the number density of cloud particles as a function of atmospheric height on the east and west limbs of a model hot jupiter with T$_\mathrm{eq}$ = 1000 K. Each color indicates a different cloud species: gray indicates KCl, purple indicates Mg$_2$SiO$_4$, green indicates Al$_2$O$_3$, blue indicates TiO$_2$, orange indicates Cr, and red indicates Fe. } 
   \label{fig:num_eastwest_1000}
\end{figure} 

\begin{figure} 
   \centering
   \includegraphics[width=0.49\textwidth]{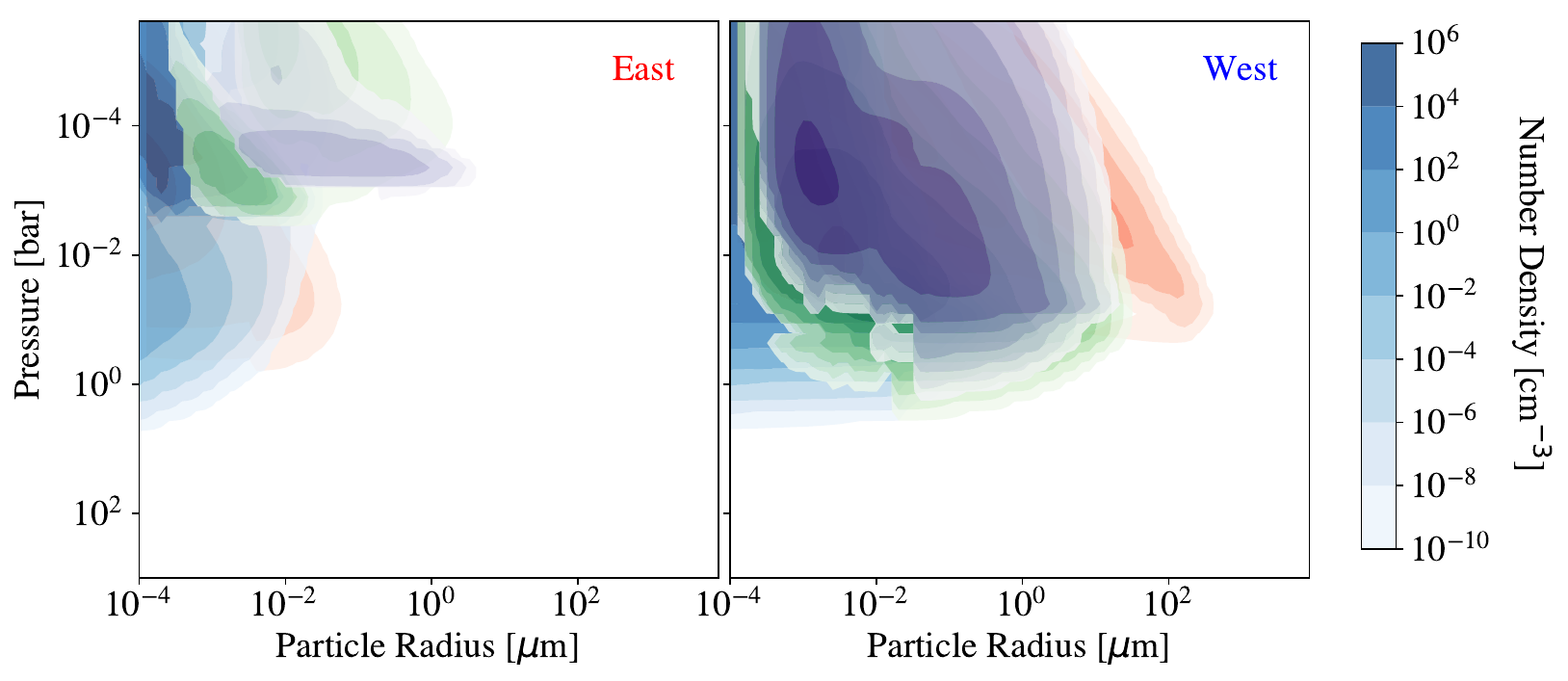}
   \caption{For planets with an inhomogeneous covering of clouds, the differences in cloud particle size distributions are significant at various locations on the same planetary atmosphere. This figure depicts the same as Figure \ref{fig:num_eastwest_2100} but for a model hot jupiter with T$_\mathrm{eq}$ = 2100 K. }
   \label{fig:num_eastwest_2100}
\end{figure} 

\subsection{Cloud Particle Size Distributions}\label{nom_distributions}

In the previous section, we discussed trends and properties evidenced in the mass distribution of cloud particles as a function of atmospheric height and longitude. However, our bin-scheme microphysical cloud model further calculates the cloud particle size distribution as a function of height for each of the 64 planetary longitudes sampled for each of the 12 model hot Jupiters. To illustrate the differences in cloud particle size distribution as a function of planetary location, we examine the number density as a function of atmospheric pressure and particle radius on the east and west limbs for the coolest (Figure \ref{fig:num_eastwest_1000}) and hottest (Figure \ref{fig:num_eastwest_2100}) planets in our sample. 

For the coolest planet in our sample (T$_\mathrm{eq}$ = 1000 K) the cloud particle size distribution on the east and west limbs show several similarities as well as a clear difference as shown in Figure \ref{fig:num_eastwest_1000}. The clearest difference between the two limbs is the presence of very small (radius $\sim10^{-4}$ micron) high altitude KCl cloud particles that preferentially form in the upper atmosphere where KCl is the most supersaturated. The other cloud species have fairly similar cloud particle size distributions on both the east and west limbs although the species form in somewhat higher abundance in the upper atmosphere on the western limb. This effect is particularly noticeable for titanium and silicate clouds, which are higher in abundance at larger particle sizes in the upper atmosphere on the western limb of the planet.

For the hottest planet in our sample (T$_\mathrm{eq}$ = 2100 K) the cloud particle size distributions on the east and west limbs are significantly different for each of the cloud species as shown in Figure \ref{fig:num_eastwest_2100}. Thus, even when horizontal advection is considered, hotter hot Jupiters with equilibrium temperatures greater than $\sim$1700K have inhomogeneous clouds on the east/west limbs as predicted in \citet{2019ApJ...887..170P}. On the warmer east limb, clouds form at significantly higher altitudes in the atmosphere and are not able to grow to sizes larger than $\sim1\;\mu$m. Interestingly, for some of the species that form via heterogeneous nucleation, particularly Mg$_2$SiO$_4$ and Al$_2$O$_3$, cloud particles are not stable at the smallest sizes (less than 10$^{-3}\;\mu$m) as particles of these sizes would quickly evaporate. These species are instead only able to nucleate onto the TiO$_2$ seed particles once they have reached larger sizes. For more discussion of the size effect operating here see \citet{2022NatAs...6.1147P}. The lower supersaturations on the eastern limb also preferentially limit the formation of species that heterogeneously nucleate to the uppermost atmospheric regions where their supersaturations are the most extreme. This requirement is less extreme for species that can homogeneously nucleate, such as TiO$_2$ and Fe, which are able to form in low abundances deeper in the atmosphere. 

In contrast to the cloud-sparse eastern limb, the western limb of the hottest planet in our sample is covered in a substantial layer of clouds throughout the upper atmosphere. The clouds on the west limb span from the smallest particle sizes we consider to very large particles $\sim$100 $\mu$m in size. We also find that clouds extend in low abundances to below each species' specific cloud base due to the finite time it takes for clouds to evaporate below this point. Thus, despite the addition of longitudinal cloud particle transport, the warmer planets in our sample have significantly different cloud particle size distributions, which exacerbates the already inhomogeneous nature of these atmospheres \citep[e.g.,][]{2019ApJ...887..170P}. 

\begin{figure*} 
   \centering
   \includegraphics[width=0.32\textwidth]{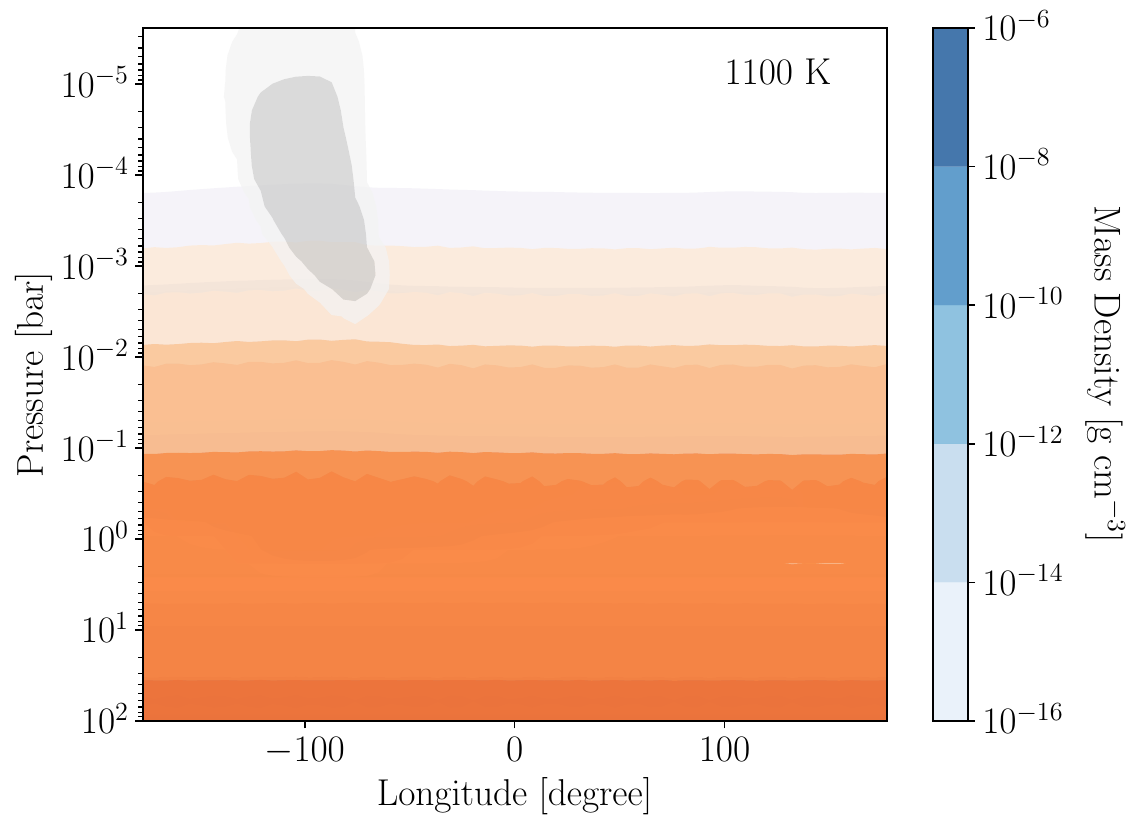}
   \includegraphics[width=0.32\textwidth]{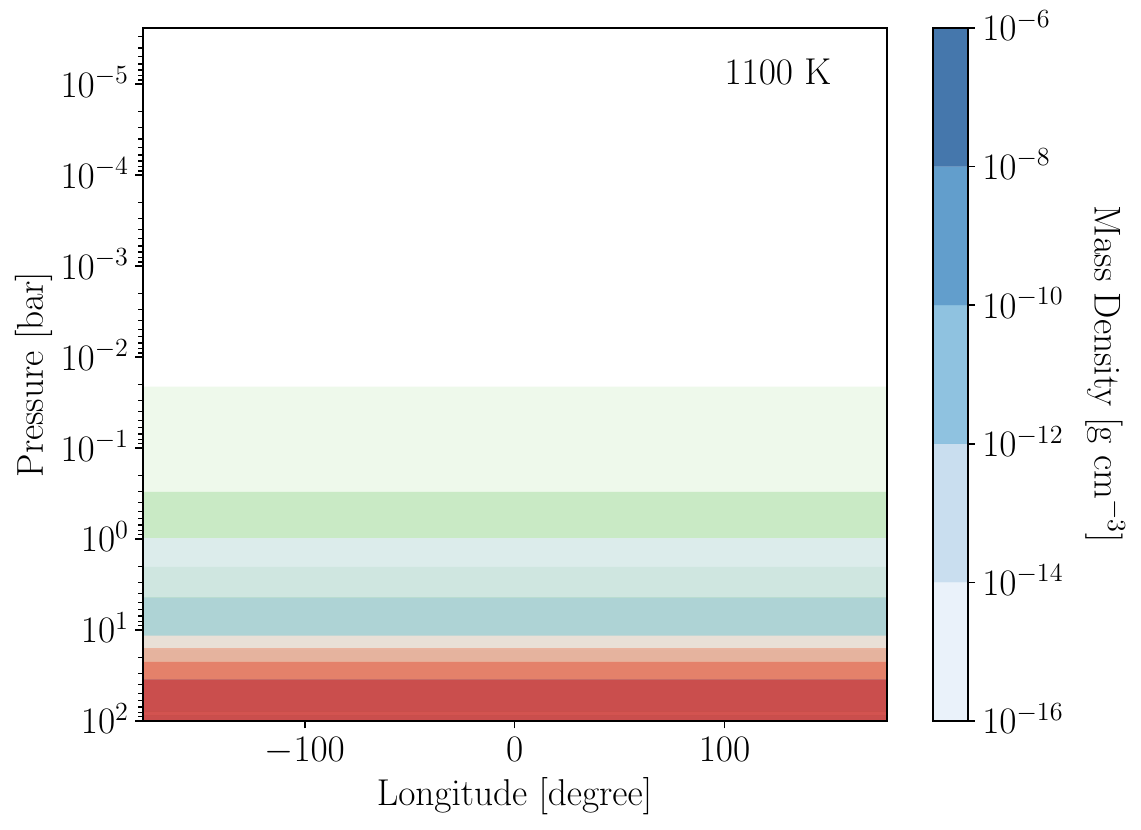}
   \includegraphics[width=0.32\textwidth]{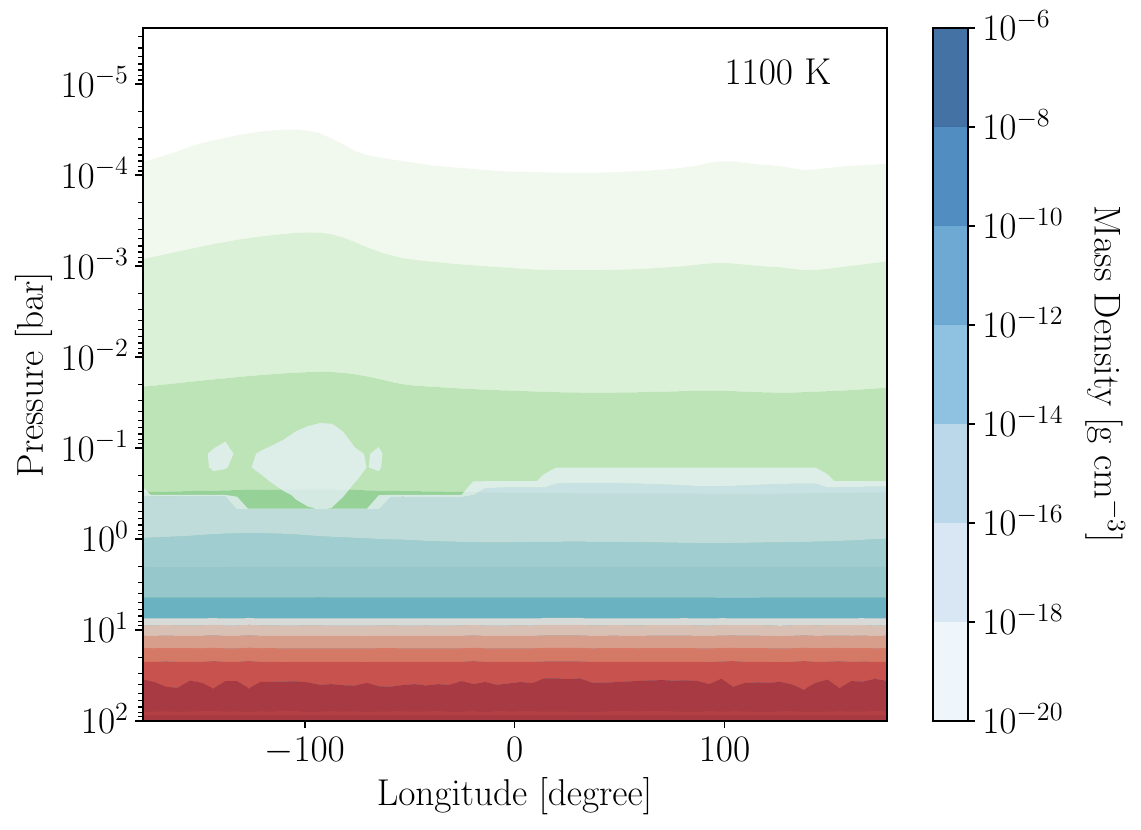}
      \includegraphics[width=0.32\textwidth]{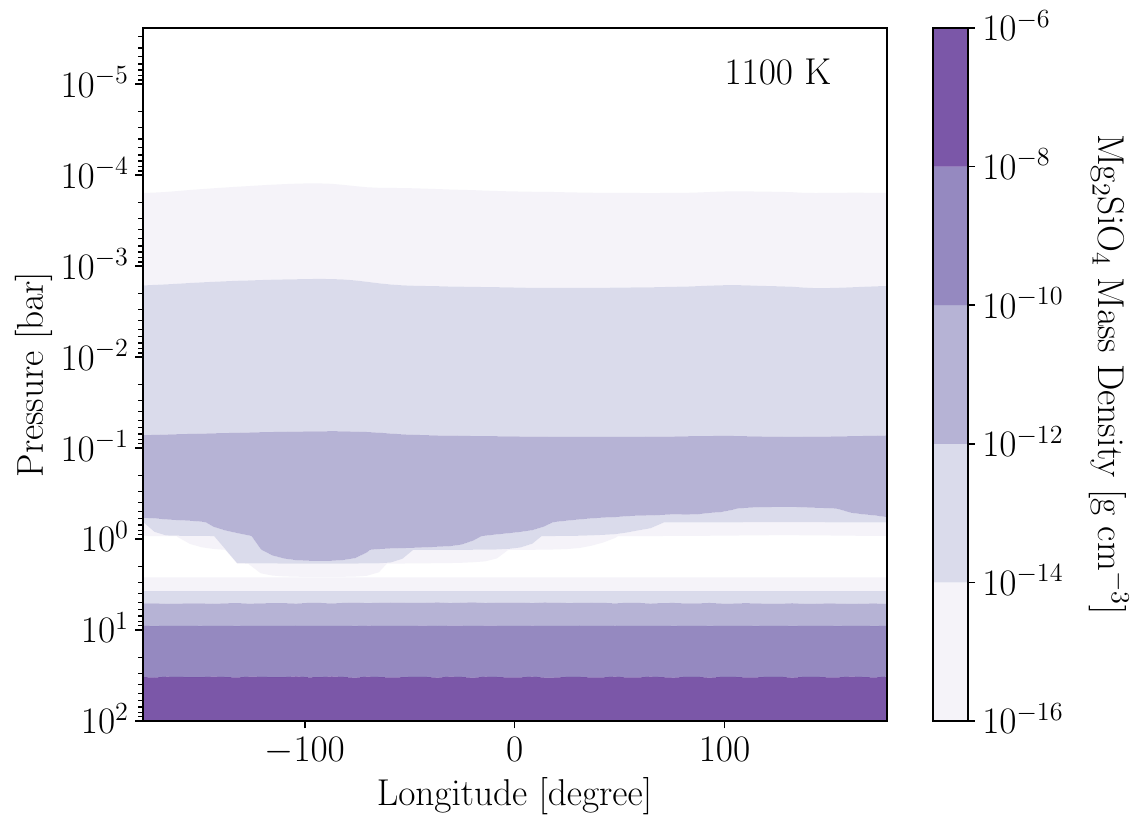}
      \includegraphics[width=0.32\textwidth]{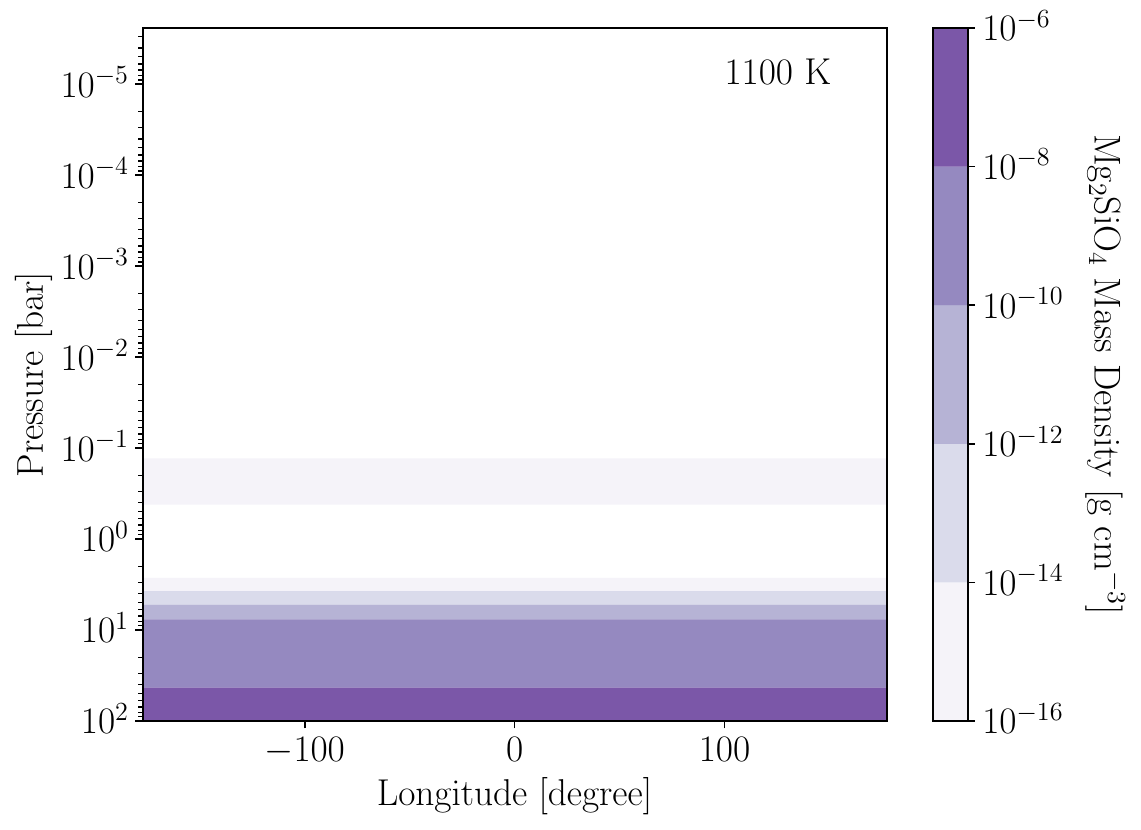}
   \includegraphics[width=0.32\textwidth]{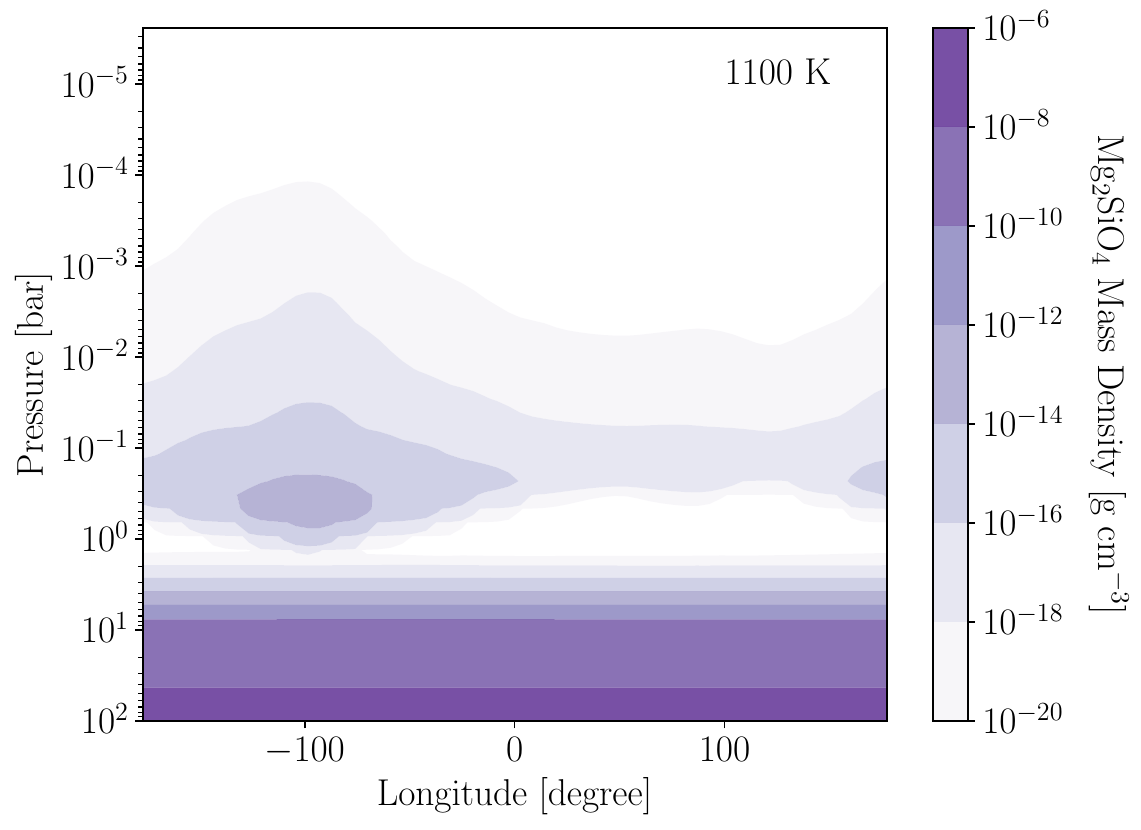}
   \caption{The mass density of cloud particles as a function of planetary longitude and atmospheric pressures varies significantly when horizontal advection is considered (2D-ExoCARMA, left panels) as compared to when horizontal advection is not included (1D CARMA, middle and right panels). Here we show the mass density of all cloud particles (top) where the colors for each cloud species are the same as those in Figure \ref{fig:num_eastwest_1000} and the mass density of silicate clouds in particular (bottom panels). The middle column shows the 1D CARMA cloud distributions with the same colorbar as the 2D models. Note that the 1D models have substantially less cloud mass. The right column shows the 1D models with colorbars that extend to lower cloud mass densities to highlight the cloud distributions in the upper atmosphere.}
   \label{fig:1d2d_11_comp}
\end{figure*}

\section{The Interplay Between Horizontal Advection and Cloud Formation}\label{horiz_advec}

Clouds are strongly dependent on nearly all atmospheric and planetary properties, which vary significantly with atmospheric location in the case of highly irradiated planets like hot Jupiters. We now investigate the specific impact that horizontal mixing, which transports clouds to regions of the atmosphere with different properties (see Section \ref{atm_struc}), has in shaping the cloud distributions discussed in Section \ref{nom_distributions}.

In order to demonstrate the impact that horizontal mixing has on cloud formation and the resultant cloud properties, we compare the 2D-ExoCARMA results with the traditional 1D cases without horizontal mixing. To do that, we model each of the hot jupiters in our sample using 1D CARMA. We thus model each of the 64 longitudinal grid points individually without considering horizontal advection. While we have simulated all of the 12 hot Jupiters in our modeling grid (not shown), here we focus on two illustrative cases to demonstrate the significant effect of horizontal mixing in cloud formation processes and the resultant cloud particle distributions across the atmosphere. We again note that we are presenting time-averaged solutions after the models have reached a steady state with the caveat that small changes that occur due to microphysical variability may occasionally minimally exacerbate or minimize the differences presented here. 

\subsection{The Impacts of Horizontal Transport for ``Cool" Hot Jupiters}\label{cooldiff}

\begin{figure} 
   \centering
   \includegraphics[width=0.4\textwidth]{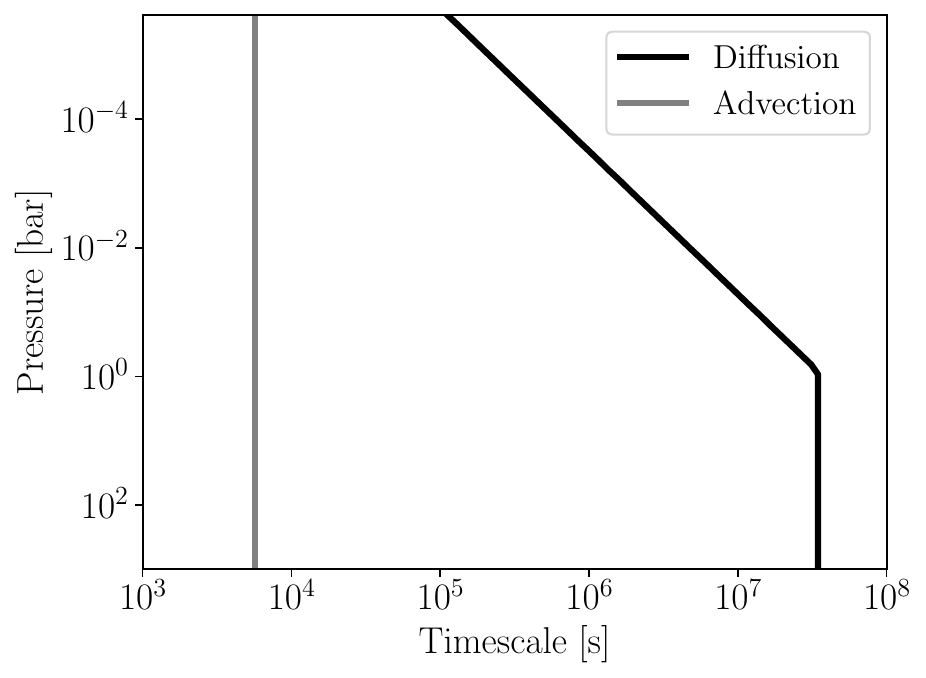}
   \includegraphics[width=0.49\textwidth]{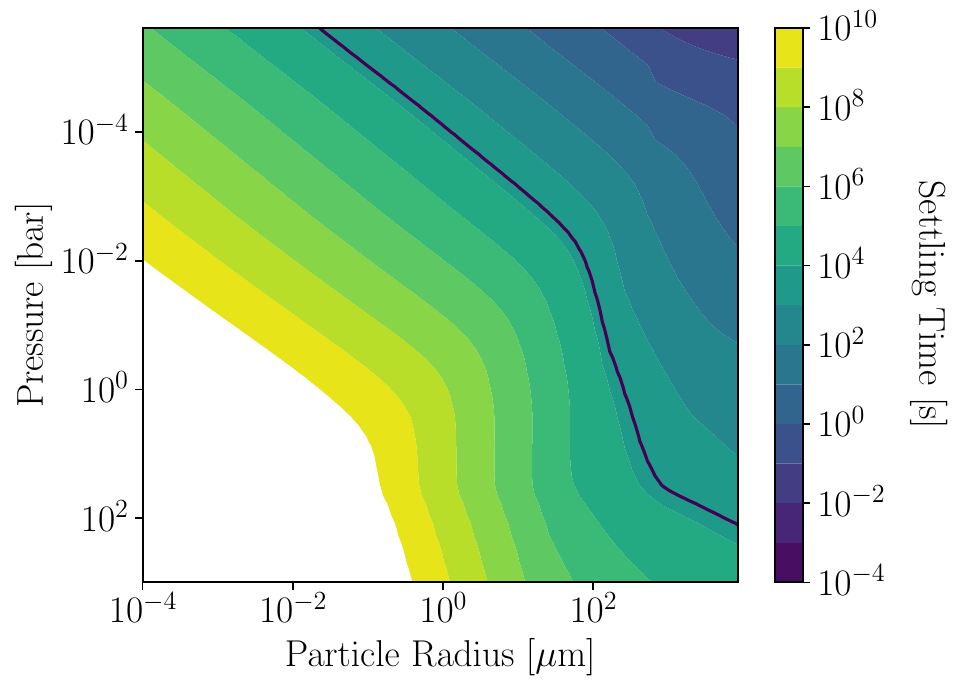}
   \caption{The vertical diffusion timescale of both gas and particles (top) is substantially larger than the horizontal advection timescale here for the case of T$_\mathrm{eq} = 1100$ K. The settling timescale of particles (bottom) varies significantly with particle size and atmospheric pressure such that large particles settle faster than they can be advected horizontally (horizontal advection timescale indicated by the black line) while small particles settle slowly and can experience significant horizontal advection.}
   \label{fig:11_timescales}
\end{figure}

We first examine a cool atmosphere case with an equilibrium temperature of 1100K. We show the resultant cloud mass distributions as a function of planetary longitude and height in the case with and without horizontal advection in Figure \ref{fig:1d2d_11_comp} for all cloud species and for silicate clouds in particular as silicate clouds likely dominate the cloud opacity for planets in this range of equilibrium temperatures. We find that there are significant differences in the distribution of cloud mass in the case of 2D-ExoCARMA versus 1D CARMA. For this planet, these differences are most notable in the upper atmosphere. In the 1D CARMA case, most of the cloud species are confined to the lower atmosphere, below $\sim$10$^{-1}$ bar. However, in the 2D case, more species are more readily lofted into the upper atmosphere. This is particularly evident for Fe and Cr clouds, which are only present in the lower atmosphere in the 1D CARMA case but extend throughout the upper atmosphere in the 2D-ExoCARMA case. Cloud formation in the upper atmosphere is also significantly more efficient in the 2D-ExoCARMA case (note the difference in Figure \ref{fig:1d2d_11_comp}) such that clouds are present at significantly higher masses. 

To understand the differences in the 1D and 2D-ExoCARMA cases, we calculate the cloud formation and condensible gas transport timescales as compared to the horizontal advection timescale. We first note, for the 1100 K case, that the transport timescales of condensible gases as well as cloud particles have timescales that are often comparable to the horizontal advection timescale. This can be seen in Figure \ref{fig:11_timescales} where the vertical diffusion timescale of both gas and cloud particles is longer than the horizontal advection timescale and the settling timescale of cloud particles is significantly longer than the advection timescale for small particles and shorter than the avection timescales for very large particles. We calculate the vertical diffusion timescale as $\tau_\mathrm{diff} = H^2/K_{zz}$, the settling timescale as $\tau_\mathrm{setl} = H/v_\mathrm{fall}$, and the horizontal advection timescale as $\tau_\mathrm{adv} = D/64/U$ for the time that it takes to cross one of the 64 grid cells, $H$ is the atmospheric scale height, $v_\mathrm{fall}$ is the particle fall velocity, $D$ is the planetary diameter (each planet in the sample has a Jupiter radius), and $U$ is the horizontal wind speed.  

In the case of cooler hot jupiter atmospheres, because the vertical diffusion timescale is significantly longer than the horizontal advection timescale, there is not a steady diffusion of gas to the upper atmosphere as the vertical column is advected around the planet such that the gas will reach a steady state profile in each vertical column. Instead, the vertical gas distribution evolves as the gas is advected horizontally around the planet. For this picture of diffusion timescale to change, we would need to consider K$_{zz}$'s that are several orders of magnitude larger than those considered at these temperatures. Because the particle settling timescale is longer than the horizontal advection timescale for small cloud particles, these particles can be continuously transported across the planet before they settle. However, large cloud particles will settle out of the atmosphere before they can be transported horizontally, effectively setting an upper size limit on the cloud particle size distribution. 

\begin{figure} 
   \centering
   \includegraphics[width=0.47\textwidth]{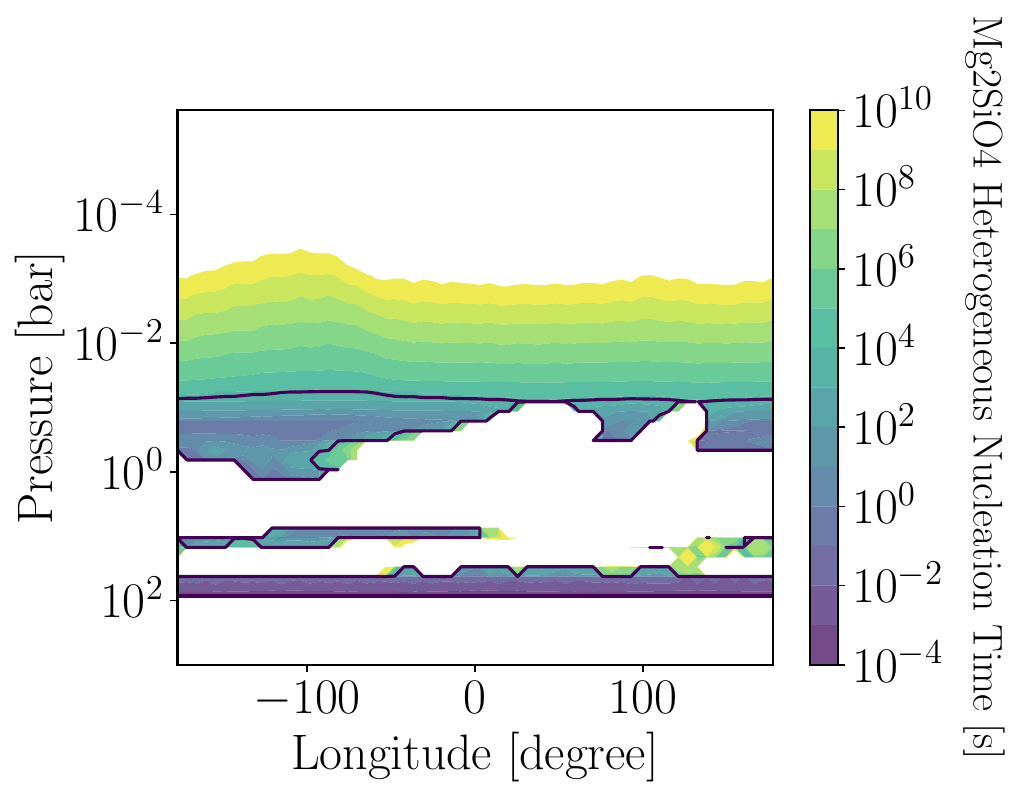}
   \includegraphics[width=0.47\textwidth]{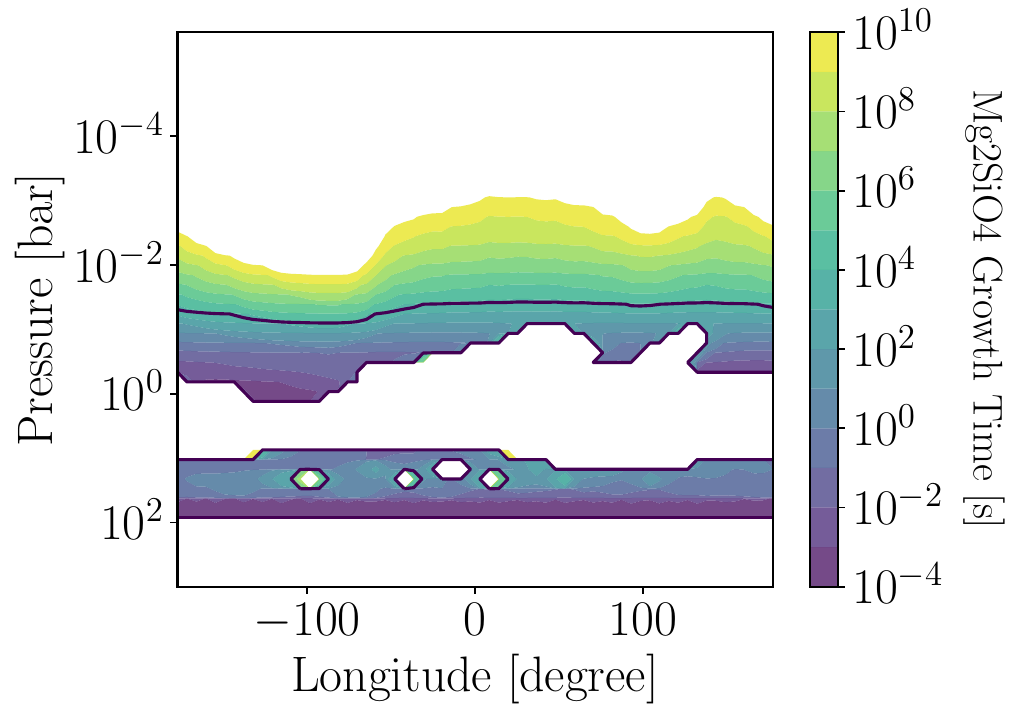}
   \includegraphics[width=0.47\textwidth]{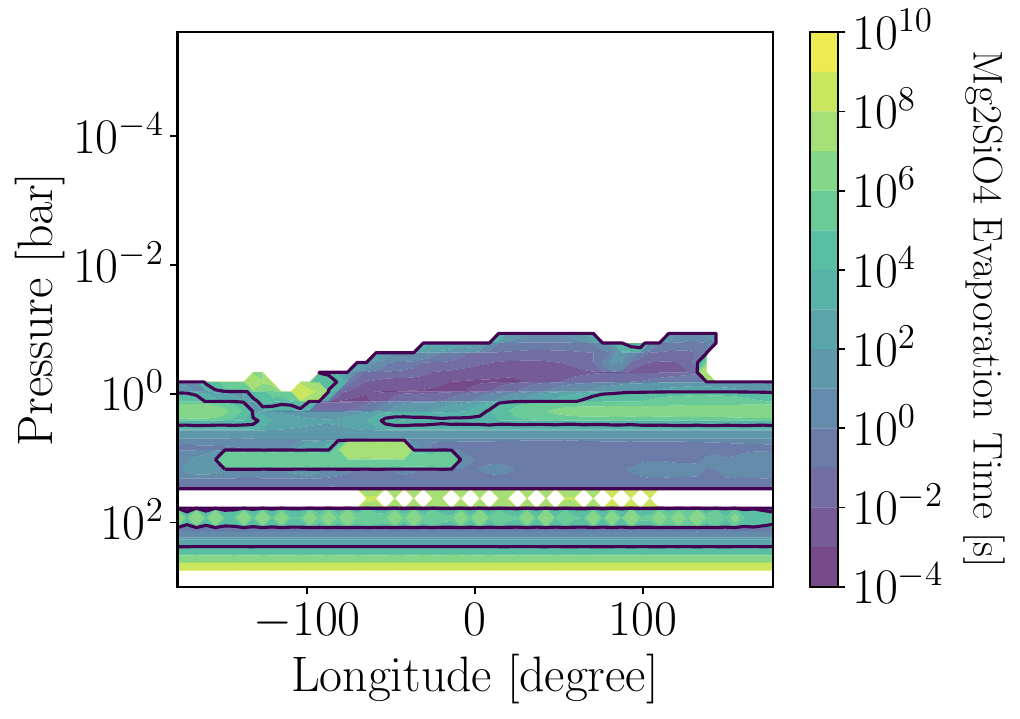}
   \caption{The microphysical processes of cloud formation vary as a function of atmospheric location. Here we show the nucleation, growth, and evaporation timescales for Mg$_2$SiO$_4$ cloud particles in 2D-ExoCARMA for a planet with an equilibrium temperature of 1100 K. The black line indicates the horizontal advection timescale. }
   \label{fig:mg_1100_time}
\end{figure}

\begin{figure} 
   \centering
   \includegraphics[width=0.47\textwidth]{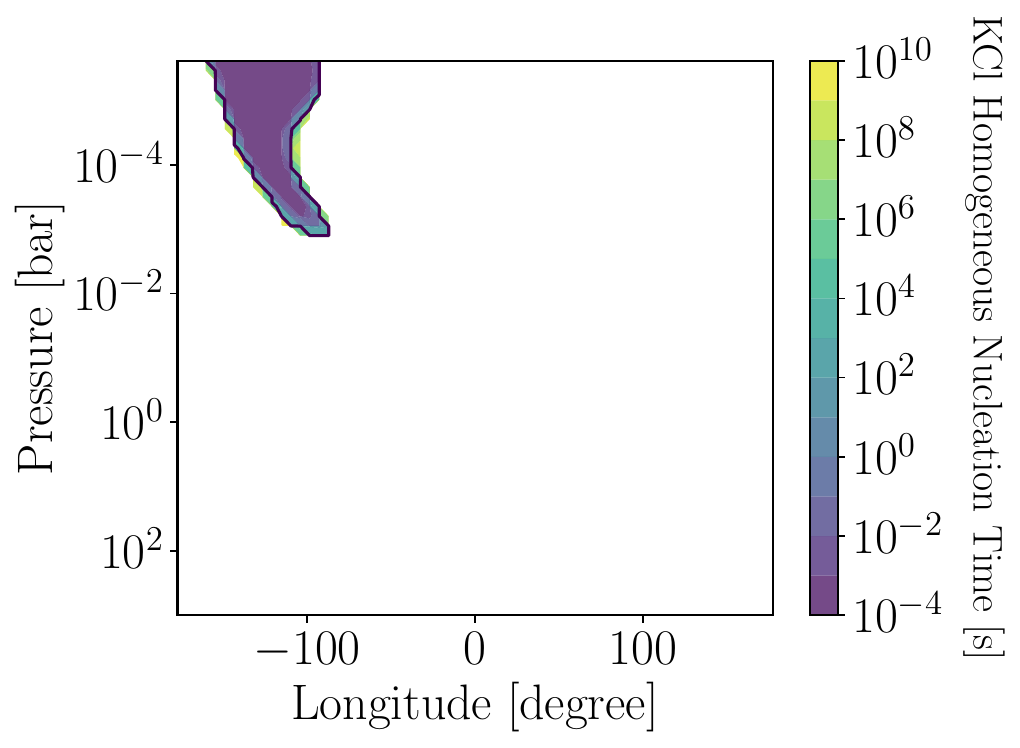}
   \includegraphics[width=0.47\textwidth]{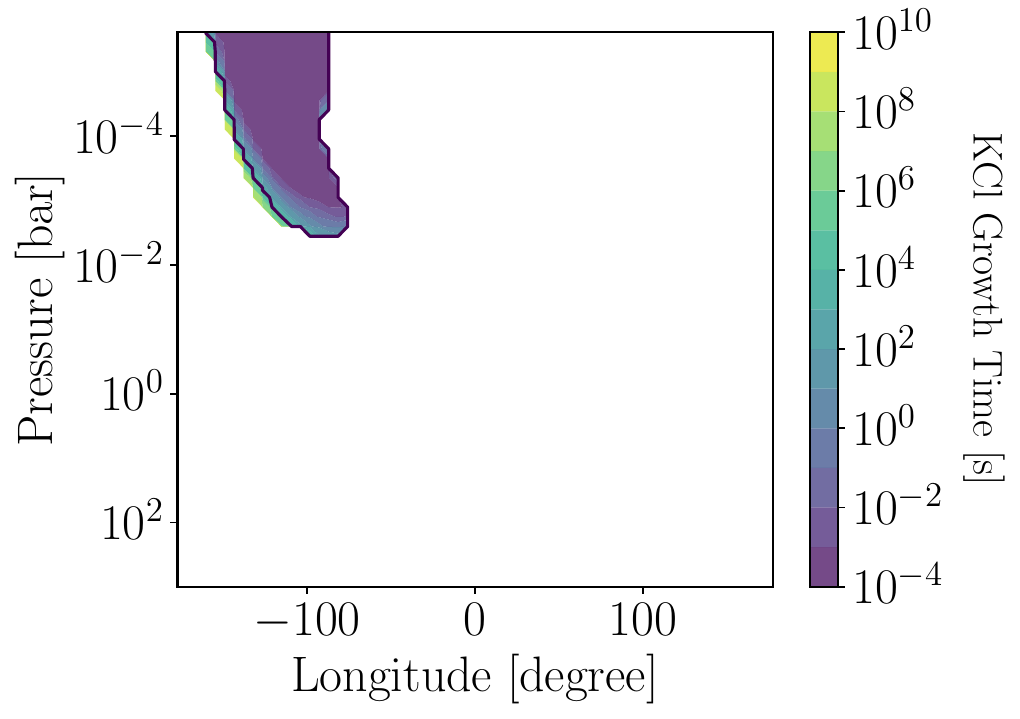}
   \includegraphics[width=0.47\textwidth]{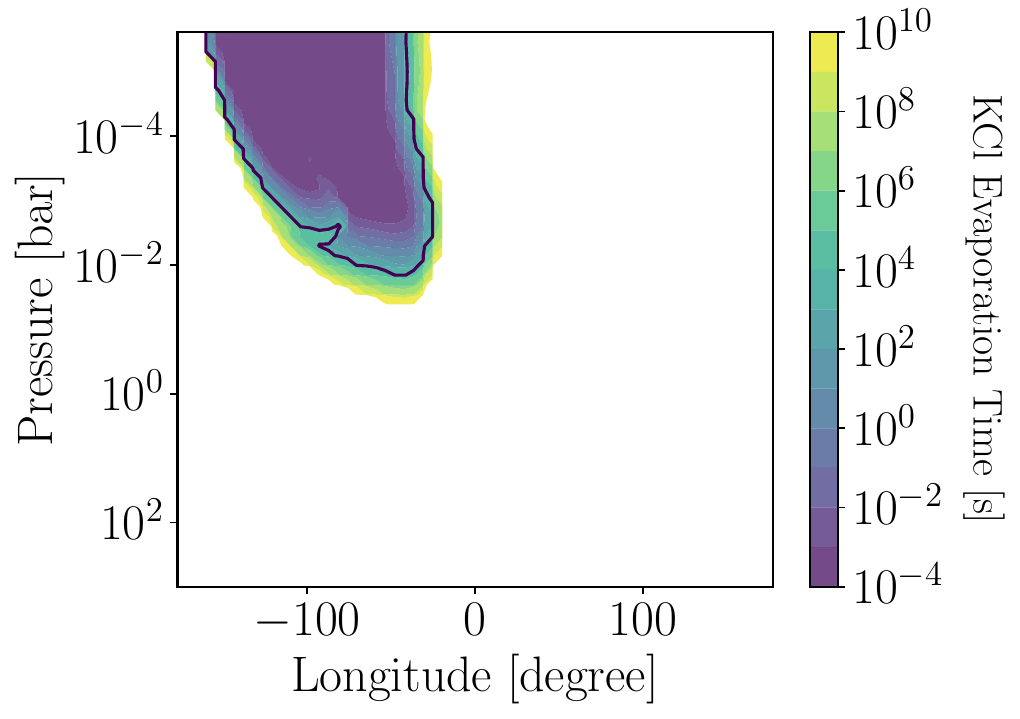}
   \caption{The same as Figure \ref{fig:mg_1100_time} but for cloud particles comprised of KCl. }
   \label{fig:kcl_1100_time}
\end{figure}

\begin{figure*} 
   \centering
   \includegraphics[width=0.32\textwidth]{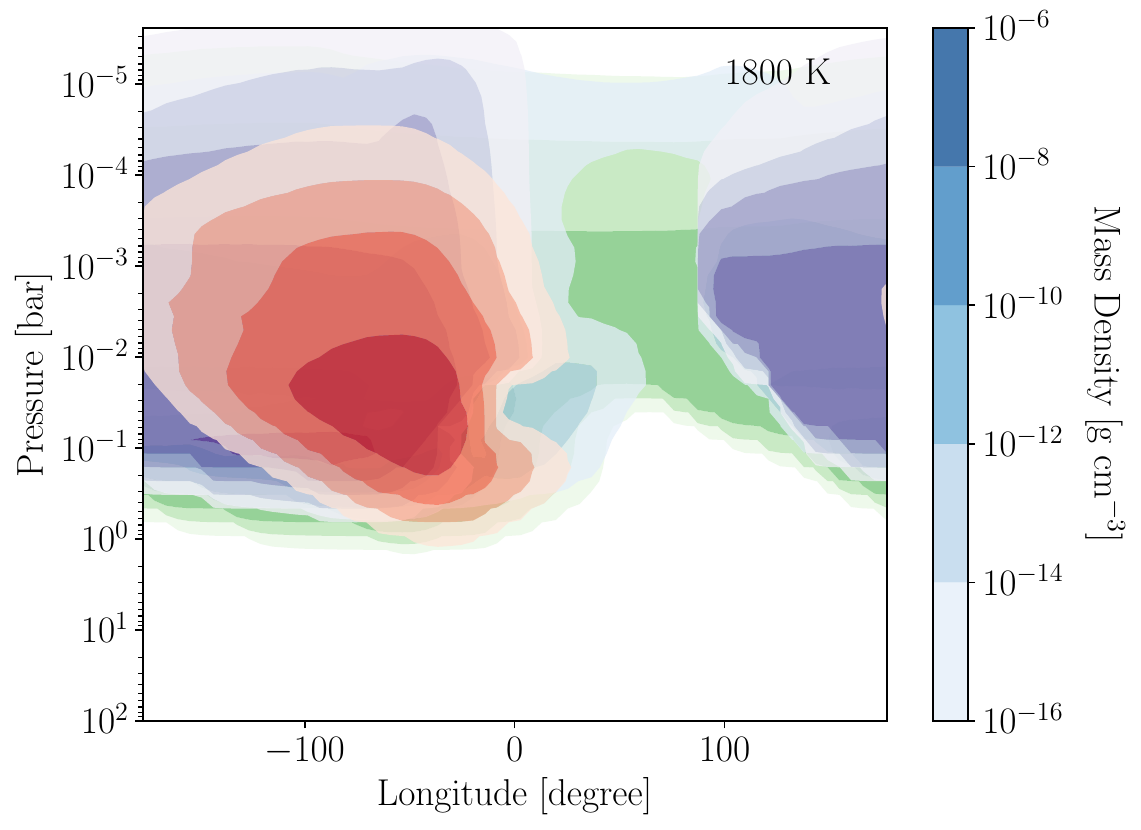}
   \includegraphics[width=0.32\textwidth]{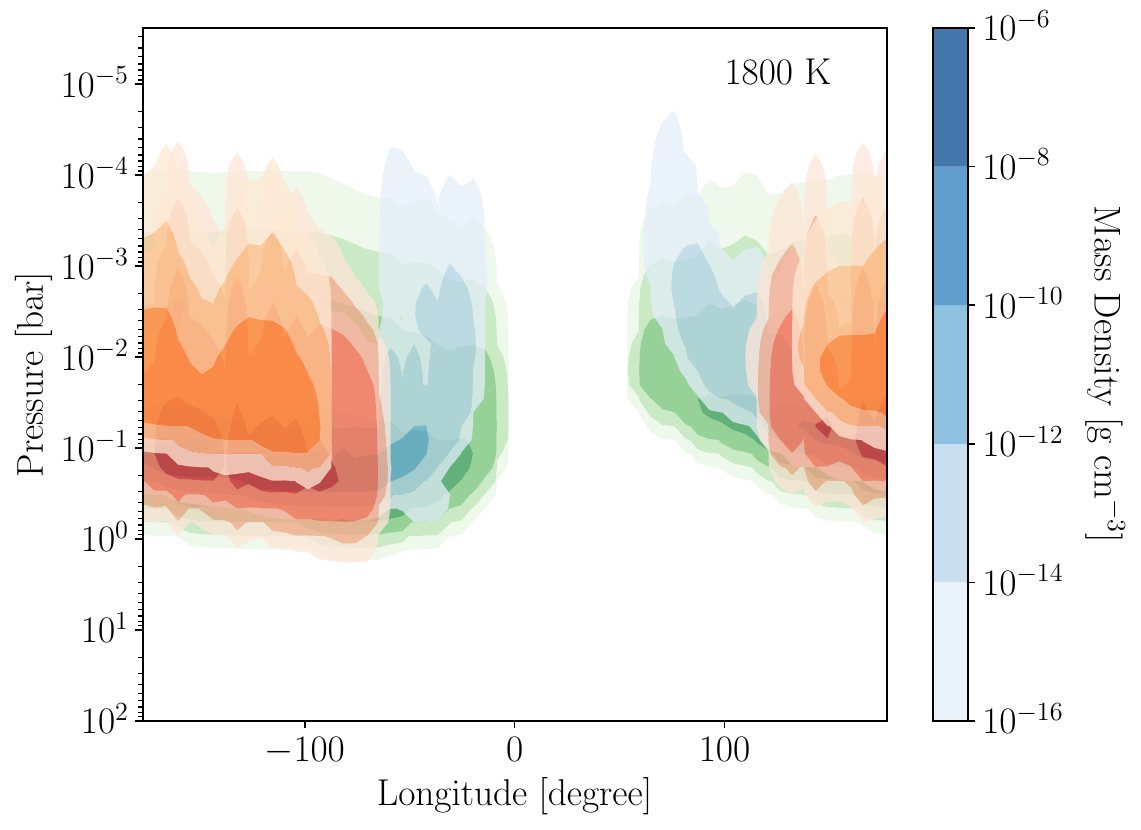}
   \includegraphics[width=0.32\textwidth]{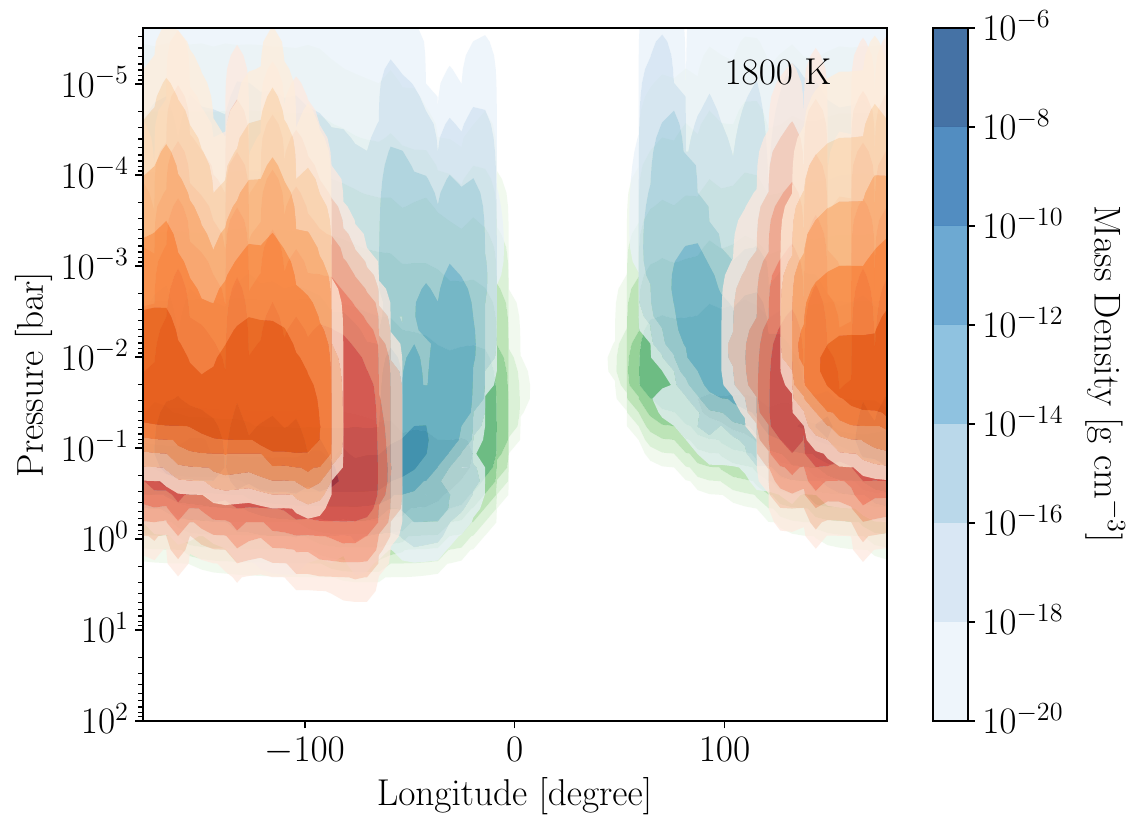}
    \includegraphics[width=0.32\textwidth]{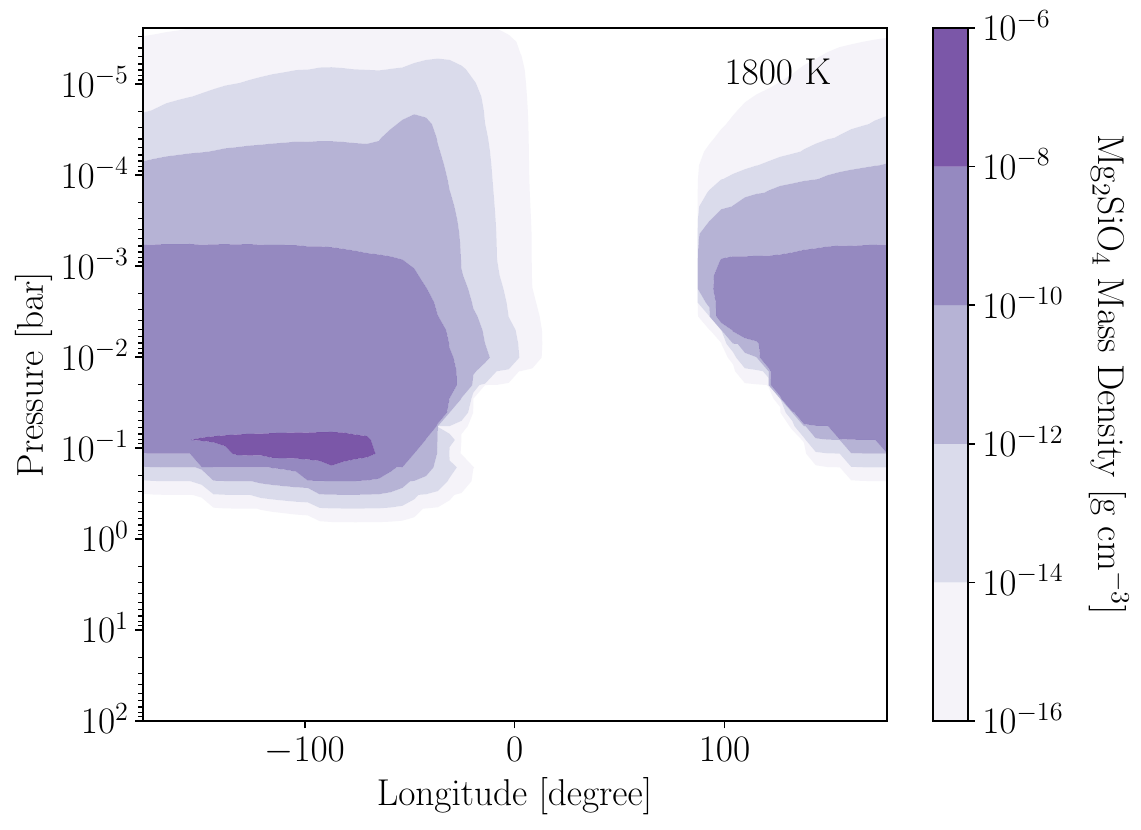}
    \includegraphics[width=0.32\textwidth]{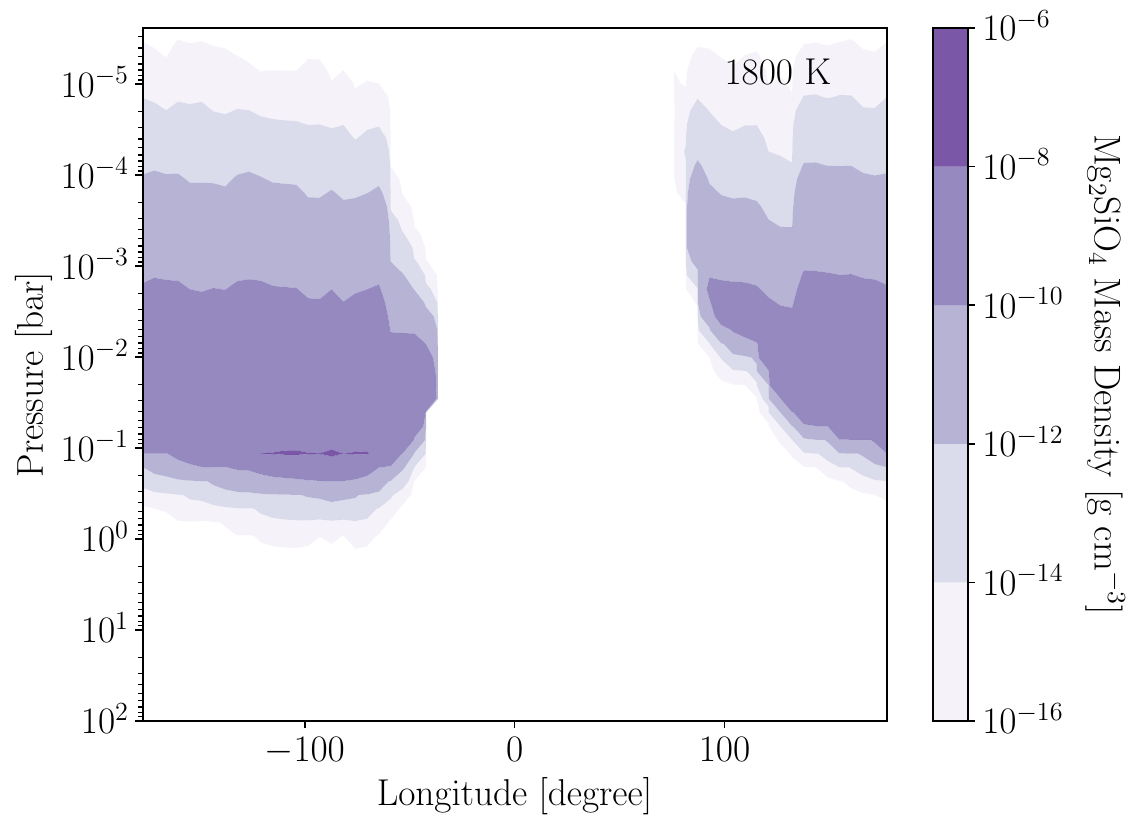}
   \includegraphics[width=0.32\textwidth]{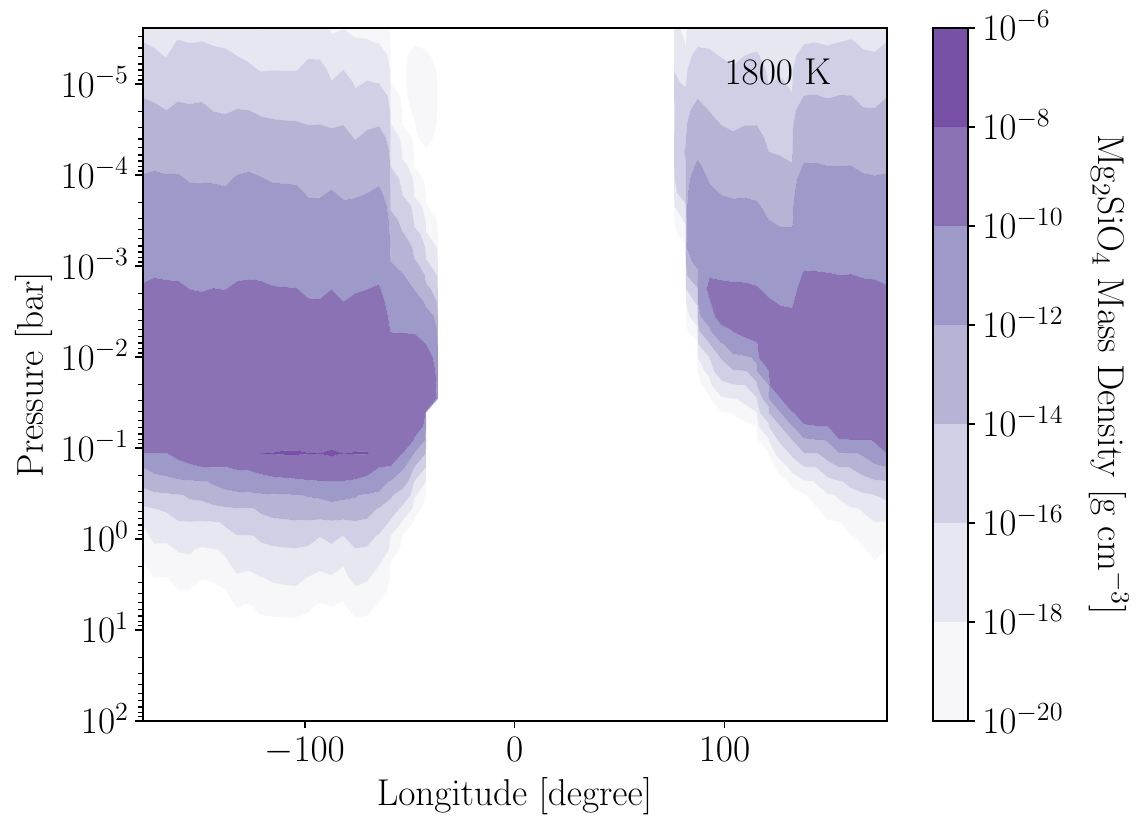}
   \includegraphics[width=0.32\textwidth]{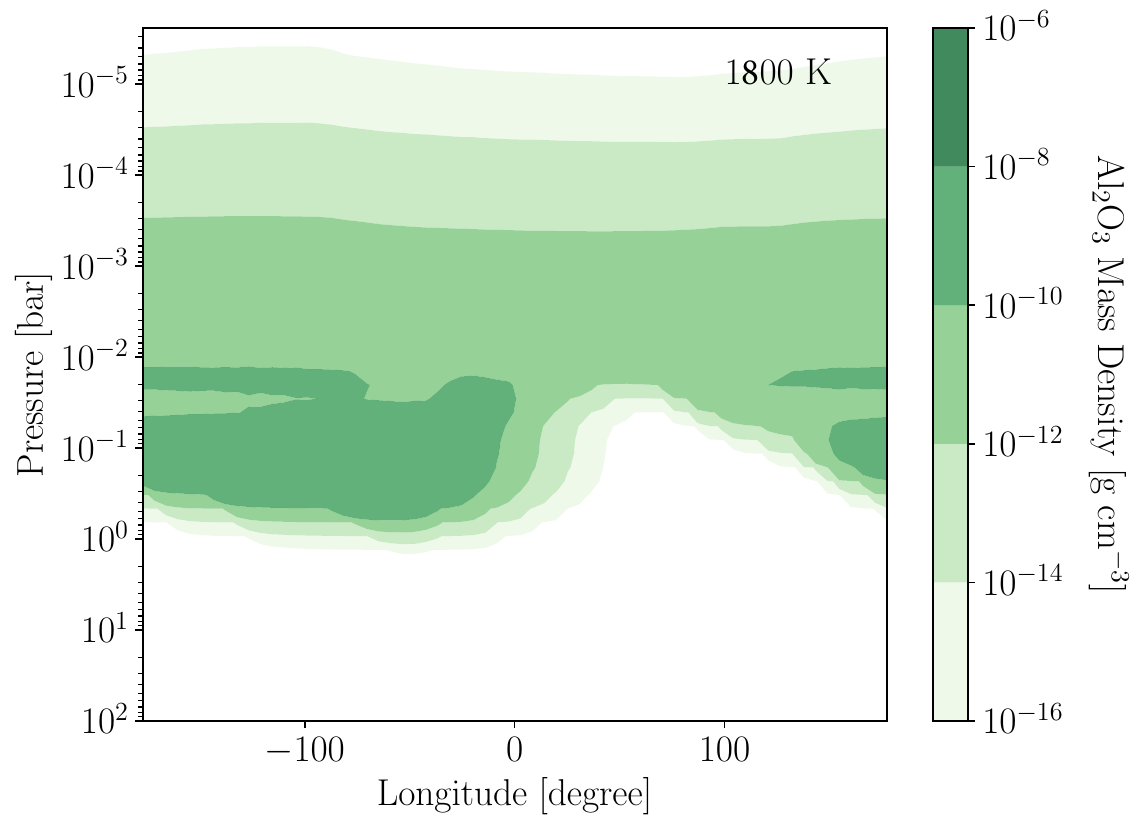}
   \includegraphics[width=0.32\textwidth]{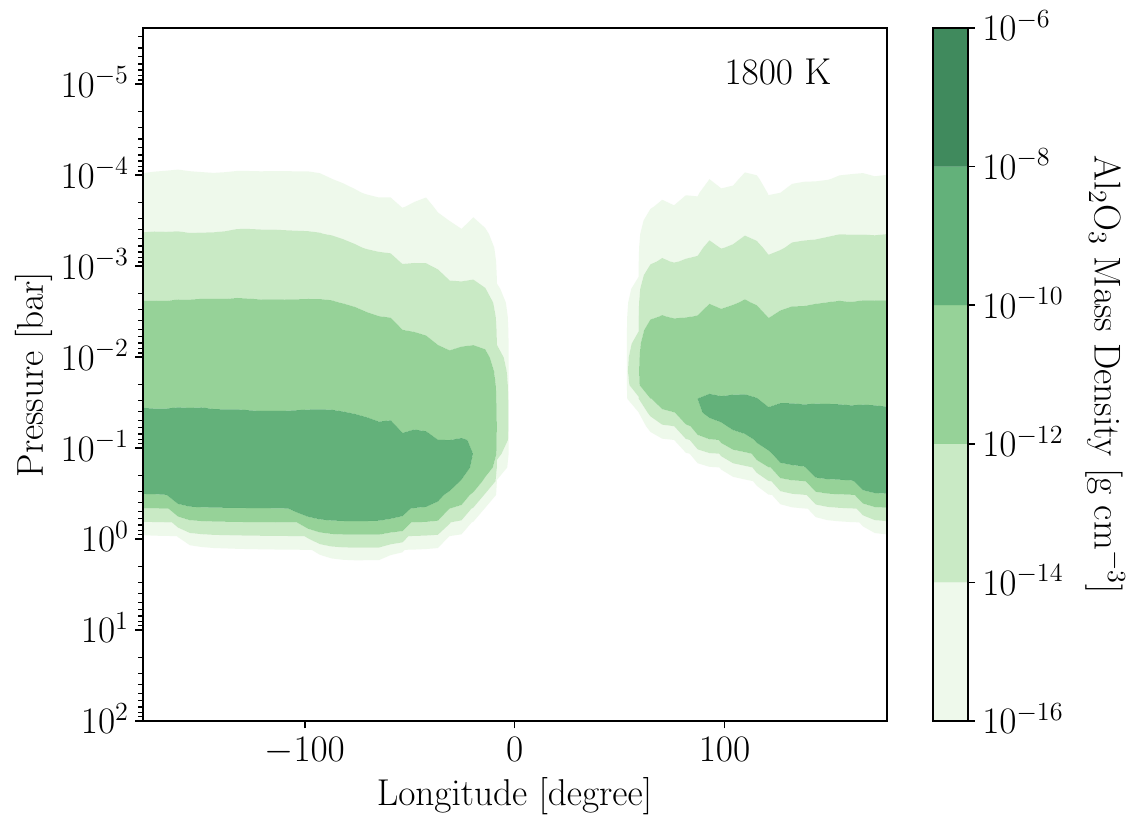}
   \includegraphics[width=0.32\textwidth]{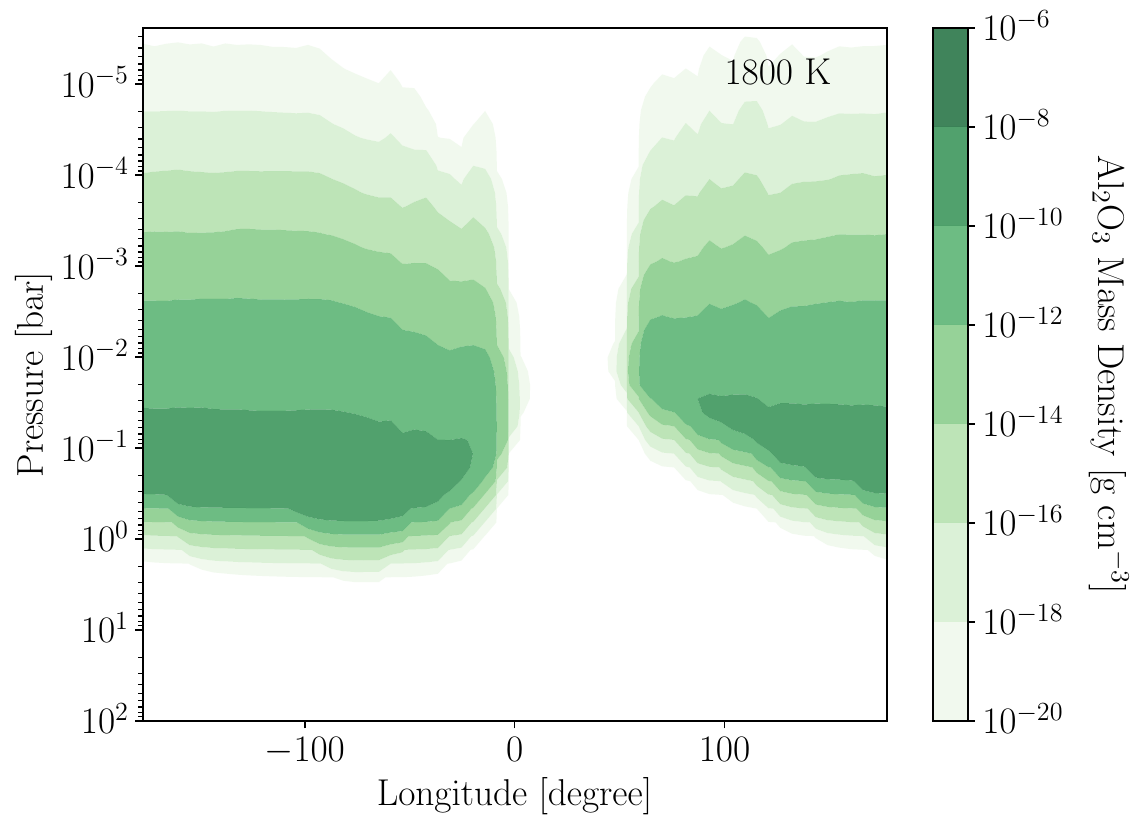}
   \caption{Same as Figure \ref{fig:1d2d_11_comp} with the addition of a comparison of the Al$_2$O$_3$ cloud mass (bottom row) and for a planet with an equilibrium temperature of 1800 K..}
   \label{fig:1d2d_18_comp}
\end{figure*}

In the case of silicate clouds, we find that the cloud distribution in the lower atmosphere in the 1D CARMA and 2D-ExoCARMA cases is roughly the same (Figure \ref{fig:1d2d_11_comp}). Furthermore, in both the 2D and 1D cases, there is again a cloud-free gap region of the atmosphere where silicate cloud formation is not efficient due to the isothermal region of the atmosphere where the cloud formation efficiency drops with pressure. The key differences between the 1D and 2D cases become evident in the upper atmosphere. While silicate clouds are present at similar locations in both cases, their total cloud mass is reduced in the 1D case as compared to the 2D case. Furthermore, the silicate cloud mass distribution is homogenized by horizontal transport in the 2D case such that the clouds form a nearly homogeneous layer as a function of longitude with much less spatial variability in the 1D models. These effects can be further clarified through an examination of the microphysical timescale of silicate cloud formation as shown in Figure \ref{fig:mg_1100_time}. All of the microphysical timescales that we calculate use the steady-state rates of evaporation, condensation, and nucleation from the 2D-ExoCARMA model. The timescales that we consider are the current times that it takes to nucleate, evaporate, or condense a cloud particle, averaged over all cloud particle size bins where nucleation/evaporation/condensation is actively occurring. We find that clouds in the upper atmosphere of the 1D case preferentially form in regions of the atmosphere where nucleation is most efficient. In the 2D case, however, clouds can be transported horizontally to regions where growth is more efficient than nucleation such that the cloud particles that form in the 2D case are more likely to survive and grow to larger sizes than in the 1D case. This causes the 2D case to produce more cloud mass for a given atmospheric structure. 

There are other interesting and non-intuitive impacts of horizontal advection on cloud formation. We examine the illustrative case of KCl clouds, which are present in the 2D-ExoCARMA case while they are notably absent in the 1D CARMA model. The presence of KCl clouds in 2D-ExoCARMA is primarily due to the longitudinally dependent variance in the microphysical timescales of the processes of nucleation, condensational growth, and evaporation. These cloud microphysical timescales are shown in Figure \ref{fig:kcl_1100_time} based off of the 2D-ExoCARMA model. We see that in the region of the atmosphere where clouds are able to nucleate efficiently, they also evaporate quickly. Thus, in the 1D case, the small cloud particles that form very quickly evaporate and a stable cloud layer is not present. However, in the 2D cloud case, there is a larger region of the atmosphere where the clouds are able to grow efficiently than where they are able to nucleate efficiently. Thus, some cloud particles nucleate and are transported to regions of the atmosphere where they can continue to grow. Furthermore, while evaporation happens fairly quickly, some cloud particles are able to survive even outside of the region of efficient growth such that the region where evaporation operates is the largest region of the atmosphere of all three microphysical timescales. Thus the introduction of horizontal transport and the extended region of efficient cloud growth allows for the KCl cloud particles to survive and maintain a stable cloud deck at certain atmospheric regions for planets with equilibrium temperatures of 1200 K or less (see figure \ref{fig:KcMass} and \ref{fig:KcNum}). We note that even in the 2D-ExoCARMA case KCl cloud particles tend to only grow to relatively small particle sizes of $\sim 10^{-2}\;\mu$m (see Figure \ref{fig:num_eastwest_1000}). 

\begin{figure} 
   \centering
   \includegraphics[width=0.47\textwidth]{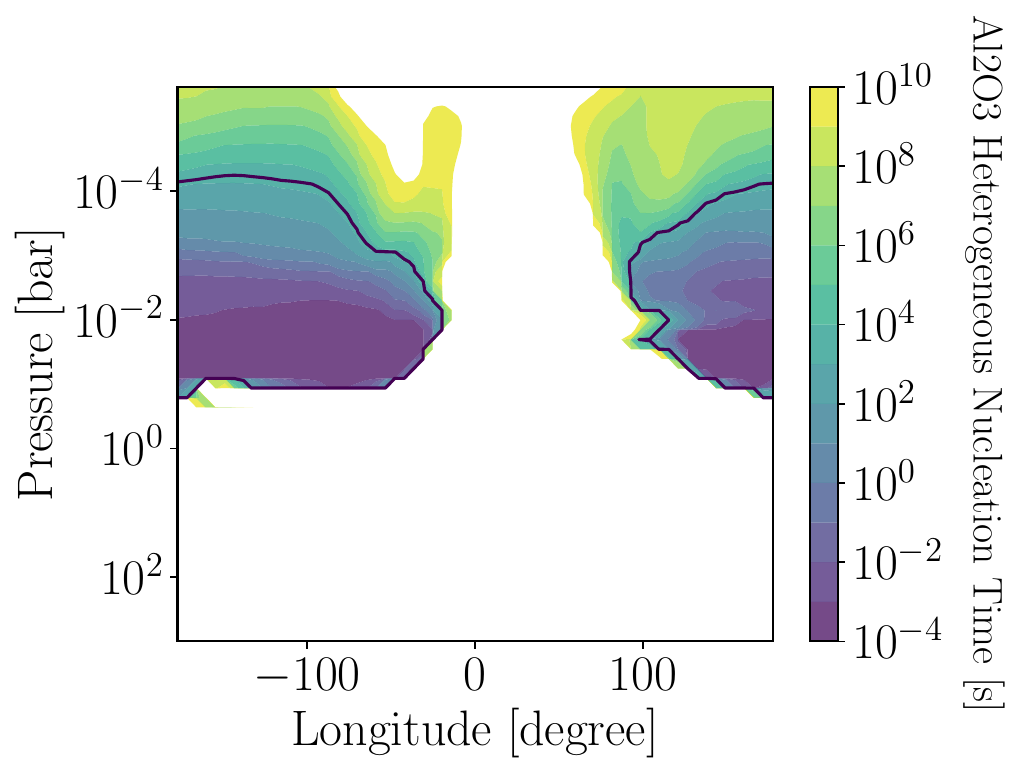}
   \includegraphics[width=0.47\textwidth]{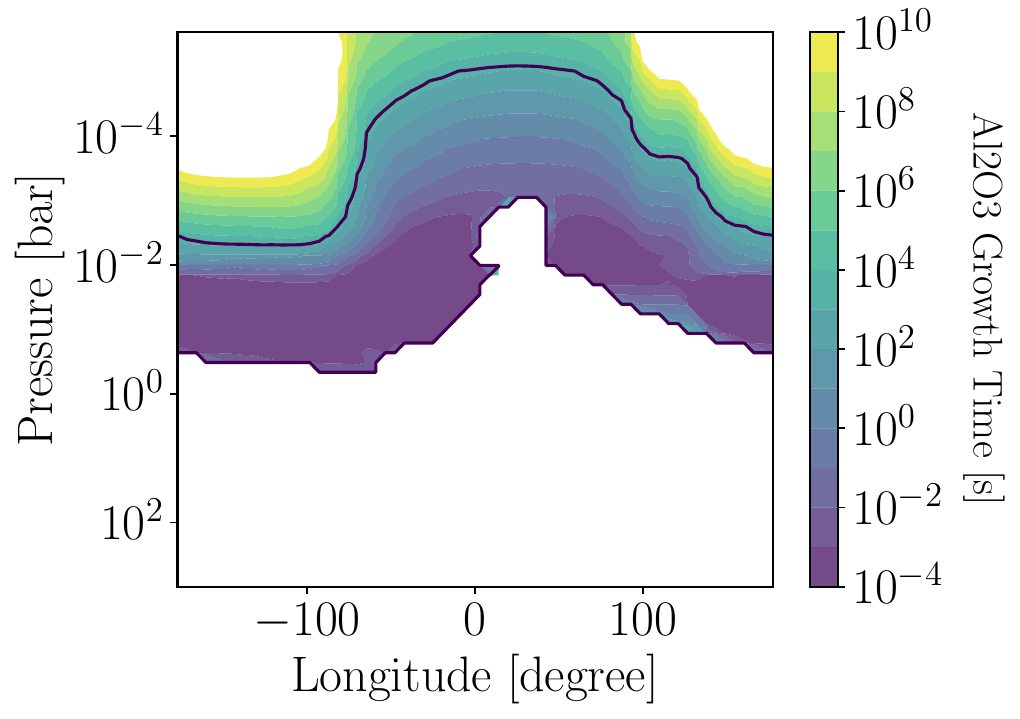}
   \includegraphics[width=0.47\textwidth]{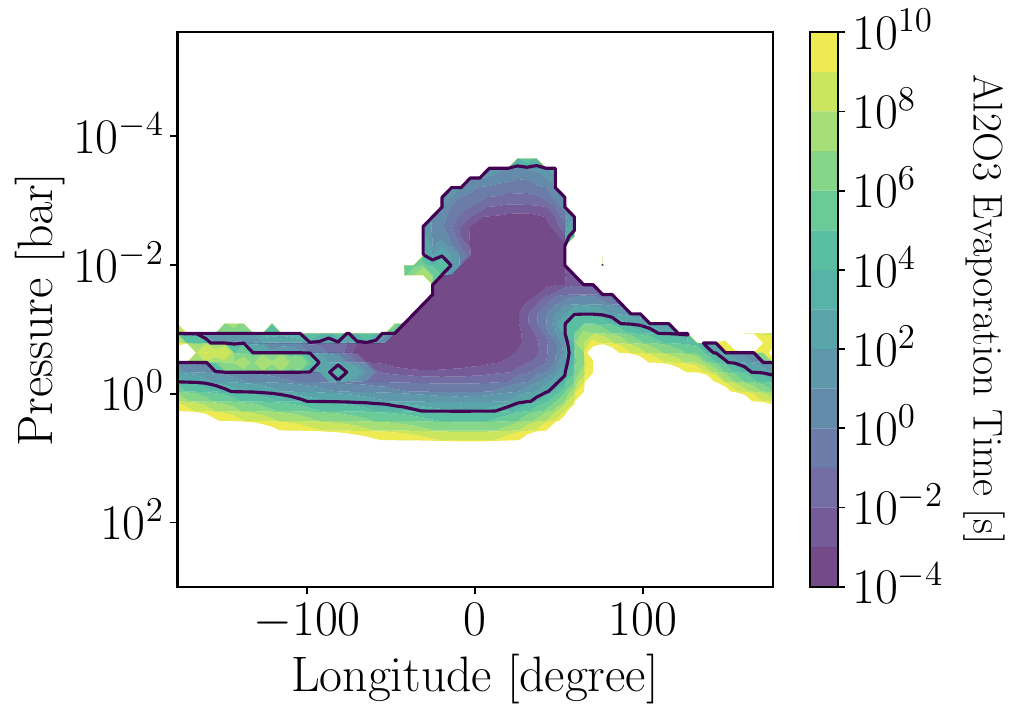}
   \caption{The same as Figure \ref{fig:mg_1100_time} but for cloud particles comprised of Al$_2$O$_3$ and for a planet with an equilibrium temperature of 1800 K.}
   \label{fig:al2o3_1800_time}
\end{figure}

Another striking difference between the 1D and 2D cloud cases for the cooler planets is the differing vertical extents of the clouds in the atmosphere for the cooler planets in our grid. This can be illustrated by the case of Fe and Cr clouds, which differ substantially in atmospheric location between the 2D and 1D cases (see Figure \ref{fig:1d2d_11_comp}). This difference arises from a similar mechanism as the KCl formation mechanism. In the 1D case, Fe and Cr are only able to form a stable cloud layer in the deep atmosphere. While they are able to nucleate in the 1D case, they quickly evaporate such that a stable cloud layer does not form. However, in the 2D case for cooler atmospheres, Fe and Cr clouds that nucleate are quickly transported to regions of the atmosphere where they can grow to sufficiently large sizes that their evaporation timescales become long enough such that lofting and horizontal transport of Fe and Cr cloud particles can occur. Thus, the abundance of Fe and Cr clouds is increased in the 2D-ExoCARMA case.

\subsection{The Impacts of Horizontal Transport for ``Hot" Hot Jupiters}\label{hotdiff}

While there are fewer differences between the 1D and 2D models in the case of hotter atmospheres as shown in Figure \ref{fig:1d2d_18_comp} and predicted in \citet{2019ApJ...887..170P}, there are still significant 2D effects that shape the cloud distributions. In terms of similarities, the clouds in both cases form in similar abundances such that the total cloud mass in both cases is roughly identical. Furthermore, some cloud species like the silicate clouds are present at similar regions in the atmosphere. As shown in Figure \ref{fig:1d2d_18_comp}, the dayside in both cases is mostly free of silicate clouds, although the silicate clouds in the 2D case extend further into the planetary dayside (again note the difference in colorbar scales between the 1D and 2D cases shown here). This is because silicate clouds can be horizontally transported to the planetary dayside for a short time before they evaporate efficiently. 

\begin{figure} 
   \centering
   \includegraphics[width=0.47\textwidth]{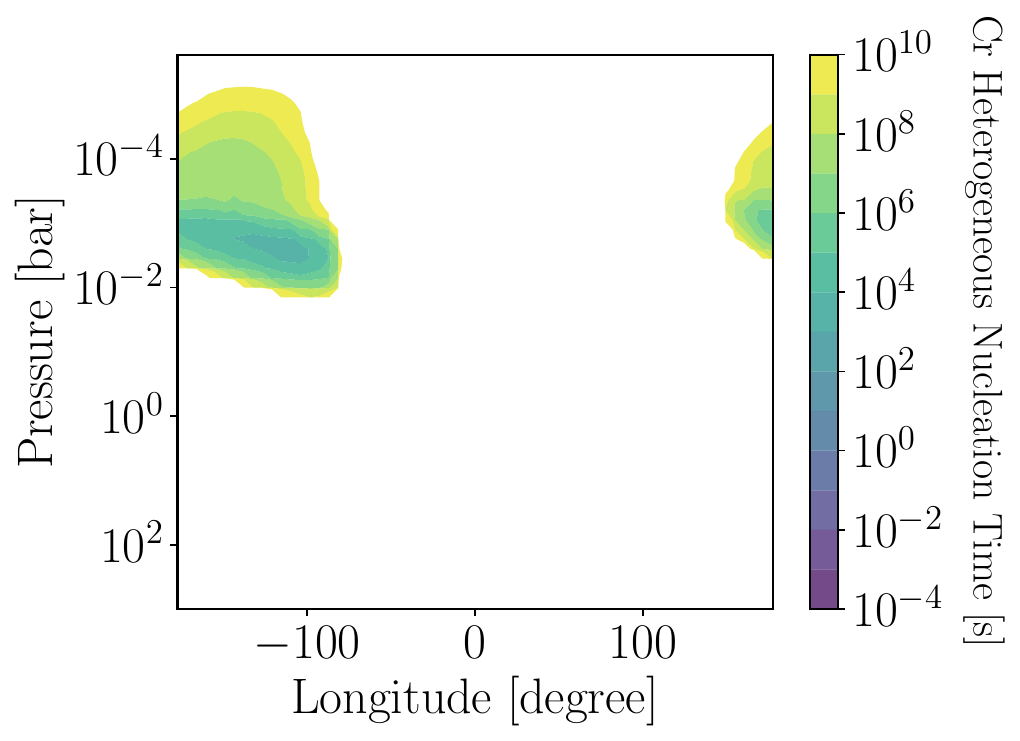}
   \includegraphics[width=0.47\textwidth]{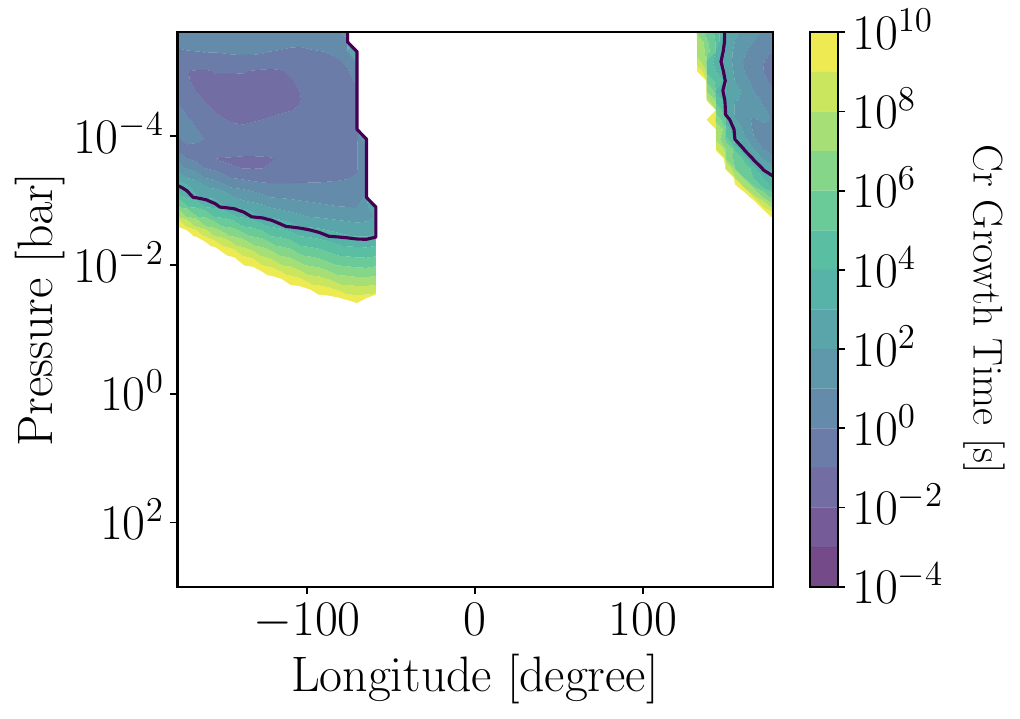}
   \includegraphics[width=0.47\textwidth]{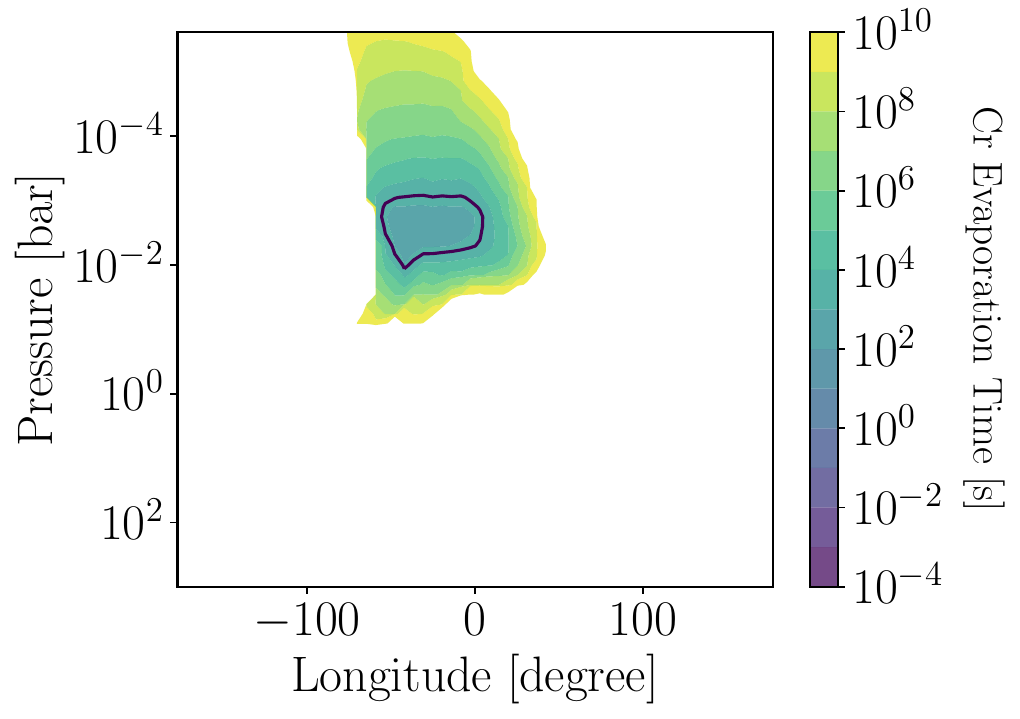}
   \caption{The same as Figure \ref{fig:mg_1100_time} but for cloud particles comprised of Cr and for a planet with an equilibrium temperature of 1800 K.}
   \label{fig:cr_1800_time}
\end{figure}

\begin{figure} 
   \centering
   \includegraphics[width=0.45\textwidth]{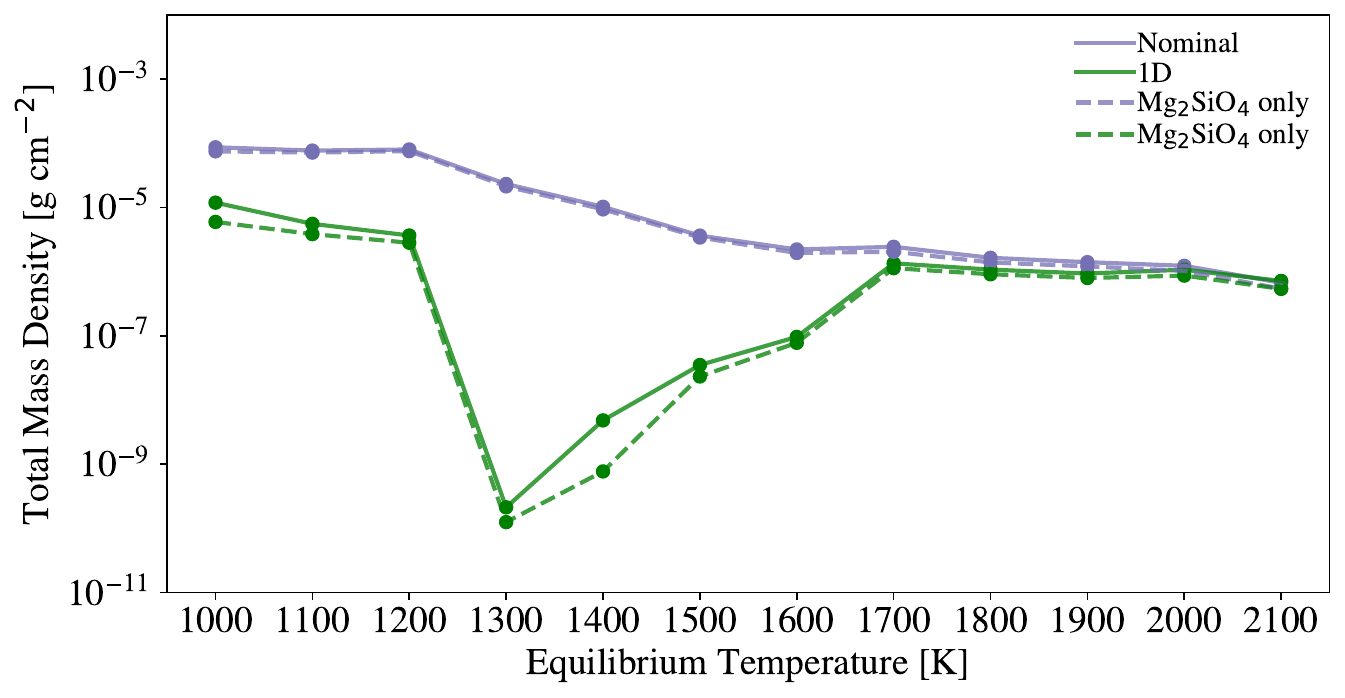}
   \caption{The total condensed cloud mass density (summed across all particle sizes, pressure-levels, and planetary longitudes) varies significantly when horizontal transport is considered (nominal case, periwinkle line) for planets with equilibrium temperatures cooler than 1700 K. Models where horizontal transport is neglected (1D) are shown in green. The dashed lines show the total Mg$_2$SiO$_4$ cloud mass which dominates the total cloud mass density.}
   \label{fig:1dmass}
\end{figure}

A notable difference is that the location of the clouds in the atmosphere varies significantly between the two cases for certain cloud species. This difference is particularly striking for Al$_2$O$_3$ clouds (though we note that the same effect occurs for TiO$_2$ clouds to a lesser extent), which are not present on the dayside in the 1D case. This can be explained by the cloud microphysical timescales as shown in Figure \ref{fig:al2o3_1800_time}. Aluminum clouds are not able to nucleate efficiently on the dayside of the planet and can only do so on the nightside and planetary limbs. Thus, in the 1D case, clouds are only present in atmospheric regions where they are able to nucleate and thus cannot exist on the planetary dayside. The regions of the planet where growth can occur, however, extend across the planetary dayside such that clouds that nucleate and form on the nightside or limbs can be transported across the planet and survive on the dayside. While evaporation is more efficient on the dayside of the planet than the planetary limbs, there still exist regions of the atmosphere where the evaporation of aluminum clouds do not occur. Thus, in 2D for certain atmospheric conditions, clouds can survive the passage across the planetary dayside. 

While in almost all cases we consider the efficiency of cloud formation in 2D is enhanced compared to 1D, there is a notable exception to this trend for the case of Cr clouds in hotter atmospheres. As shown in Figure \ref{fig:1d2d_18_comp}, Cr clouds are not abundant in the 2D case (though note that some Cr clouds do form in the 2D case as shown in Figure \ref{fig:CrNum}). The reason for the decreased abundance of Cr clouds in the 2D case can be understood by the microphysical timescales of Cr cloud formation as shown in Figure \ref{fig:cr_1800_time}. While Cr clouds can nucleate both homogeneously and heterogeneously in our modeling setup, the dominant formation pathway in the majority of the model cases presented here is heterogeneous nucleation. While Cr clouds can heterogeneously nucleate in the regions of the atmosphere where they are abundant in the 1D case, the timescale of heterogeneous nucleation is relatively long (and is significantly longer than the Cr nucleation timescale in cooler atmospheres where Cr is significantly more supersaturated). Thus, in the 2D case for hotter atmospheres, very few particles are able to nucleate in the nucleation regions before they are transported to a region of the planet where evaporation can occur. Thus, relatively few Cr cloud particles nucleate and grow in the 2D case. This is in direct contrast for the other cloud species with substantially shorter nucleation timescales. For clouds other than Cr and for Cr formation in cooler atmospheres, nucleation is sufficiently fast such that particles can nucleate and grow to larger sizes that are less vulnerable to fast evaporation. Thus, while 2D effects generally enhance cloud formation, 2D effects can also inhibit formation efficiency depending on the interplay of the microphysical and advection timescales. 

\subsection{The Impact of Horizontal Advection as a Function of Planetary Equilibrium Temperature}

We can now examine the impacts of horizontal transport on the broad grid of models. We first examine how the total cloud mass changes as a function of including horizontal advection in Figure \ref{fig:1dmass}. In all cases, the total cloud mass is dominated by Mg$_2$SiO$_4$ clouds. We find that the coolest planets in our sample, with equilibrium temperatures of 1000 - 1200 K, the total cloud mass formed in the 1D case is around an order of magnitude less than in the 2D case due to the decreased formation efficiency of silicate clouds in the cooler planets in our sample as discussed in Section \ref{cooldiff}. For planets with equilibrium temperatures of 1300 - 1400 K, the cloud mass in the 1D case is lower than in the 2D case by several orders of magnitude. The difference is primarily due to a lower cloud formation efficiency in the lower atmosphere in the 1D case where silicate clouds are only marginally supersaturated. The boost in cloud formation efficiency in the 2D models, however, allows for substantial silicate cloud formation in the narrow region of pressures in the lower atmosphere. For planets with equilibrium temperatures of 1500 - 1600 K, there is a lowered total cloud mass due to the efficient transport and survival of silicate clouds in the 2D case across the dayside of the planet that is not possible in the 1D case without horizontal advection. Finally, for planets with equilibrium temperatures of 1700 - 2100 K, the total condensed cloud mass is roughly equivalent between the 1D and 2D cases although the distribution of the cloud mass across the planet varies significantly between these two cases. 

The more detailed differences between the 2D and 1D cases can be seen in a comparison of the cloud mass distribution of the nominal 2D case in Figures \ref{fig:nom_mg} and \ref{fig:nom_all} and the 1D case in Figures \ref{1D_mg} and \ref{1D_all}. The differences in distributions between the 2D and 1D cases for planets with equilibrium temperatures of 1000-1200 K can be understood by the discussion presented in Section \ref{cooldiff}. Though we note that the increased abundance of Fe clouds on the west limb of the 1200 K hot Jupiter in the 1D case (see Figure \ref{1D_all}) is due to the sufficiently low temperatures around the west limb allowing for the efficient heterogeneous nucleation of Fe in the upper atmosphere. Similarly, the differences in the 2D and 1D cloud mass distributions for planets with equilibrium temperatures of 1700 - 2100 K can be understood by the discussion presented in Section \ref{hotdiff}. 

For planets with equilibrium temperatures of 1300-1400 K, several effects contribute to lowering the cloud formation efficiency. Fe and silicate clouds are not able to form efficiently in the lower atmosphere in the 1D case (Figures \ref{1D_mg} and \ref{1D_all}) because there exists only a narrow region of the atmosphere ($\sim$10 bar) where these clouds can nucleate. In the 2D case, the cloud particles that do nucleate in the deep atmosphere are transported and able to grow and survive across the entire planet. In the case of Fe clouds, which are relatively resilient to evaporation once they reach large particle sizes, these clouds can even survive below the 10 bar cloud base due to a relatively slow evaporation timescales. Furthermore, all cloud formation efficiencies are decreased in the upper atmosphere when horizontal transport is neglected. In the 2D case, clouds that form in the upper atmosphere can be horizontally transported away from their nucleation regions to regions where growth is efficient such that the distribution of clouds throughout atmosphere is homogenized. In the 1D case, however, this can not occur and the resultant cloud distributions are only abundant in regions where nucleation and growth are particularly efficient. 

With the increase in vertical transport efficiency at higher planetary equilibrium temperatures, the differences between 1D and 2D in the total cloud mass density begin to diminish as the formation of clouds in the upper atmosphere becomes more efficient even without the addition of horizontal transport. However, the lack of horizontal advection for planets with equilibrium temperatures of 1500-1600 K causes there to be a decreased abundance of Fe and Cr clouds on the planetary dayside, which lowers the cloud mass in the 1D case compared to the 2D case. 

\textit{Thus, the dominant effect of horizontal transport in hot Jupiter atmospheres is an increase in cloud formation efficiency for cloud species with relatively short formation timescales (see the note on Cr cloud formation in 2D in Section \ref{hotdiff}) a longitudinal homogenization of cloud properties.} 

\begin{figure*} 
   \centering
   \includegraphics[width=0.65\textwidth]{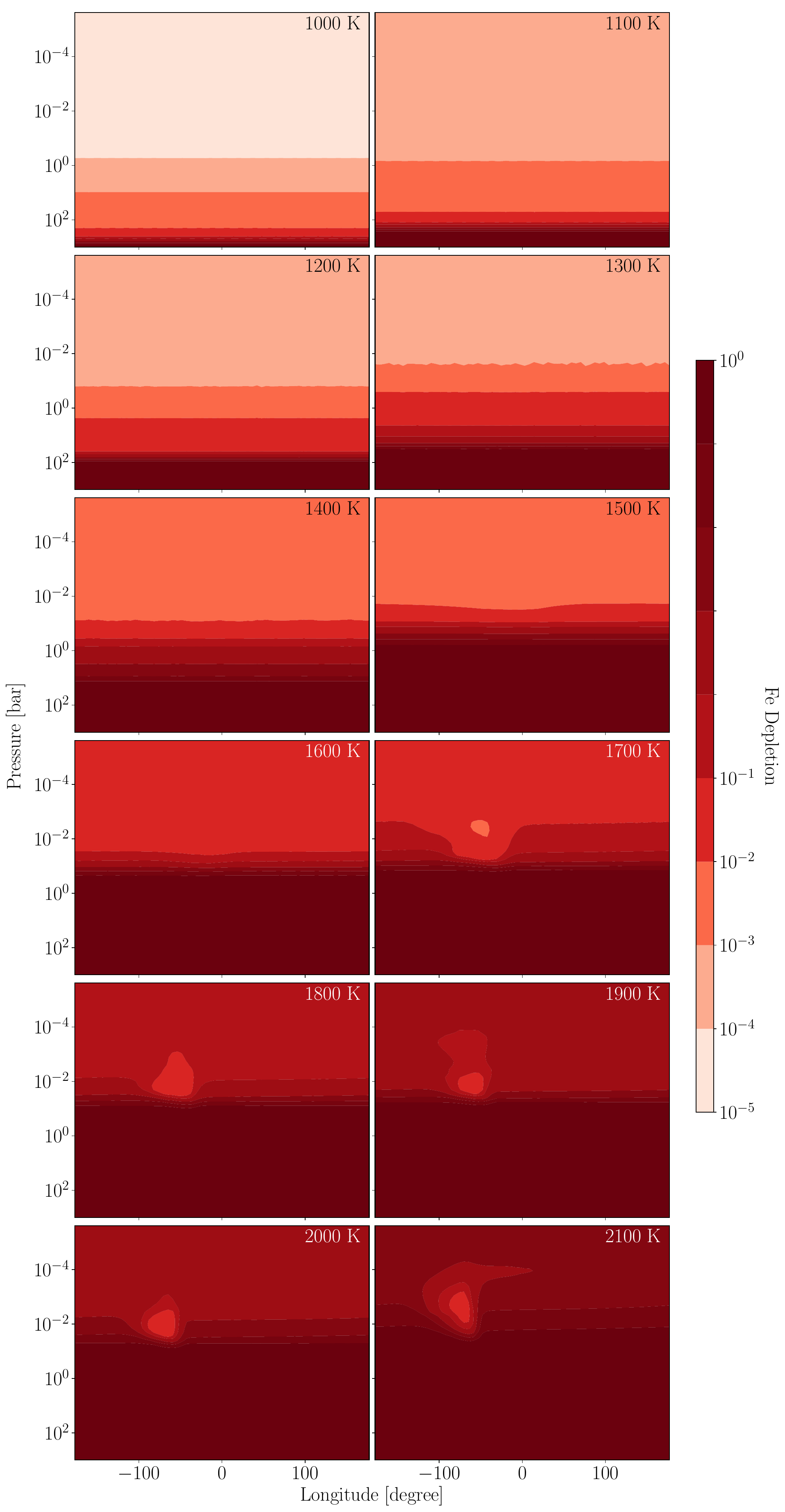}
   \caption{The depletion (the current partial pressure depleted by cloud formation divided by the initial partial pressure) of Fe gas is vertically inhomogeneous and varies longitudinally for the hotter planets in our model grid. Lighter colors indicate an increased depletion of Fe.}
   \label{fig:FeDeplete}
\end{figure*}

\section{Cloud Formation Induced Inhomogeneous Trace Gas Depletion}\label{deplete}
A natural consequence of the formation of condensible clouds is the depletion of condensible gases. The depletion of condensible gases directly impacts which gas-phase species will be visible in atmospheric spectra and has important consequences for gas-phase atmospheric chemistry \citep[e.g.,][]{visscher-etal-2010}. For example, if a gaseous species is not depleted to the level predicted by equilibrium chemistry then we may expect that the resultant gas-phase chemistry will be altered with time and that there may even be other condensible species that have favorable formation conditions. With 2D-ExoCARMA we are able to determine the non-equilibrium depletion of gaseous species as a function of atmospheric pressure and planetary longitude. Here we focus on the case of Fe gas depletion as a representative case and show the depletion of the remaining condensible gas species in Figures \ref{fig:TiDeplete} - \ref{fig:AlDeplete}. Here we define gas depletion as the current partial pressure of the condensible gas divided by the initial partial pressure of that gas (for Fe the initial condition is a constant solar abundance mixing ratio with pressure, see Section \ref{themodel} for the other assumed initial gas compositions). 

As shown in Figure \ref{fig:FeDeplete}, at the coolest equilibrium temperatures, Fe gas is significantly depleted in the upper and mid atmosphere across all planetary longitudes. The depletion of Fe gas remains homogeneous across the upper atmosphere until $\sim 1700$ K. Interestingly, while the distribution of Fe clouds becomes inhomogeneous as a function of planetary longitude at equilibrium temperatures of 1500 K and higher (see Figures \ref{fig:FeMass} and \ref{fig:FeNum}), the level of gas depletion in the atmospheres remains homogeneous as a function of planetary longitude at hotter equilibrium temperatures until eventually becoming inhomogeneous at equilibrium temperatures of 1700 K. For the hottest planets in our grid, Fe gas is only marginally depleted around the western limb of the planets. A similar trend holds for the other condensible species that we consider.

\section{Discussion}\label{discuss}

\subsection{Sensitivity to Cloud Microphysical Parameters and the Importance of Material Properties}\label{micro_sense}

To understand the impact of our choice of microphysical parameters that are poorly constrained, such as the desorption energy of each species and the contact angle for each species, we run a model with a different reasonable choice of these parameters. We use the same setup as our nominal model but choose a desorption energy of 0.1 eV (instead of 0.5 eV, see Section \ref{themodel}), which is roughly representative of the minimum desorption energy, which is seen for small molecules as they desorb from silicate grains \citep{1983Ap&SS..94..177S,2015MNRAS.454.3317S,2017MNRAS.472..389S}. We note that a smaller desorption energy generally leads to less efficient cloud formation as it is easier for a molecule to return to the gas phase. We further calculate a contact angle following \citet{owens_wendt} calculated as cos $\theta_c = W_{C,x}/\sigma_x -1$ where $W_{C,x}=2\sqrt{\sigma_x\sigma_C}$. This formulation for the contact angle further leads to less-efficient cloud formation than our nominal case. This case corresponds to the minimum-cloud case in \citet{2019ApJ...887..170P}. 

\begin{figure} 
   \centering
   \includegraphics[width=0.45\textwidth]{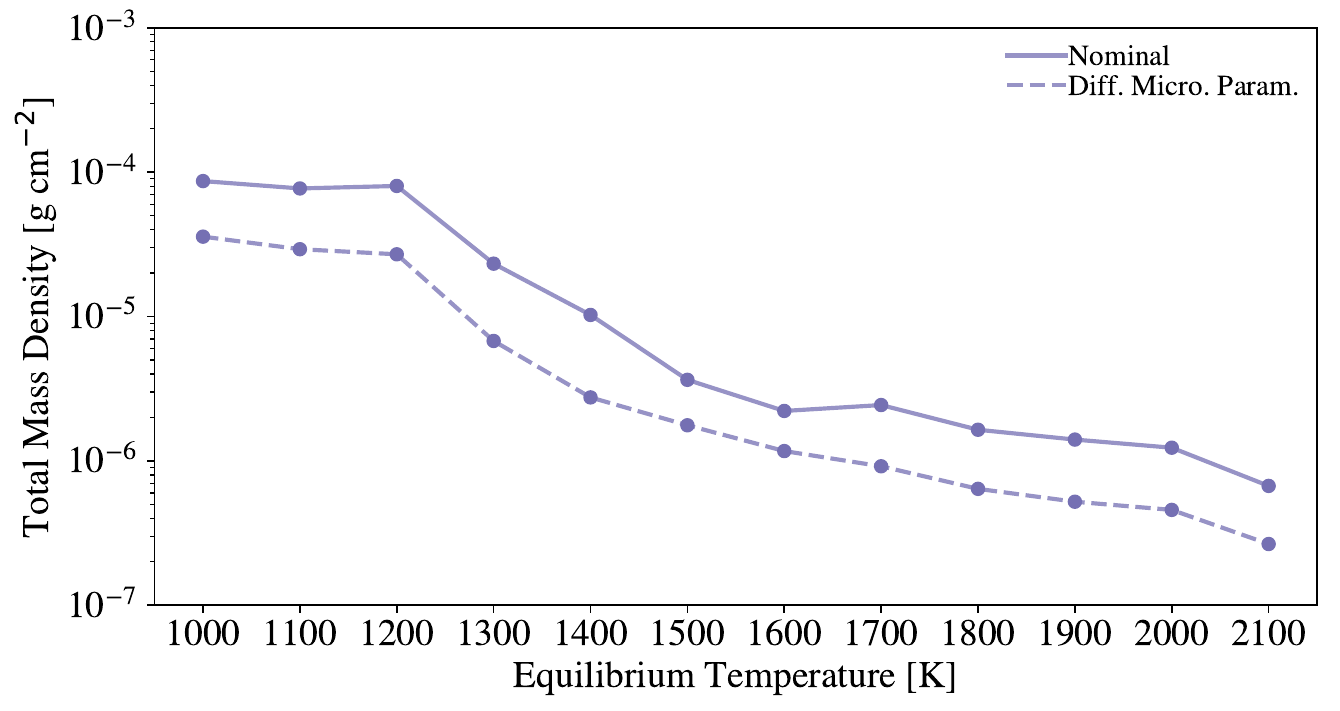}
   \caption{Same as Figure \ref{fig:1dmass} but a comparison of the nominal model runs (solid line) and runs with different microphysical parameters (dashed line). }
   \label{fig:micromass}
\end{figure}

We find that our choices in the nominal case and this minimum-cloud case do not result in substantially different cloud formation behavior. As shown in Figure \ref{fig:micromass}, we see that the general trend in total cloud mass is the same for both the nominal case and the case with differing microphysical parameters although the minimum-clouds case does produce $\sim$70 \% fewer clouds in terms of the total cloud mass. We thus note, that for these relatively unconstrained microphysical parameters, variations in these values are likely to change the efficiency of cloud formation without radically changing the general cloud formation trends. 

As such, the uncertainties in cloud microphysical parameters on the general behavior of cloud formation appear to be substantially less than the uncertainties that arise from an uncertain atmospheric thermal structure or the uncertainties in cloud transport both vertically and horizontally across the planet.

\subsection{Uncertainties in Atmospheric Dynamics}\label{uncertain}
While 3D GCMs tend to predict roughly similar behavior across different models, the numerical values of wind speeds and the efficiency of tracer transport varies significantly across different models with different assumptions about the atmosphere, planetary properties, and external environment. We thus consider the level of horizontal winds and vertical mixing to be a source of significant uncertainty in this work.

Our estimated K$_{zz}$ values are based on the transport of tracer particles, which may prove to be a more robust method of tracing the vertical transport of clouds as compared to calculating mixing based off of wind speeds alone \citep[e.g.,][]{parmentier20133d}. However, the values that we use for the horizontal advection of cloud particles are representative of the equatorially averaged horizontal wind speed at the silicate cloud base. The coupling of cloud particles to these winds may prove to be size-dependent and lower in efficiency than the values for mixing presented in this work. In particular, we assume that all particles are perfectly coupled to the atmospheric wind speeds. However, previous work that examines the vertical mixing of tracer particles often finds that their mixing is reduced as compared to simplified assumptions based on the gas vertical velocities alone \citep[e.g.,][]{parmentier20133d,2019ApJ...887..170P}. We thus consider the horizontal wind speeds used here to be a roughly approximate upper-limit in the efficiency of cloud particle horizontal transport. 

The dominant atmospheric circulation pattern for the planets in our sample consists of a super-rotating equatorial jet \citep{parmentier2016transitions}. This circulation pattern on hot Jupiters and other highly-irradiated tidally-locked planets is a common outcome across a broad range of GCMs \citep[e.g.,][]{showman-guillot-2002,rauscher-menou-2010,mayne2014unified} and indications of this feature have been seen in phase-curve observations \citep[e.g.,][]{knutson-2007,knutson-etal-2009a,zellem-etal-2014}. We note, however, that in the case of significant atmospheric drag (which may be more likely for the hotter planets in our sample) the atmospheric flow pattern can shift to a bulk dayside to nightside flow \citep{showman-etal-2013}. Such a flow structure would alter the pattern of horizontal advection and likely change the resultant cloud properties. 

We note that such a flow pattern is likely for Wasp-18b based off of the thermal emission spectrum using the NIRISS instrument on JWST \citep{Coulombe2023}. Wasp-18b has an equilibrium temperature of $\sim$2000 K, which is comparable to the hottest planets in our grid. Further observations such as these will be essential in informing modeling efforts moving forward and in understanding the dominant atmospheric flow pattern in hot Jupiters across a range of equilibrium temperatures. 

There are other uncertainties of the wind pattern on hot Jupiters because the atmospheric circulation strongly depend on the atmospheric drag, planetary rotation rate, atmopsheric temperature regime, metallicity, etc. Different wind pattern could occur at different regimes \citep[e.g., see Figure 14 in][]{zhang2020}. Future theoretical work with different wind patterns from GCMs in across a broad parameter space should be conducted to systematically investigate the behavior of cloud formation on hot Jupiters. 

\subsection{The Lack Atmospheric Cold Traps}\label{nom_distributions}

The presence or lack of an atmospheric cold-trap could have a significant impact on whether or not clouds are observable in an atmosphere. An atmospheric ``cold trap" can occur when when the process of gravitational settling is more efficient than upward vertical mixing (see \citet{2018ApJ...860...18P,parmentier20133d} for a more detailed description). In this case, cloud particles will efficiently settle after formation and, depending on the strength of the cold trap, the cloud particles may only be abundant near the cloud base. There can also be a ``deep cold trap", which occurs when there are two thermodynamically favorable regions in the atmosphere for cloud formation, either due to an atmospheric thermal inversion or due to an isothermal region of the atmosphere where cloud formation becomes inefficient. 

In this study, the deep atmospheric temperature is high because we have focused on inflated hot Jupiters with hot interiors. In our nominal case, we find that vertical mixing is efficient such that a cold trap does not occur and cloud particles are present across many orders-of-magnitude in pressure space throughout the atmosphere. Thus, for reasonable assumptions regarding vertical mixing, an atmospheric cold-trap sufficiently strong to suppress cloud formation in the upper atmosphere is not likely for the planets in our model grid. 


\subsection{Comparison to Other Cloud Models}

A large grid of models using a different microphysical framework without the horizontal transport of clouds was produced in \citet{Helling2022} where they find that cloud coverage transitions from a homogeneously cloud covered regime, to a regime with an inhomogeneously covered planetary dayside, to a cloud-free dayside regime. We similarly find that the planets in our grid differentiate into these rough regimes although they do so at different planetary temperatures. In our models we find that the planets covered with a homogeneous layer of clouds extends to hotter planetary equilibrium temperatures of $\sim$1500K, a patchy dayside cloud coverage for planets with equilibrium temperatures less than 2100 K and a nearly cloud-free dayside for planets with equilibrium temperatures of 2100 K or greater. However, we note that when we focus on silicate clouds, which are likely to dominate the atmospheric opacity for many of the planets in this temperature range \citep[e.g.,][]{2020NatAs...4..951G,2019ApJ...887..170P,2021ApJ...918L...7G} we find that there are effectively two regimes in planetary cloud cover: planets with temperatures less than $\sim$ 1600 K with a nearly homogeneous coverage of silicate clouds and planets with temperatures of 1700 K or more that have nearly cloud-free planetary daysides. 

In this work we focus on the interaction between cloud formation and horizontal transport and choose this hierarchical modeling approach to facilitate understanding. We do not focus on the fully-3D interaction between clouds and atmospheric mixing, as in e.g., \citet{2016A&A...594A..48L,2019MNRAS.488.1332L,lee2023}, due to computational constraints, which make a large grid of such models unfeasible. We also do not yet couple our microphysical model to a model of radiative feedback (as in \citet{roman2019} for non-microphysical clouds in 3D). 

Upcoming work will focus on the observational impact of the cloud models presented in this work in detail. We thus reserve comparisons to the conclusions found in \citet{2020NatAs...4..951G} and \citet{2021ApJ...918L...7G} for future work. 


\section{Summary and Conclusions}\label{conclude}
We present the first two-dimensional bin-scheme microphysical model of cloud formation on hot Jupiters. This framework predicts the detailed properties of cloud particles from first principles with the inclusion of how these properties are shaped by horizontal advection. We demonstrate that horizontal advection fundamentally shapes cloud properties and will be essential to consider when interpreting observations of planetary atmospheres. 

In this work, we summarize the model updates we use to consider the horizontal advection of clouds. We consider cloud formation in a representative grid of planets from 1000 - 2100 K and discuss the cloud properties for the planets in this sample. We discuss the interplay between horizontal advection and cloud formation through comparing our 2D models with 1D models without horizontal advection. We consider how the timescales of material transport and cloud microphysical processes interact to shape cloud distributions in 2D. We also discuss how cloud formation depletes the condensible gas phase species and can do so inhomogeneously across the planet. We finally discuss sensitivities of our model and differences between this model and other modeling approaches. Our main conclusions are summarized below. 

\begin{enumerate}
    \item Cloud properties are strongly sensitive to dynamical transport.
    \item For hot Jupiters in our grid with equilibrium temperatures of 1000 - 1400 K, the cloud distributions are relatively homogeneous across the planetary atmosphere and the bulk of the cloud mass is located in the lower atmosphere. KCl clouds are able to form near the west limb for the coolest planets in our sample. 
    \item For planets with equilibrium temperatures of 1500 K or higher, the cloud mass is located entirely in the upper atmosphere and cloud cover becomes increasingly inhomogeneous with increased planetary equilibrium temperature. 
    \item The cooler planets in our sample have cloud particle size distributions that are roughly homogeneous on the east and west limbs of the planet, with the exception of a population of KCl clouds that are only present on the west limb of the planet. 
    \item The hotter planets in our sample have very inhomogeneous cloud particle size distributions on the east and west limbs. 
    \item Horizontal advection shapes the cloud properties in hot Jupiter atmospheres primarily by increasing the cloud formation efficiency and longitudinally homogenizing the cloud properties across the planet. 
    \item The impacts of horizontal advection on the cloud mass density and the distribution of cloud particles varies as a function of planetary equilibrium temperature with the most noticeable effects present for the cooler planets in the sample. 
    \item Cloud formation depletes the condensible gas species inhomogeneously in the atmosphere as a function of height and, for the hotter planets in our sample, as a function of longitude. 
    \item Uncertainties in microphysical properties give rise to uncertainties in cloud properties. However, these uncertainties shape cloud properties less than differences in atmospheric thermal structure or atmospheric mixing. 
\end{enumerate}

This work reveals the necessity of gas and cloud transport in shaping cloud properties. Our future works in this series of papers will calculate how these cloud properties shape the transmission, emission, and reflection spectra of this grid of hot Jupiters. 
\\ \\ \\
\indent We thank Peter Gao and Karin \"{O}berg for their useful comments and insightful discussion. D.P. acknowledges support from NASA (the National Aeronautics and Space Administration) through the NASA Hubble Fellowship grant HST-HF2-51490.001-A awarded by the Space Telescope Science Institute, which is operated by the Association of Universities for Research in Astronomy, Inc., for NASA, under contract NAS5-26555. X.Z. acknowledges support from the National Science Foundation grant AST2307463, NASA Exoplanet Research grant 80NSSC22K0236, and the NASA Interdisciplinary Consortia for Astrobiology Research (ICAR) grant 80NSSC21K0597.

\bibliography{references}{}
\bibliographystyle{aasjournal}

\appendix
\section{Cloud Distributions for Individual Condensible Species}
\begin{figure*} 
   \centering
   \includegraphics[width=0.65\textwidth]{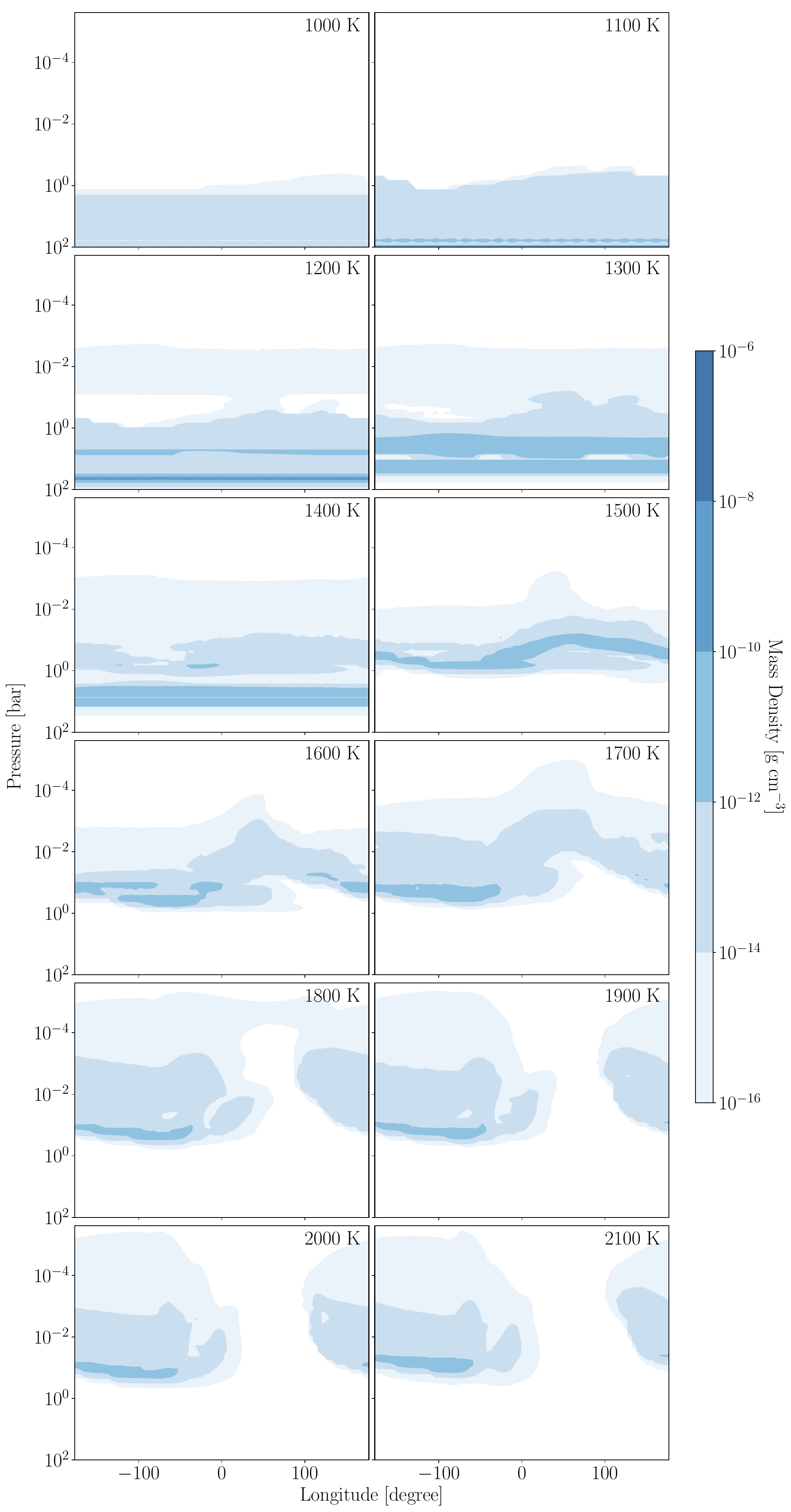}
   \caption{Same as Figure \ref{fig:nom_mg} but for TiO$_2$ clouds. }
   \label{fig:TiMass}
\end{figure*}

\begin{figure*} 
   \centering
   \includegraphics[width=0.65\textwidth]{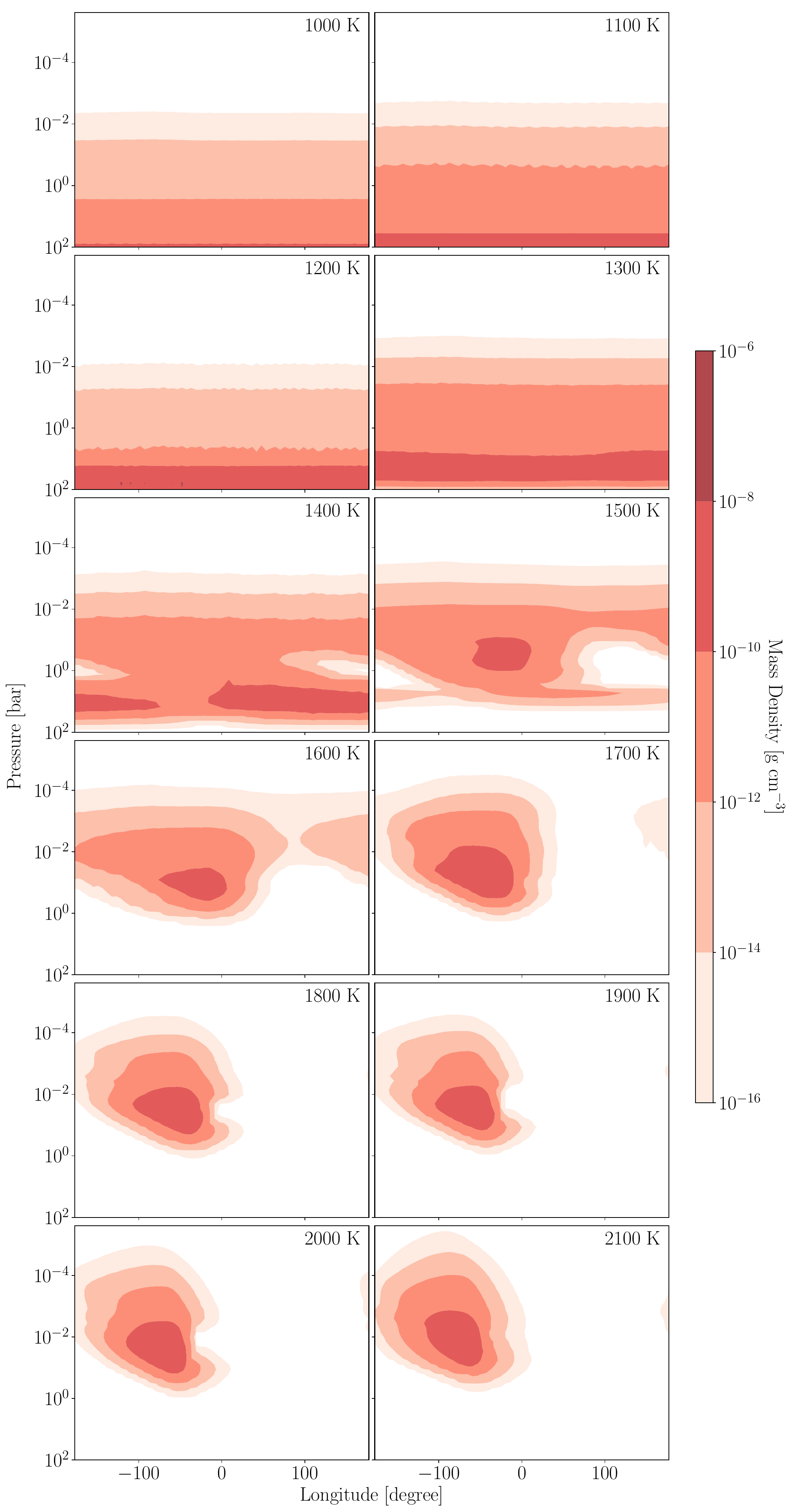}
   \caption{Same as Figure \ref{fig:nom_mg} but for Fe clouds.}
   \label{fig:FeMass}
\end{figure*}

\begin{figure*} 
   \centering
   \includegraphics[width=0.65\textwidth]{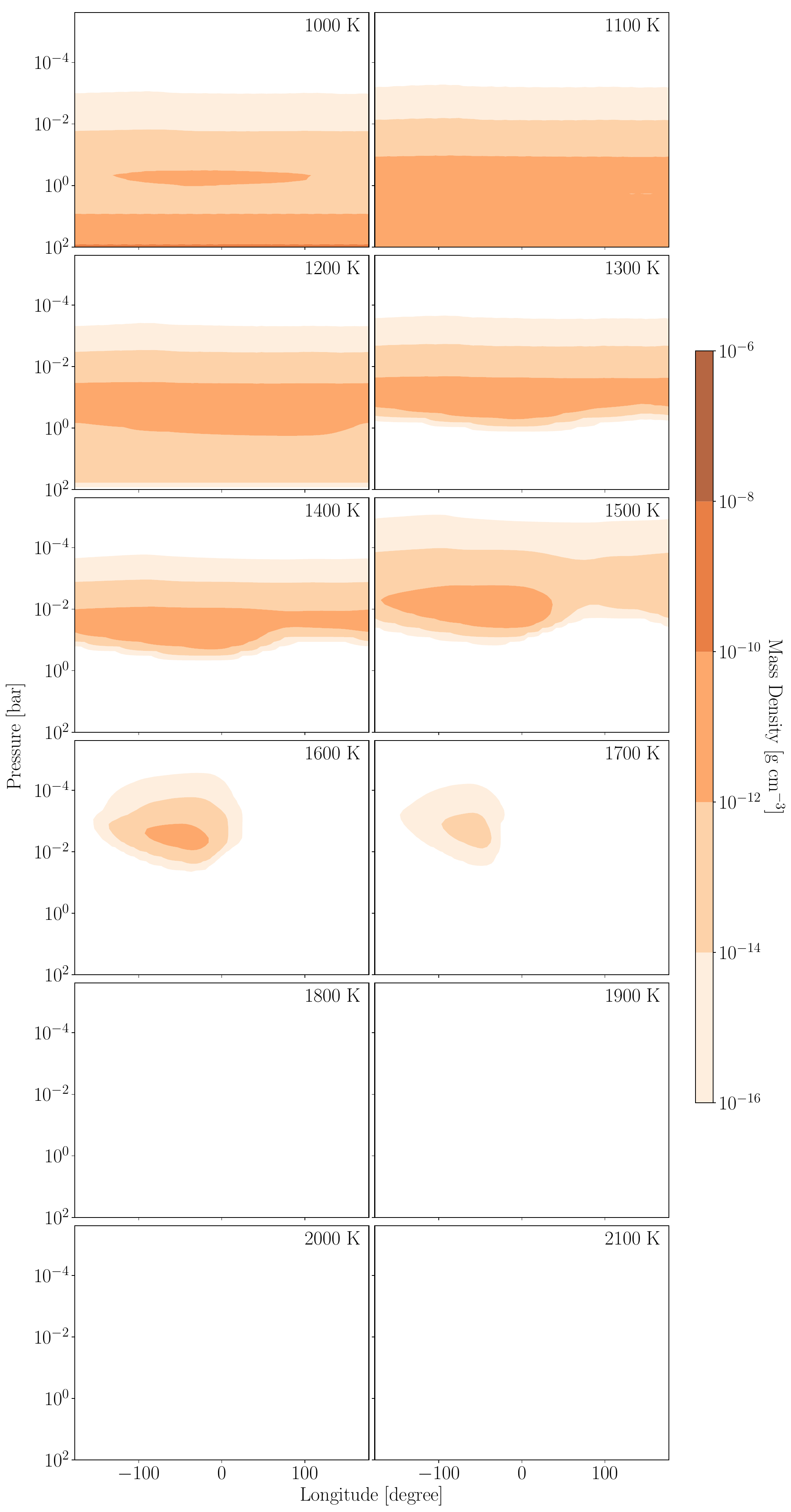}
   \caption{Same as Figure \ref{fig:nom_mg} but for Cr clouds.}
   \label{fig:CrMass}
\end{figure*}

\begin{figure*} 
   \centering
   \includegraphics[width=0.65\textwidth]{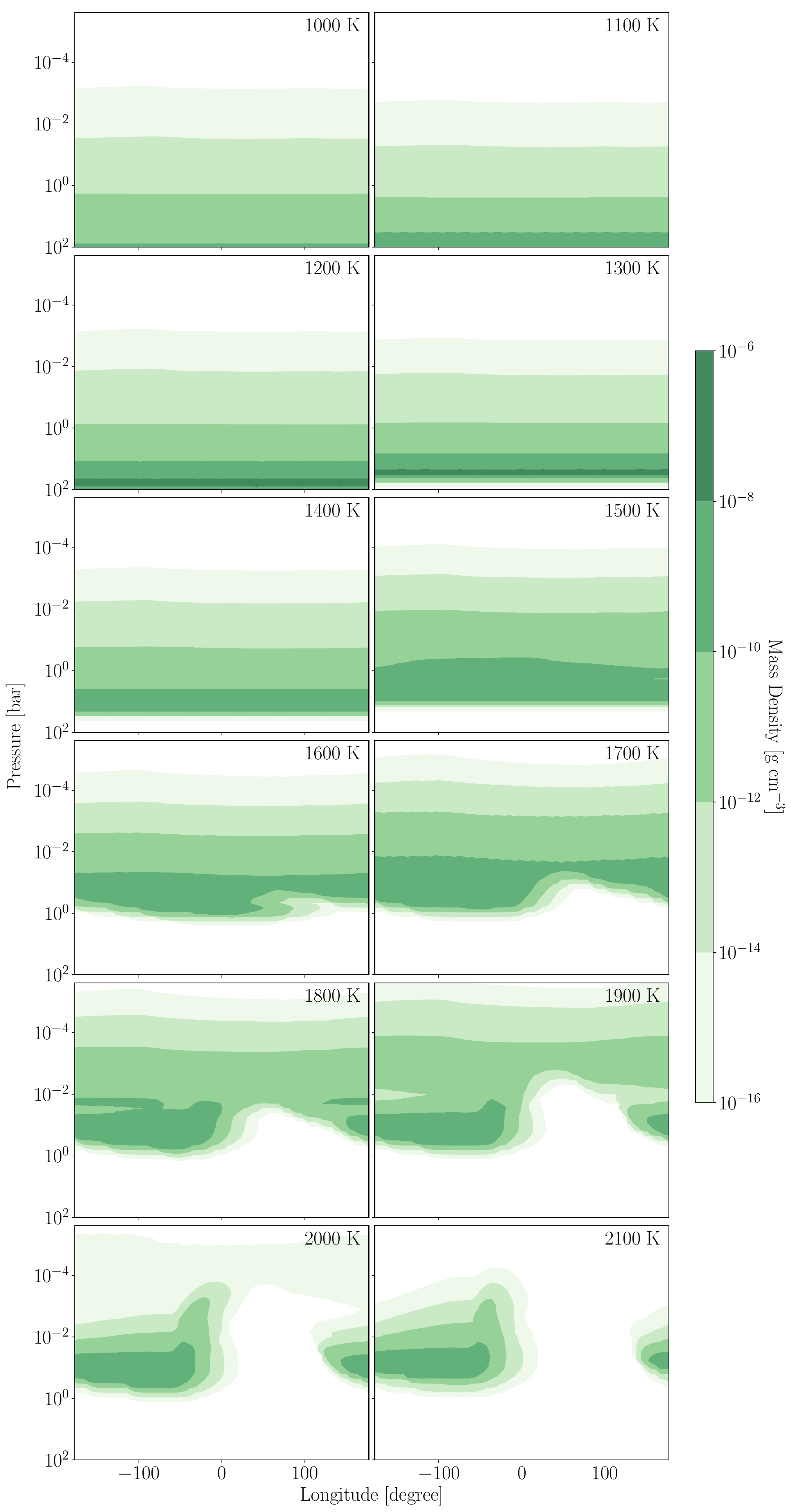}
   \caption{Same as Figure \ref{fig:nom_mg} but for Al$_2$O$_3$ clouds.}
   \label{fig:AlMass}
\end{figure*}

\begin{figure*} 
   \centering
   \includegraphics[width=0.65\textwidth]{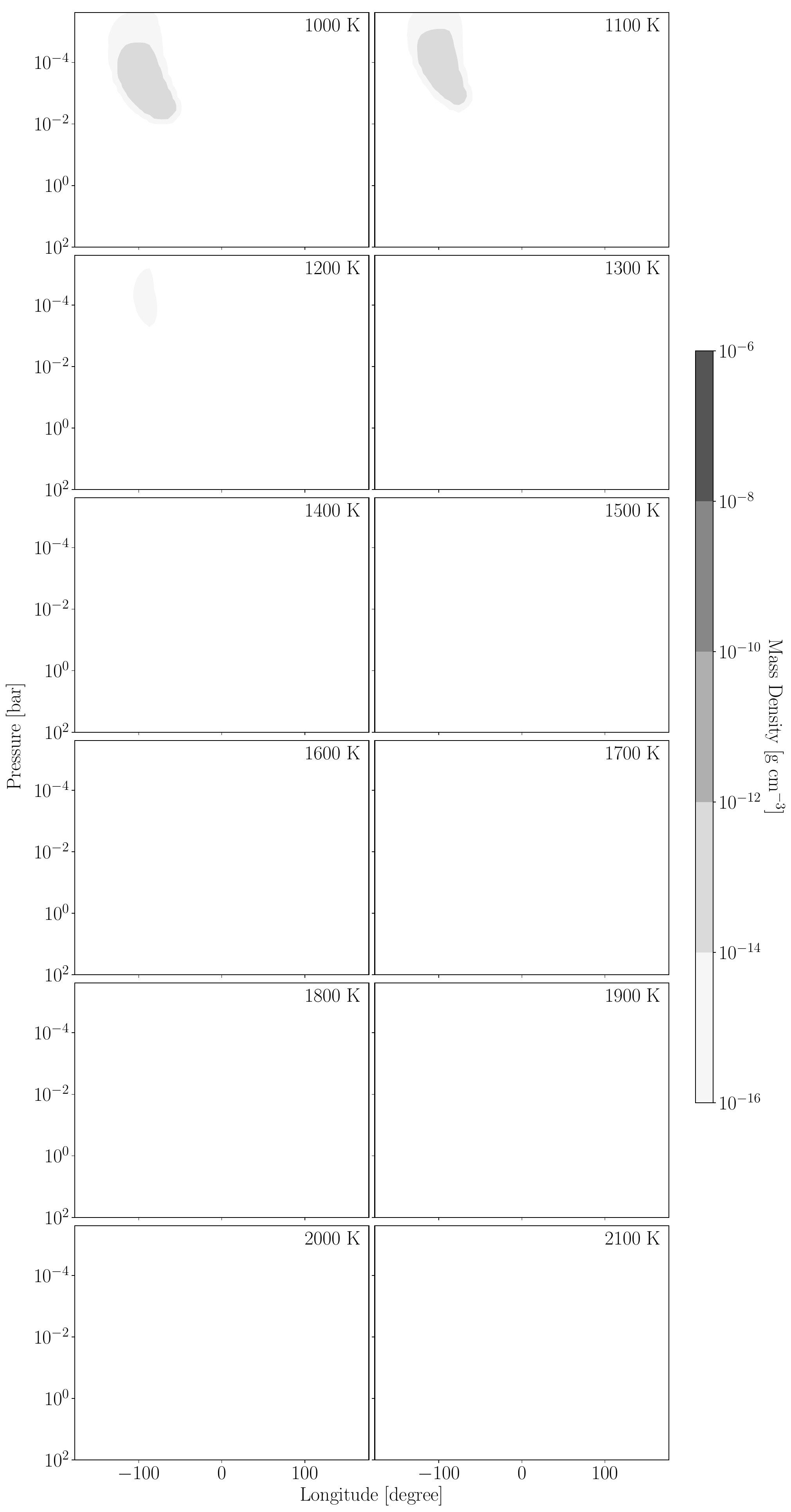}
   \caption{Same as Figure \ref{fig:nom_mg} but for KCl clouds.}
   \label{fig:KcMass}
\end{figure*}

\begin{figure*} 
   \centering
   \includegraphics[width=0.65\textwidth]{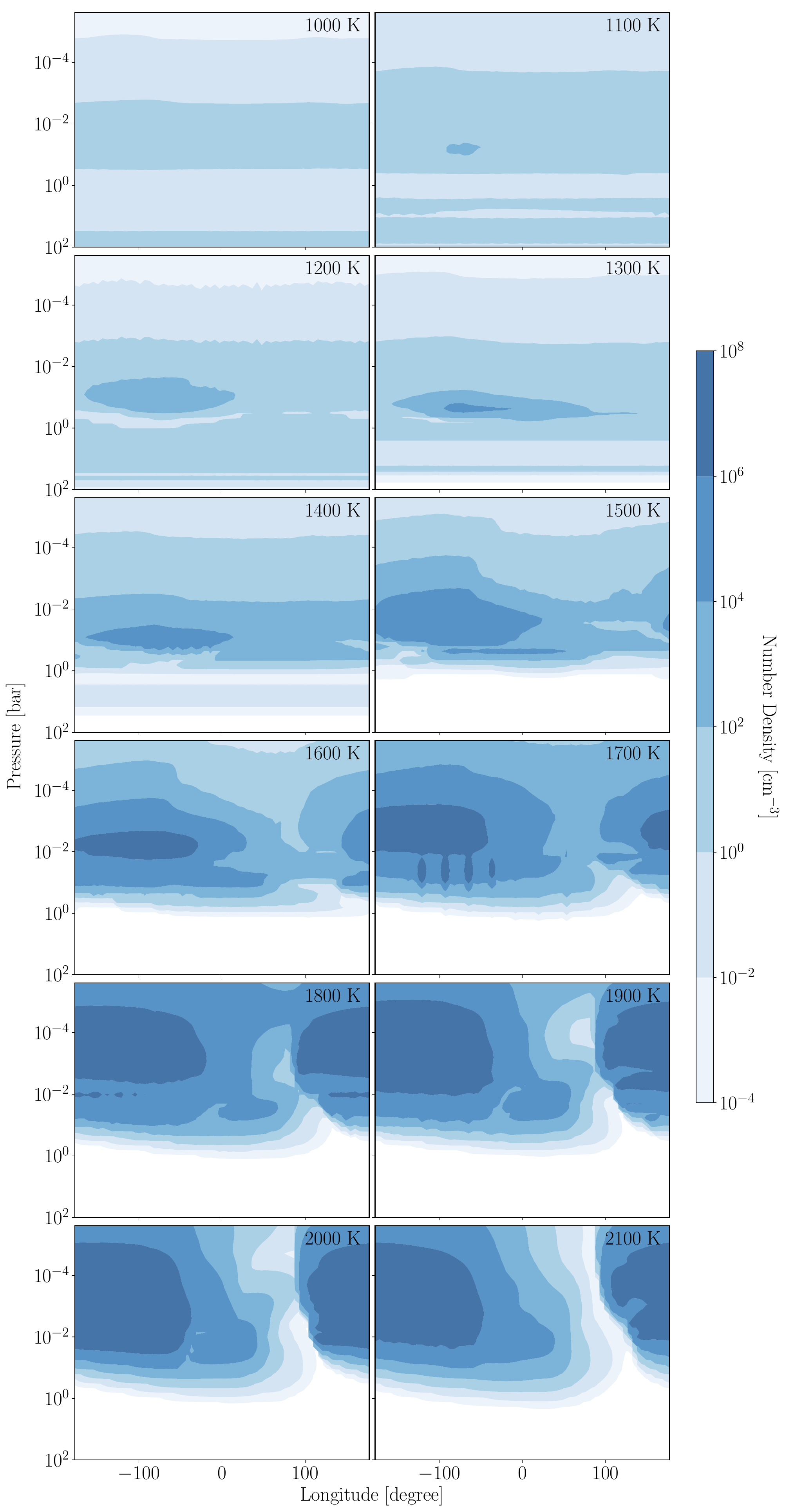}
   \caption{The number density of TiO$_2$ cloud particles varies significantly as a function of planetary equilibrium temperature. Here we show the time-averaged number density of TiO$_2$ clouds as a function of planetary longitude and atmospheric pressure. While 2D-ExoCARMA calculates the full cloud particle size distribution, here we have summed the number density over all cloud particle sizes. }
   \label{fig:TiNum}
\end{figure*}

\begin{figure*} 
   \centering
   \includegraphics[width=0.65\textwidth]{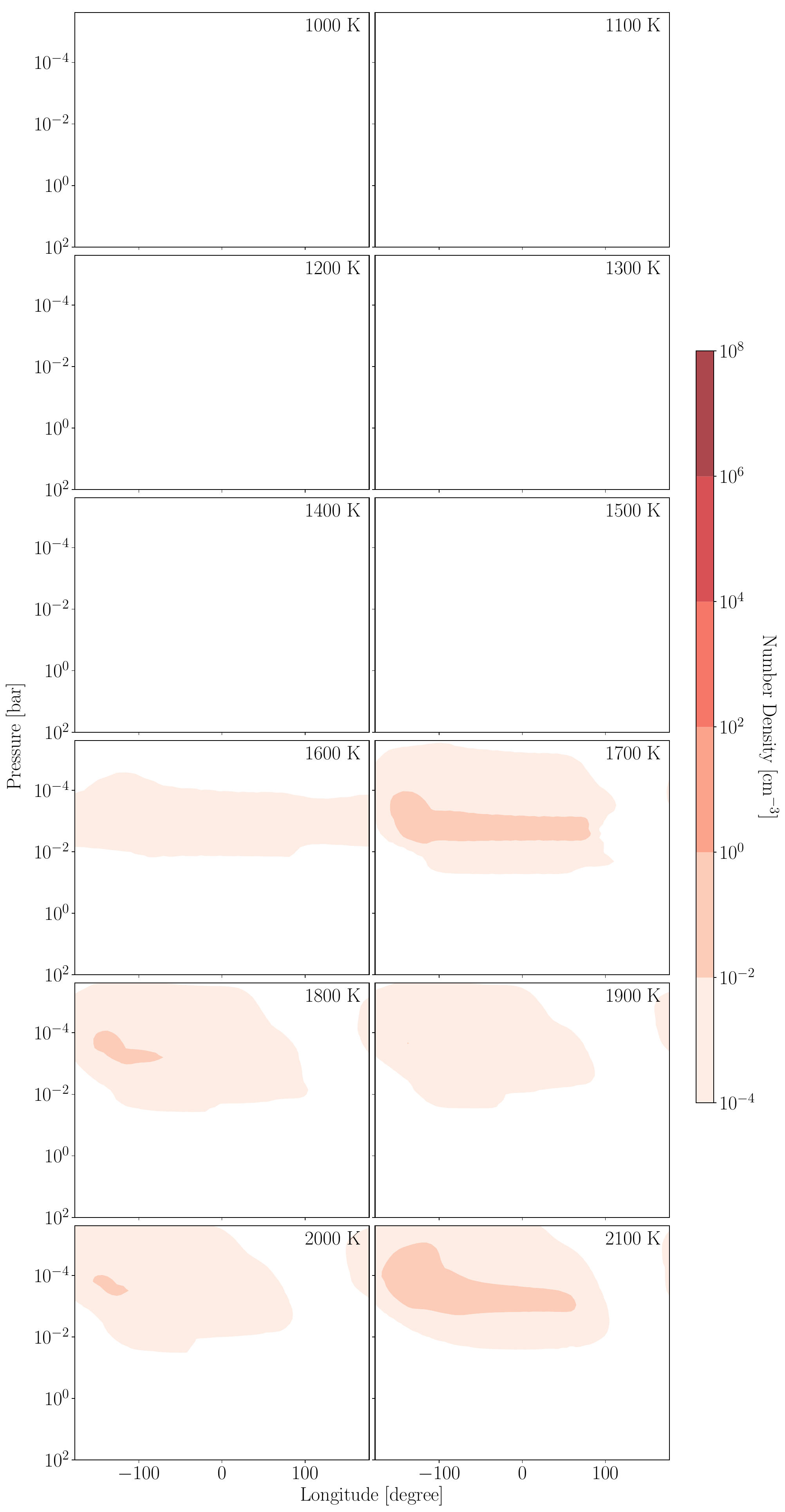}
   \caption{The same as Figure \ref{fig:TiNum} but for Fe clouds. }
   \label{fig:FeNum}
\end{figure*}

\begin{figure*} 
   \centering
   \includegraphics[width=0.65\textwidth]{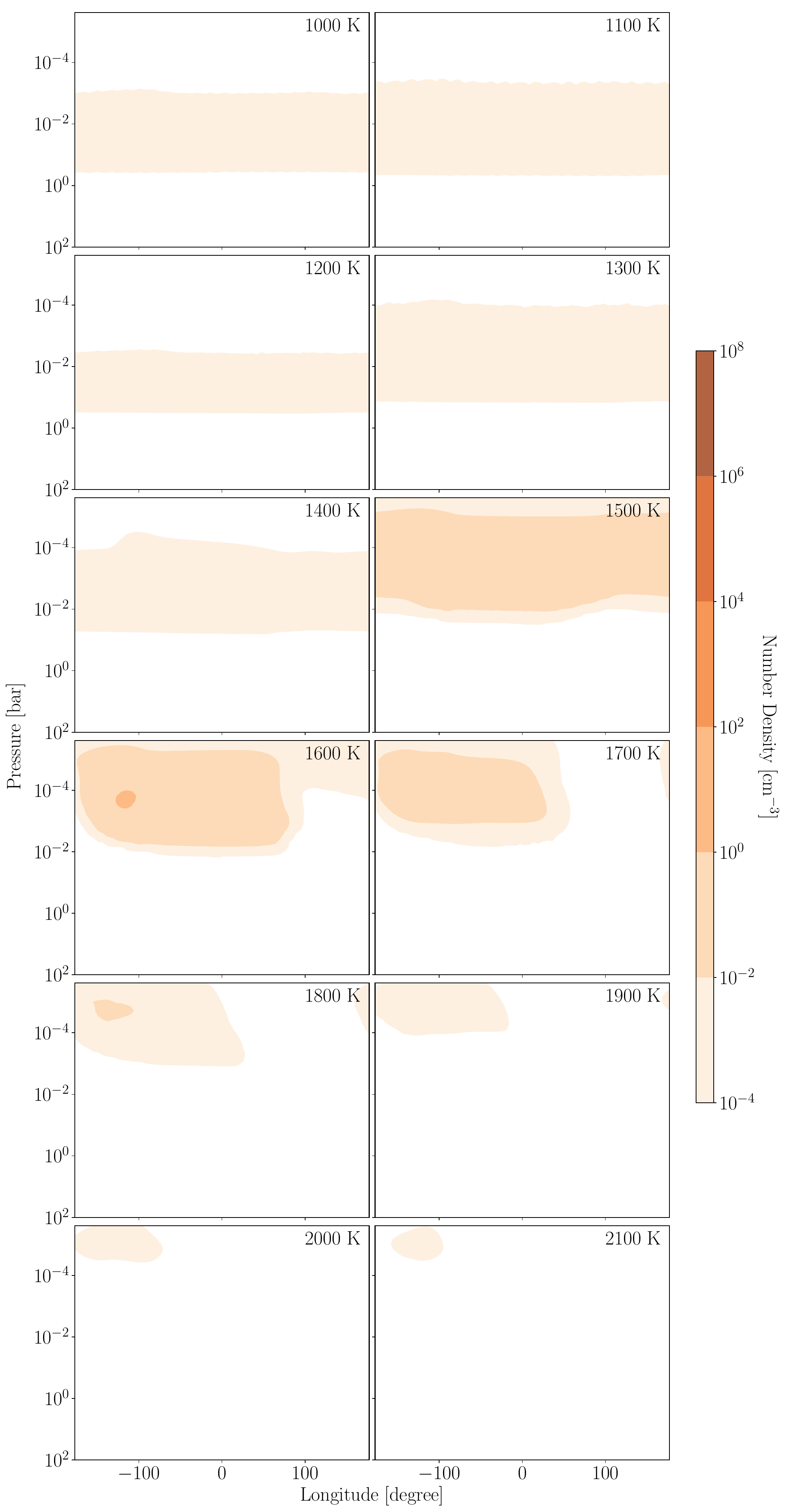}
   \caption{The same as Figure \ref{fig:TiNum} but for Cr clouds.}
   \label{fig:CrNum}
\end{figure*}

\begin{figure*} 
   \centering
   \includegraphics[width=0.65\textwidth]{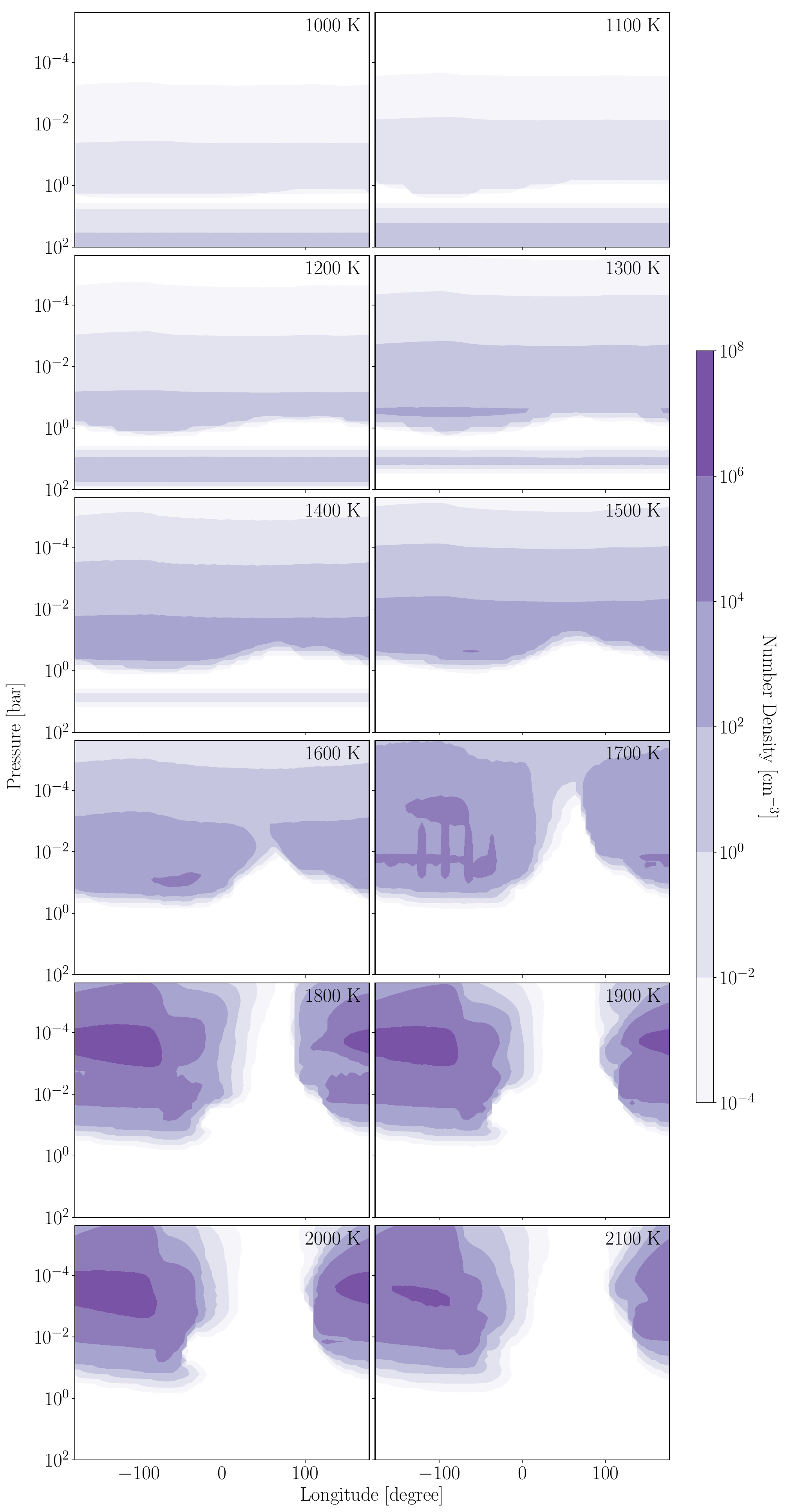}
   \caption{The same as Figure \ref{fig:TiNum} but for Mg$_2$SiO$_4$ clouds.}
   \label{fig:MgNum}
\end{figure*}

\begin{figure*} 
   \centering
   \includegraphics[width=0.65\textwidth]{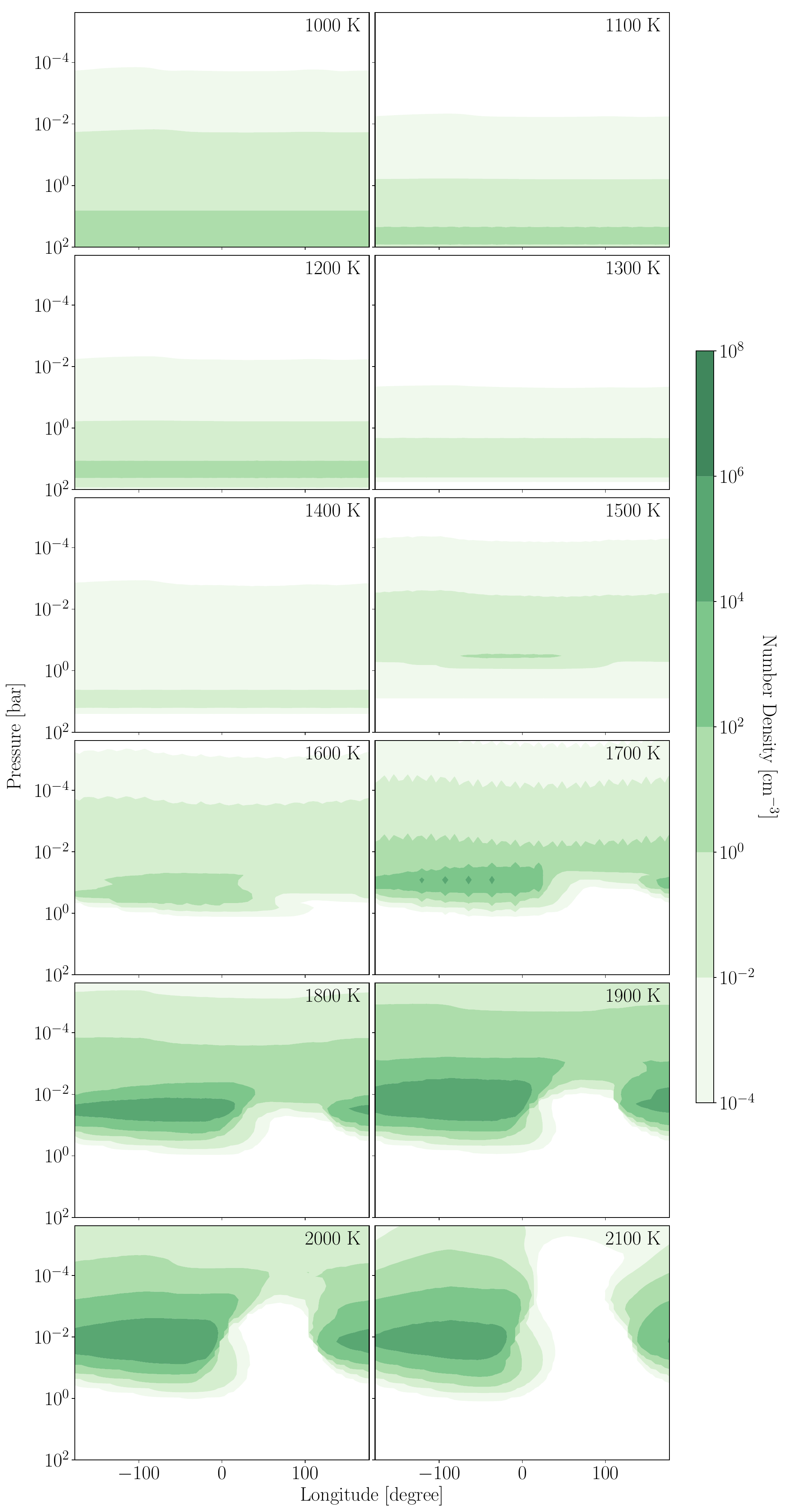}
   \caption{The same as Figure \ref{fig:TiNum} but for Al$_2$O$_3$ clouds.}
   \label{fig:AlNum}
\end{figure*}

\begin{figure*} 
   \centering
   \includegraphics[width=0.65\textwidth]{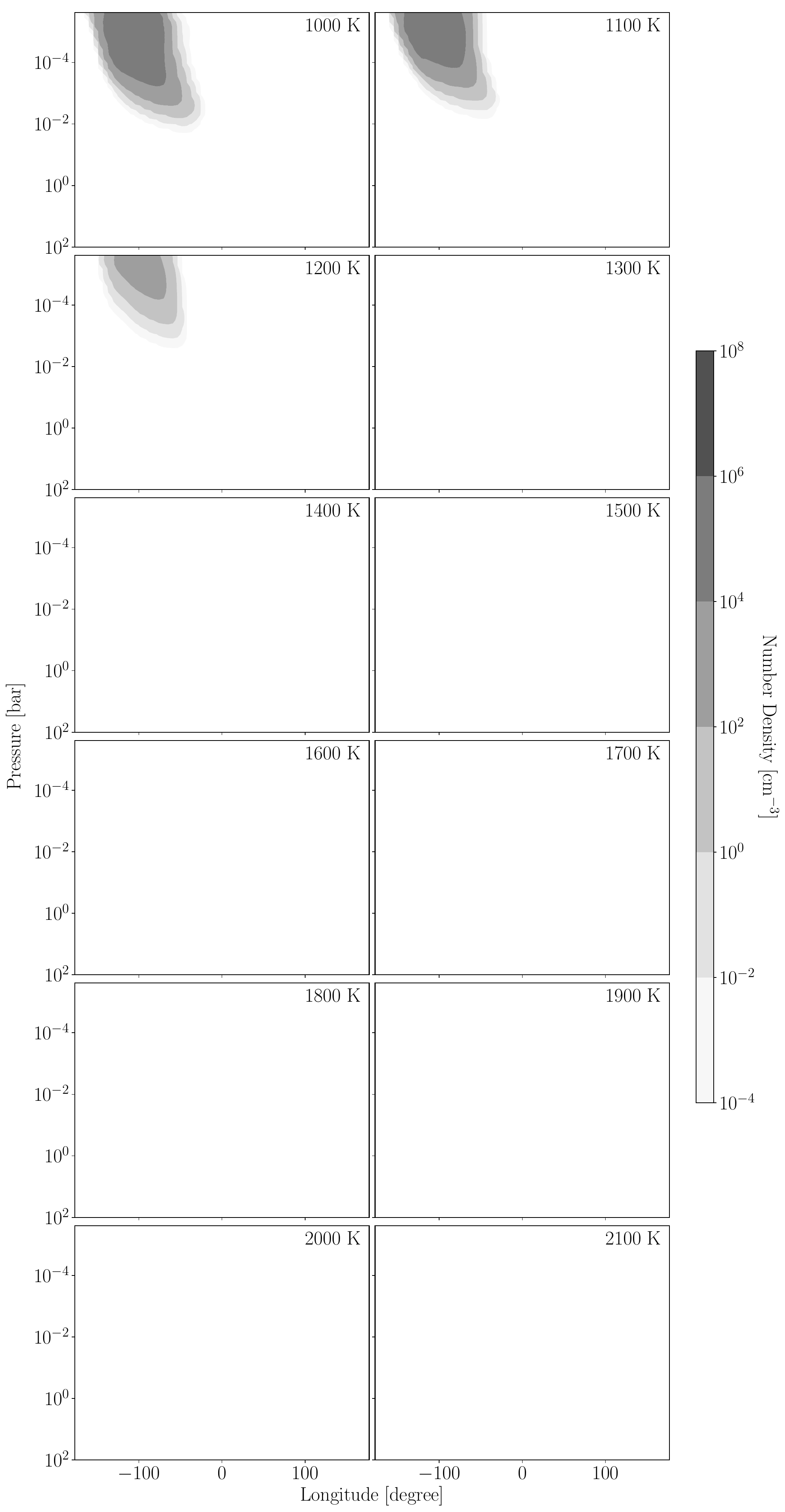}
   \caption{The same as Figure \ref{fig:TiNum} but for KCl clouds.}
   \label{fig:KcNum}
\end{figure*}

Here we present the total cloud mass and number density distributions for each individual cloud species. The mass density distributions of the different cloud species that form are shown in Figures \ref{fig:TiMass}-\ref{fig:KcMass}. The number density distributions of the different cloud species that form as a function of longitude and pressure for every planet in our sample are shown in Figures \ref{fig:TiNum}-\ref{fig:KcNum}.

\section{1D Cloud Distributions Without Horizontal Advection}
\begin{figure*} 
   \centering
   \includegraphics[width=0.65\textwidth]{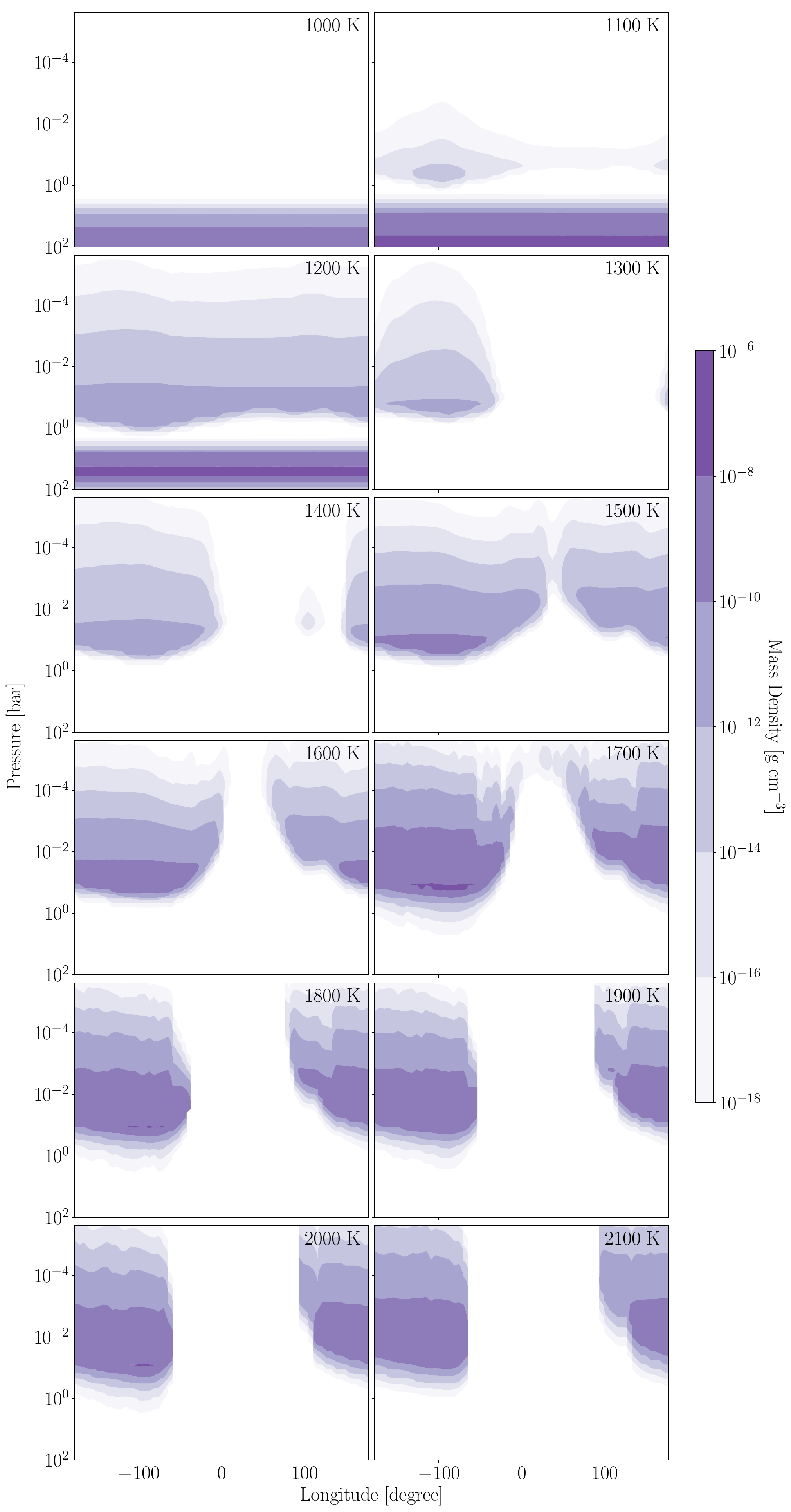}
   \caption{The same as Figure \ref{fig:nom_mg} but for the 1D case without horizontal advection.}
   \label{1D_mg}
\end{figure*}

\begin{figure*} 
   \centering
   \includegraphics[width=0.65\textwidth]{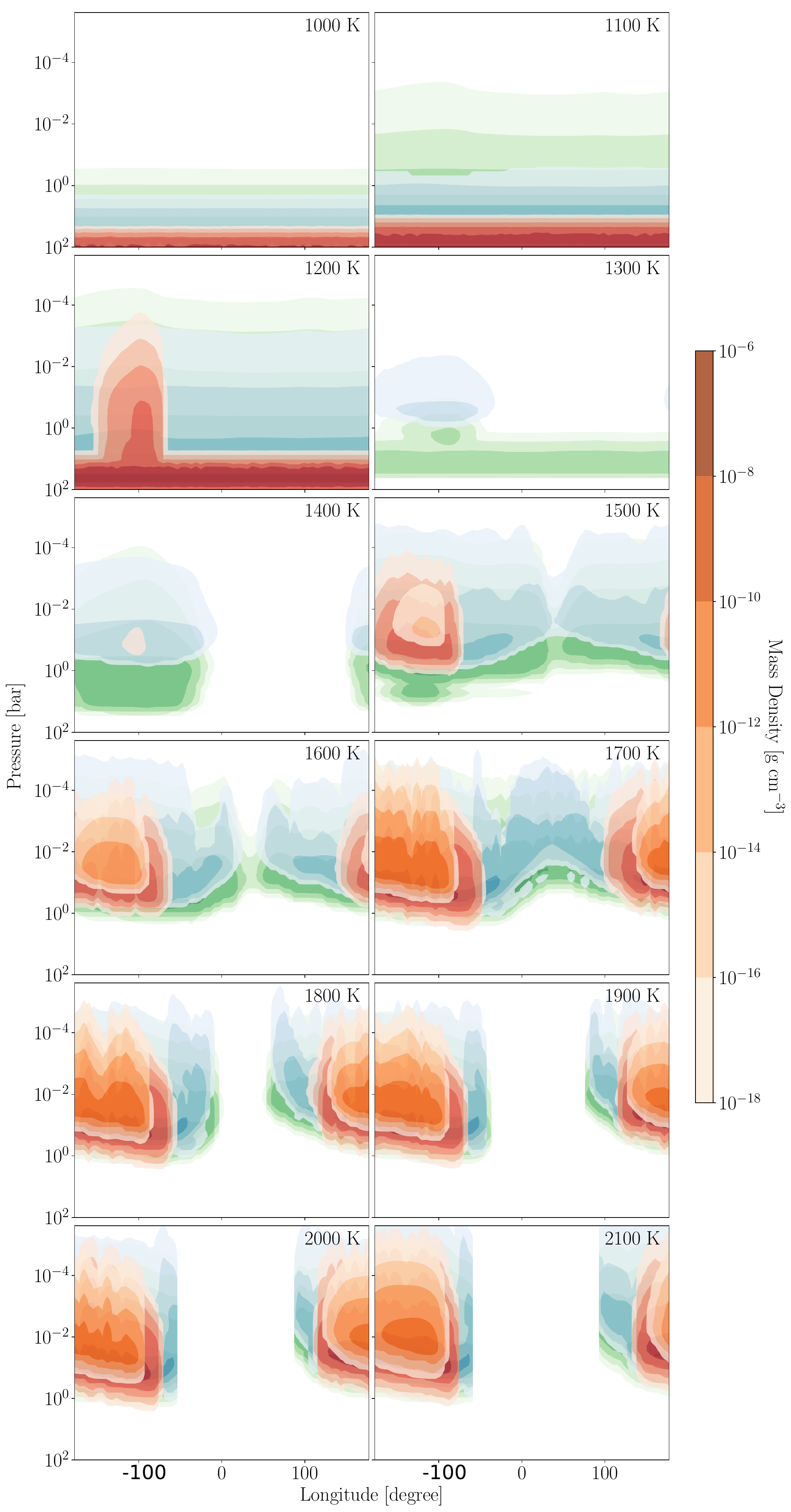}
   \caption{The same as Figure \ref{fig:nom_all} but for the 1D case without horizontal advection.}
   \label{1D_all}
\end{figure*}

Here we present the cloud mass distributions for the same models as in our nominal case but without horizontal transport in the model. The cloud mass distribution of silicate clouds is shown in Figure \ref{1D_mg} and the cloud mass distribution of the remaining cloud species is shown in Figure \ref{1D_all}. These 1D cases and their differences from the nominal 2D case presented in this work are discussed in more detail in Section \ref{horiz_advec}. 

\section{Depletion of Condensible Gas Species}
\begin{figure*} 
   \centering
   \includegraphics[width=0.65\textwidth]{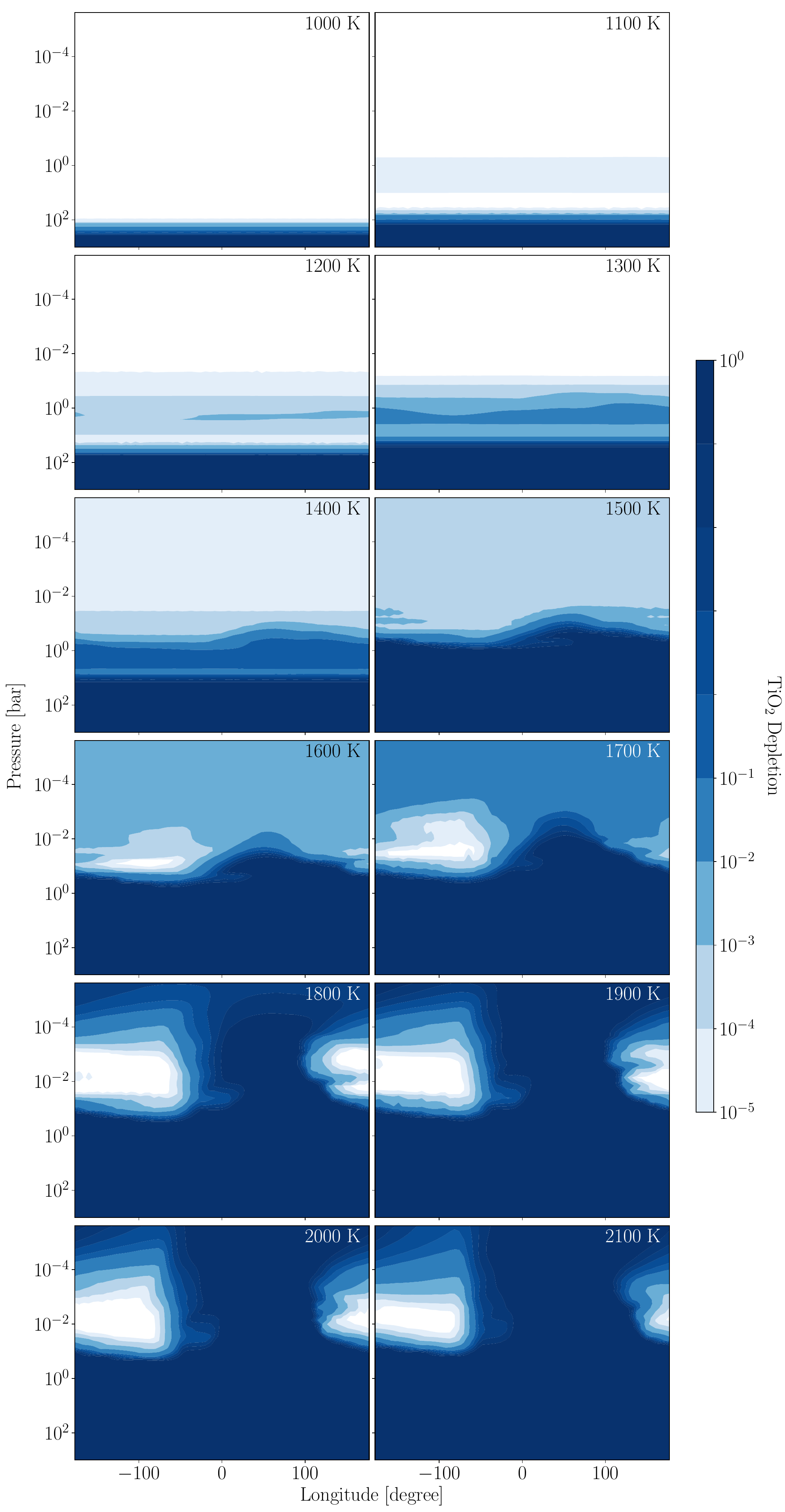}
   \caption{Same as Figure \ref{fig:FeDeplete} but for TiO$_2$ gas. }
   \label{fig:TiDeplete}
\end{figure*}

\begin{figure*} 
   \centering
   \includegraphics[width=0.65\textwidth]{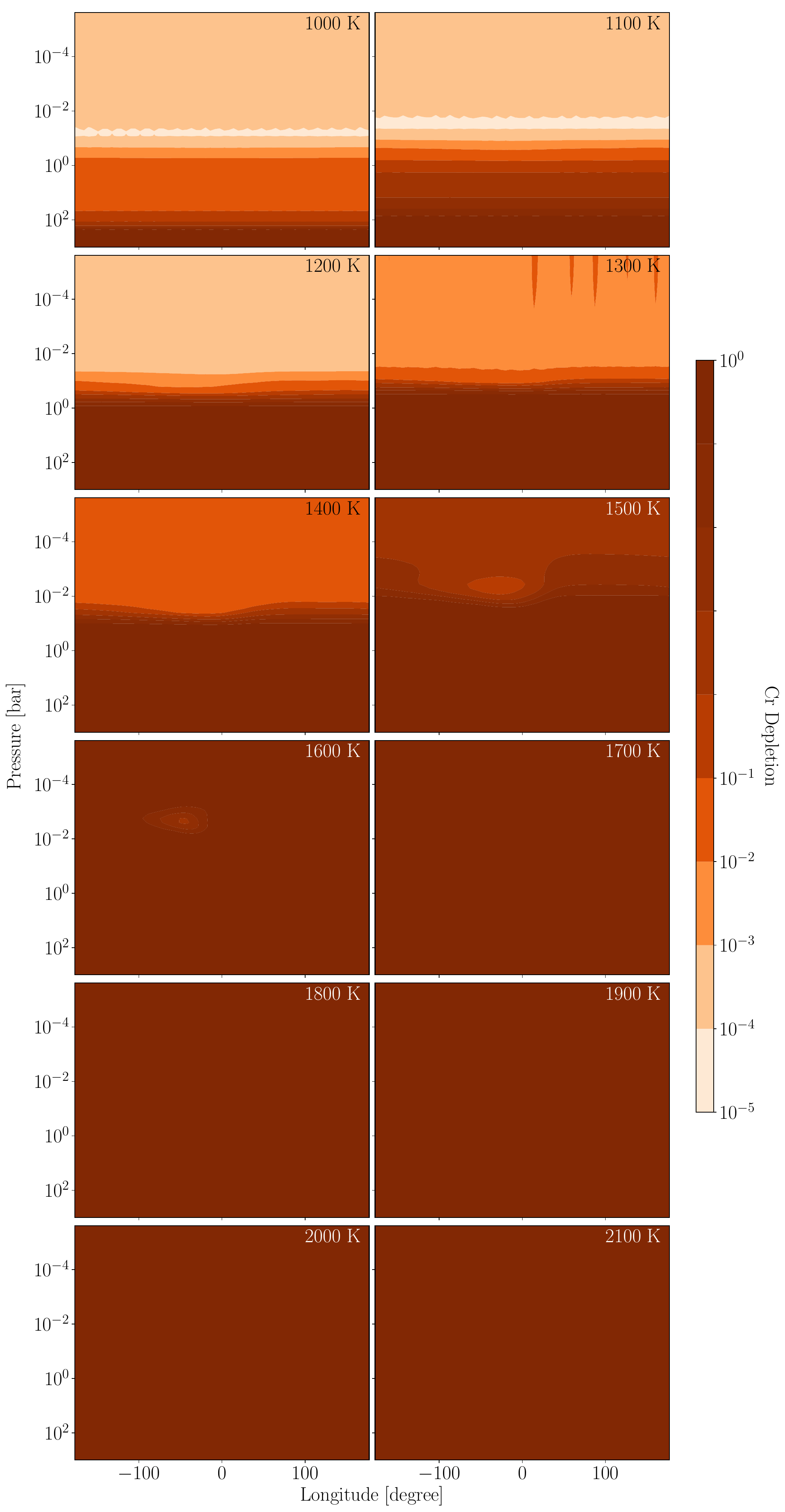}
   \caption{Same as Figure \ref{fig:FeDeplete} but for Cr gas.}
   \label{fig:CrDeplete}
\end{figure*}

\begin{figure*} 
   \centering
   \includegraphics[width=0.65\textwidth]{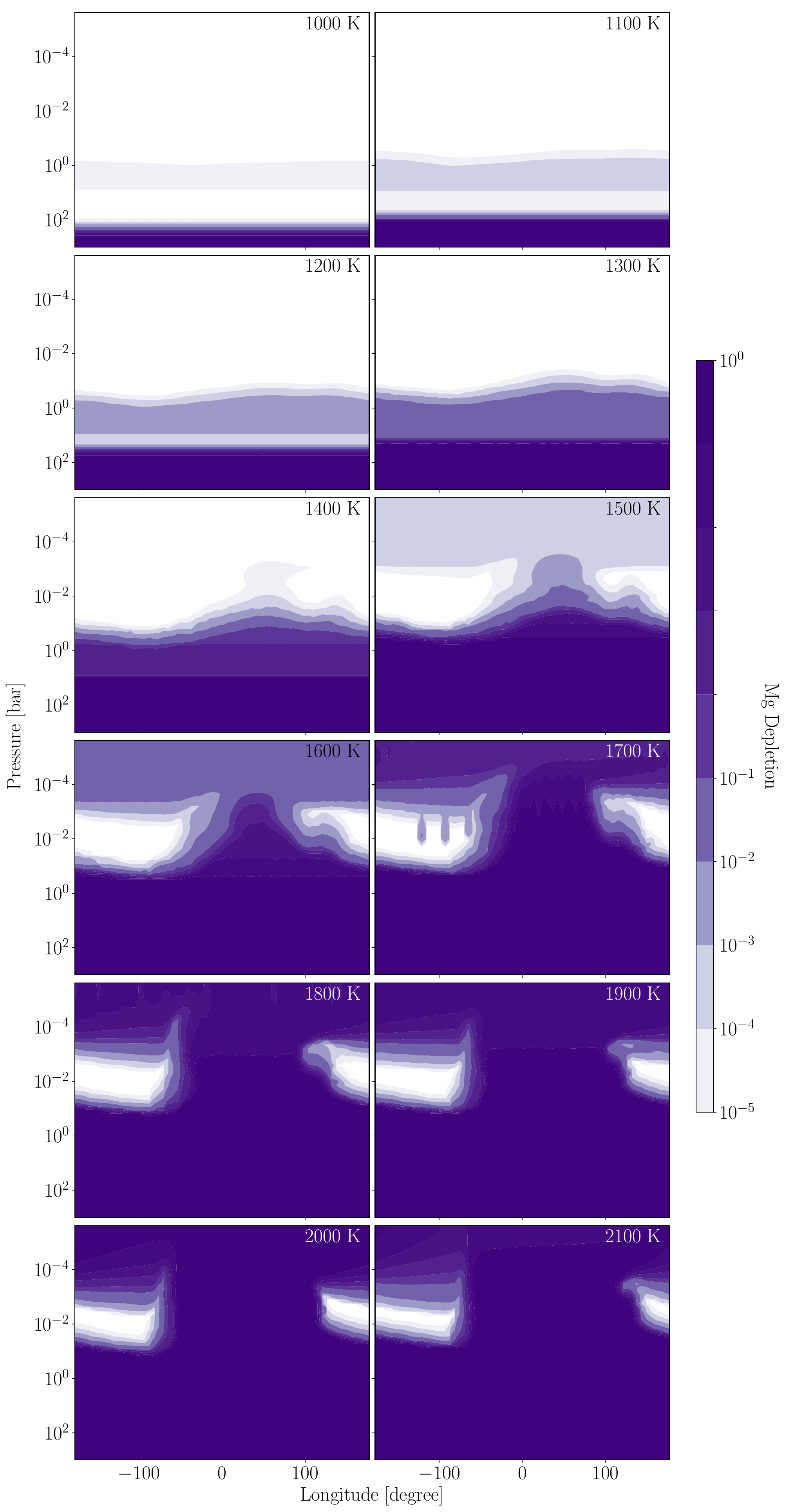}
   \caption{Same as Figure \ref{fig:FeDeplete} but for Mg gas.}
   \label{fig:MgDeplete}
\end{figure*}

\begin{figure*} 
   \centering
   \includegraphics[width=0.65\textwidth]{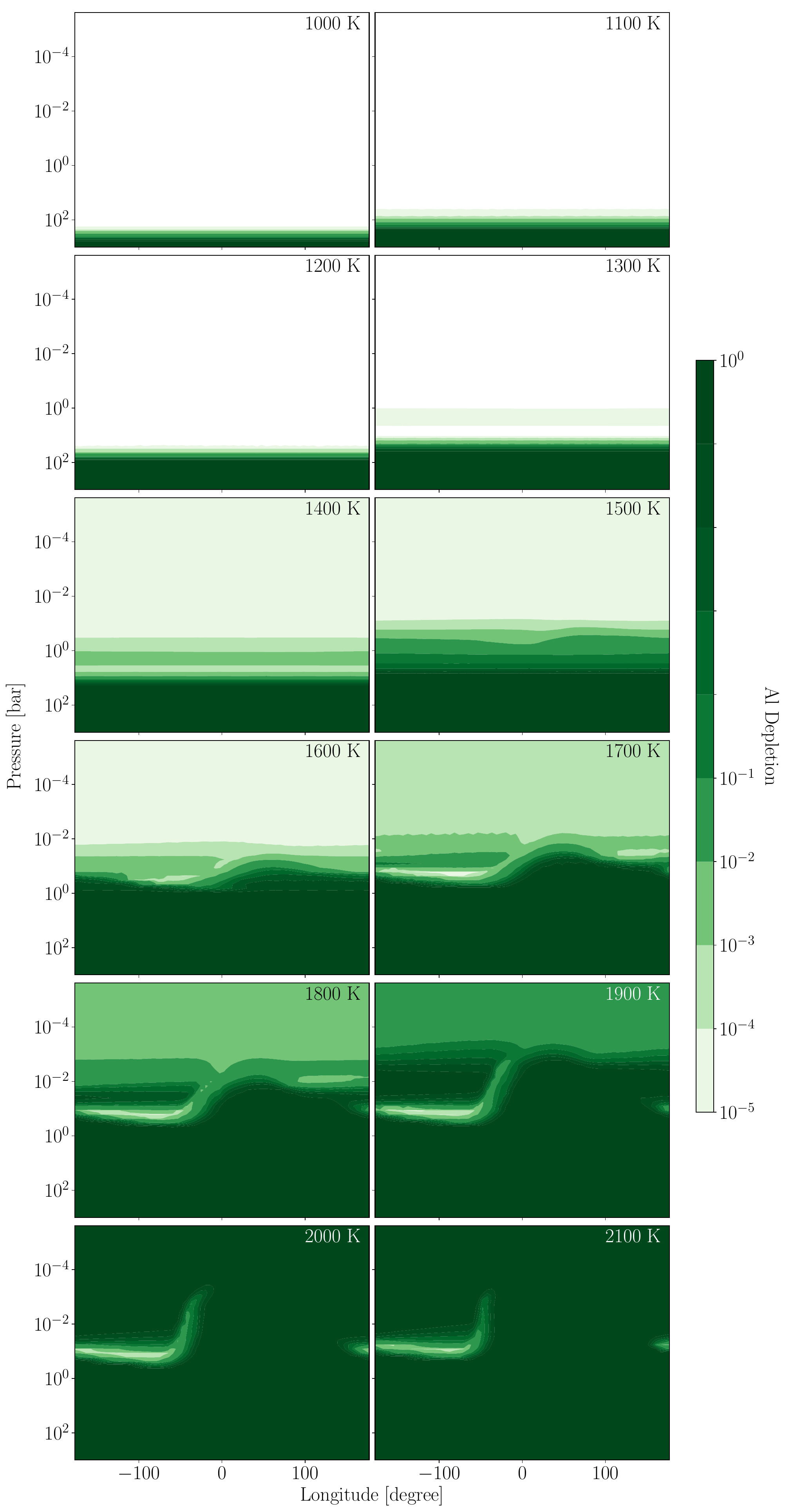}
   \caption{Same as Figure \ref{fig:FeDeplete} but for Al gas.}
   \label{fig:AlDeplete}
\end{figure*}

Here we present the depletion of condensible gas species as a function of planetary longitude and atmospheric pressure for each individual gaseous species. The depletion of various gas constitutents is shown in Figures \ref{fig:TiDeplete} - \ref{fig:AlDeplete}. 

\end{document}